\documentclass[12pt]{article}
\usepackage{amsfonts}
\usepackage{amsmath}
\usepackage{extarrows}
\usepackage{bbm}

\usepackage{mathrsfs}
\title{\textbf{Trigonal curves and algebro-geometric solutions 
to soliton hierarchies 
}}
\date{}
\author{
Wen-Xiu Ma\thanks{E-mail address: mawx@cas.usf.edu}
\\
{\small
Department of Mathematics and Statistics, University of South Florida, }\\
{\small Tampa, FL 33620-5700,
USA}
}
\begin{document} \maketitle

\newtheorem{prop}{Proposition}[section]
\newtheorem{lem}{Lemma}[section]
\newtheorem{thm}{Theorem}[section]
\newtheorem{defn}{Definition}[section]

\numberwithin{equation}{section}

\setlength{\baselineskip}{18pt}
\def \part {\partial}
\def \be {\begin{equation}}
\def \ee {\end{equation}}
\def \bea {\begin{eqnarray}}
\def \eea {\end{eqnarray}}
\def \ba {\begin{array}}
\def \ea {\end{array}}
\def \si {\sigma}
\def \al {\alpha}
\def \la {\lambda}
\def \D {\displaystyle}

\newcommand{\ju}[4]{\mbox{$
  \left[\begin{array}{cc}{#1} & {#2}\vspace{2mm} \\{#3} &{#4}  \end{array}\right]$}}

\begin{abstract}
Using linear combinations of 
Lax matrices of soliton hierarchies, 
we introduce trigonal curves by 
their characteristic equations, and determine Dubrovin type equations  
for zeros and poles of meromorphic functions defined as ratios of 
the Baker-Akhiezer functions.
We straighten out all
 flows in soliton hierarchies under the Abel-Jacobi coordinates 
associated with Lax pairs, 
and generate algebro-geometric solutions to soliton hierarchies in terms of the Riemann theta functions, through observing asymptotic behaviors of the Baker-Akhiezer functions.   
We analyze the four-component AKNS soliton hierarchy in such a way  
that it leads to a general theory of trigonal curves applicable to construction of 
algebro-geometric solutions of an arbitrary soliton hierarchy.


\end{abstract}

\section{Introduction}

Algebro-geometric solutions to soliton
equations are
one important class of 
exact solutions, which describe periodic and quasi-periodic nonlinear phenomena in 
physical and engineering sciences
\cite{NovikovMPZ-book1984,BolokolosBEIM-book1994,GesztesyH-book2003}.
With the development of solitons and 
finite-gap solutions to the Korteweg-de Vries equation,
the mathematical theory of 
algebro-geometric solutions has been 
systematically 
developed since the early 1970s, 
particularly for
the Korteweg-de Vries, modified Korteweg-de Vries, nonlinear Schr\"odinger, sine-Gordon, 
Kadomtsev-Petviashvili, 
Toda lattice
and Camassa-Holm equations (see, e.g., \cite{NovikovMPZ-book1984}-\cite{EilbeckEH-RSLPSA2003}).

Dubrovin and Krichever 
proposed
a systematic method of 
algebraic geometry
to integration of nonlinear partial differential equations, which aims at 
constructing periodic and almost periodic 
solutions in terms of the Riemann theta functions for well-known integrable equations including 
the Korteweg-de Vries equation and the Kadomtsev-Petviashvili 
equation \cite{Krichever-SMD1976,Krichever-FAA1977,Dubrovin-RMS1981}. Cao and Geng 
 made use
of the nonlinearization technique of Lax pairs to generate algebro-geometric solutions
of finite-dimensional integrable Hamiltonian systems and combined systems from lower-dimensions to higher-dimensions \cite{CaoWG-JMP1999,CaoGW-JPA1999,GengW-JMP1999}, 
and later, the nonlinearization technique was applied to constructing algebro-geometric solutions of a great number
of soliton equations in both (1+1)- and (2+1)-dimensions (see, e.g., 
\cite{Zhou-JMP1997}-\cite{HouFQW-JNS2015}). 
Gesztesy et al. established an alternative approach for constructing quasi-periodic solutions 
to soliton hierarchies associated with 
$2\times 2$ matrix spectral problems \cite{GesztesyH-book2003,GesztesyR-RMP1998,GesztesyHMT-book2008}, and by this approach, quasi-periodic solutions to many continuous and discrete soliton hierarchies have been constructed within a different kind of formulation using the Riemann theta functions (see, e.g., \cite{GesztesyR-RMP1998,DicksonGU-MN1999,GesztesyH-RMI2003}).

The study of algebro-geometric solutions has opened up
 a new vista in the analysis of nonlinear partial differential equations.
The adopted algebro-geometric techniques 
brought innovative ideas and led to inspiring results in soliton theory as well as algebraic geometry, for example, a solution of the Riemann-Schottky problem \cite{GesztesyH-book2003,Krichever-RMS2008}.
The successful idea in constructing algebro-geometric
solutions is to employ the theory
of algebraic curves associated with Lax pairs 
producing soliton hierarchies to represent the Baker-Akhiezer functions 
\cite{Baker-PRSLA1928,Akhiezer-SMD1961}
 in terms of the Riemann theta function \cite{ItsM-TMP1976,ItsM-FAA1975}. The obtained algebro-geometric solutions satisfy 
a class of stationary counterparts of soliton equations, called Novikov type equations 
\cite{Novikov-FAA1975}.  
It is also noticed 
that symmetry constraints pay a way of separation of variables for soliton equations and the corresponding canonical variables solving the associated Jacobi inversion problems provide the so-called 
characteristic variables in the Riemann theta function presentation of algebro-geometric solutions \cite{ZengM-JMP1999,MaZ-ANZIAM2002}.  
There are primarily two types of research on algebro-geometric solutions.
One is to
explore asymptotics of 
the Baker-Akhiezer functions
to construct algebro-geometric solutions to given nonlinear equations, and the other is to connect the Baker-Akhiezer functions possessing given asymptotics with potential nonlinear equations and their algebro-geometric solutions. 

Very recently, Geng et al. successfully attempted a few $3\times 3 $ matrix spectral problems and constructed algebro-geometric solutions to the associated soliton hierarchies, including the modified Boussinesq hierarchy, the Kaup-Kupershmidt hierarchy and the hierarchy of three-wave resonant interaction equations (see, e.g., 
\cite{GengWH-PD2011}-\cite{GengZD-AM2014}). 
In this paper, we would like to 
propose a general framework to
analyze $3\times 3 $ matrix spectral problems and their corresponding trigonal curves, and to generate algebro-geometric solutions to soliton hierarchies by observing asymptotic behaviors of the 
Baker-Akhierzer functions.
We 
analyze
the four-component AKNS soliton hierarchy, 
particularly asymptotics of the Baker-Akhiezer functions, 
in such a way  
that it yields a general theory applicable to soliton hierarchies associated with arbitrary $3\times 3 $ matrix spectral problems. 

The rest of the paper is structured as follows.
In Section \ref{sec:sh:pma-quasiperiodicsols}, with the aid of
the zero-curvature formulation and the trace identity, we rederive the
four-component AKNS soliton hierarchy and its bi-Hamiltonian structure. 
In Section \ref{sec:BAfunctions:pma-quasiperiodicsols},
we introduce
a class of trigonal curves by taking linear combinations of the Lax matrices
and analyze the corresponding Baker-Akhiezer functions.
In Section \ref{sec:Dubrovineqns:pma-quasiperiodicsols}, we first
present a general structure of 
Dubrovin type dynamical equations \cite{Dubrovin-FAA1976} of zeros and poles of meromorphic functions as the characteristic variables, and then apply the resulting general theorems to the 
four-component AKNS case.
In Section \ref{sec:AsymptoticBehaviors:pma-quasiperiodicsols}, 
we explore asymptotic behaviors for the three Baker-Akhiezer functions
in the 
four-component AKNS case
 at the points at infinity.
In Section \ref{sec:Algebro-geometricSolutions:pma-quasiperiodicsols},
we 
straighten out all the flows of the four-component AKNS soliton hierarchy under the
Abel-Jacobi coordinates, and  
construct
algebro-geometric solutions of the whole soliton 
hierarchy by use of
the Riemann theta functions according to the asymptotic properties
 of the Baker-Akhiezer functions.
In the last section, we present a few concluding remarks and open questions related to lump solitons and soliton hierarchies.

\section{Four-component AKNS soliton hierarchy}

\label{sec:sh:pma-quasiperiodicsols}

\subsection{Soliton hierarchy}

Let us recall the zero curvature formulation and 
the trace identity 
 \cite{Tu-JPA1989}.
Let $U=U(u,\lambda)$ be
a square spectral matrix belonging to a given matrix loop algebra, where 
$u$ is a potential and $\lambda $ is a spectral parameter.  
Assume that 
\be 
W=W(u,\lambda)=\sum_{k=0}^\infty W_k\lambda ^{-k}=\sum_{k=0}^\infty W_k(u)\lambda ^{-k}
\ee
solves the corresponding stationary zero curvature equation
\be W_x=[U,W]. 
\label{eq:szce:pma-quasiperiodicsols}
 \ee
Then introduce a series of Lax matrices
\be V^{[r]}=V^{[r]}(u,\lambda )= 
(\lambda ^r W)_+ +\Delta _r, 
\ee
where the subscript $+$ denotes the operation of taking a polynomial part in $\lambda$
and $\Delta _r$, $r\ge 0$, are appropriate modification terms, 
such that 
a soliton hierarchy
\be u_{t_r}=K_r(u)=K_r(x,t,u,u_x,\cdots),\ r\ge 0,  \label{eq:gsh:pma-quasiperiodicsols} \ee
 can be generated from a series of 
zero curvature equations
\begin{equation}
  U_{t_r} - V^{[r]}_x + [  U,  V^{[r]}] = 0,\ r\ge 0. \label{eq:gCZCEs:pma-quasiperiodicsols} \ee
The two matrices $U$ and $V^{[r]}$ 
are
called a Lax pair \cite{Lax-CPAM1968} of the $r$-th soliton equation 
in the hierarchy \eqref{eq:gsh:pma-quasiperiodicsols}.
The zero curvature equations in 
\eqref{eq:gCZCEs:pma-quasiperiodicsols} are the compatibility conditions of the spatial 
and temporal spectral problems
\be \psi_x=U\psi=U(u,\lambda )\psi,\ \psi_{t}=V^{[r]}\psi=V^{[r]}(u,\lambda )\psi, \ r\ge 0,
\label{eq:spofsh:pma-quasiperiodicsols}
\ee
where $\psi$ is the vector eigenfunction. 

One important task in soliton theory is
to show the Liouville integrability of soliton equations in a hierarchy. 
This 
can be usually achieved 
by establishing a bi-Hamiltonian formulation \cite{Magri-JMP1978}:
\be 
u_{t_r}=K_r=J\frac {\delta \tilde{H}_{r+1}}{\delta u}=
M\frac  {\delta \tilde{H}_{r}}{\delta u},\ r\ge 1,
\ee
where $J$ and $M$ constitute a Hamiltonian pair and $\frac {\delta}{\delta u}$
denotes the variational derivative (see, e.g., \cite{MaF-CSF1996}).  
The Hamiltonian structures can be furnished through the trace identity 
\cite{Tu-JPA1989}:
\be 
 \frac{\delta}{\delta u}\int \textrm{tr}(W\frac{\partial U}{\partial  \lambda})dx=
\lambda^{-\gamma}\frac{\partial}{\partial \lambda}\Bigl[\lambda^\gamma \textrm{tr}
(W\frac{\partial U}{\partial u})\Bigr],\ 
\gamma= -\frac {\lambda }2 \frac {d}{d\lambda }\ln |\textrm{tr}(W^2) |,
\ee 
or more generally,  
the variational identity \cite{MaC-JPA2006}:
\be 
 \frac{\delta}{\delta u}\int \langle W,\frac{\partial U}{\partial  \lambda}\rangle dx=
\lambda^{-\gamma}\frac{\partial}{\partial \lambda}\Bigl[\lambda^\gamma \langle
W,\frac{\partial U}{\partial u}\rangle\Bigr],\ 
\gamma= -\frac {\lambda }2 \frac {d}{d\lambda }\ln |\langle W,W \rangle |,
\ee 
where $\langle\cdot,\cdot\rangle$ is a non-degenerate, symmetric and ad-invariant bilinear form on the underlying matrix loop algebra \cite{Ma-NA2009}. 
The bi-Hamiltonian formulation guarantees 
the commutativity of infinitely many Lie symmetries $\{K_n\}_{n=0}^\infty$ and conserved quantities
$\{\tilde {H}_n\}_{n=0}^\infty$:
\bea &&
[K_{n_1},K_{n_2}]=K_{n_1}'[K_{n_2}]-K_{n_2}'[K_{n_1}]=0,
\\
&&
\{\tilde {\cal H}_{n_1}, \tilde {\cal H}_{n_2}\}_N=
\int \Bigl(\frac {\delta \tilde {\cal H}_{n_1} }{\delta u}\Bigr)^T N
\frac {\delta \tilde {\cal H}_{n_2} }{\delta u}
\, dx=0,
\eea   
where 
$n_1,n_2\ge 0$, $N=J$ or $M$, and 
$K'$ denotes the Gateaux derivative of $K$:
\be K'(u)[S]=\frac {\partial }{\partial \varepsilon }\Bigl.\Bigr|_{\varepsilon =0}K(u+\varepsilon 
S,u_x+\varepsilon S_x,\cdots).
 \ee 

It is known that for an evolution equation $u_t=K(u)$, $\tilde {H}=\int H\, dx$ is a conserved functional iff 
$\frac {\delta \tilde {H}}{\delta u}$ is an adjoint symmetry \cite{MaZ-JNMP2002}, and so, the Hamiltonian structures 
links conserved functionals to adjoint symmetries and further symmetries. When the underlying matrix loop algebra in the zero curvature formulation is
simple, 
the associated zero curvature equations yield classical soliton hierarchies \cite{DrinfelʹdS-DANSSSR1981}; when semisimple, the associated zero curvature equations yield a collection of different soliton hierarchies; and
when
non-semisimple, we obtain a hierarchy of integrable couplings \cite{MaXZ-PLA2006}, which needs extra care in constructing 
exact solutions.  

\subsection{Four-component AKNS hierarchy}

Let us consider a $3\times 3$ matrix spectral problem
\be \psi_x =U\psi=U(u,\lambda)\psi,\ U=(U_{ij})_{3\times 3}=\left[\ba 
{ccc}
-2 \lambda & p_1&p_2\vspace{2mm}\\
q_1& \lambda & 0
\vspace{2mm}\\
q_2& 0&\lambda 
\ea \right],\ \psi=\left[\ba {c} \psi_1\vspace{2mm}\\ \psi_2 \vspace{2mm}\\
\psi_3 \ea \right],
\label{eq:sp:pma-quasiperiodicsols}
\ee
where $\lambda $ is a spectral parameter and $u$ is a four-component potential
\be   
u=(p,q^T)^T,\ p=(p_1,p_2),\ q=(q_1,q_2)^T.
\ee
Since $U_0=\textrm{diag}(-2,1,1)$
has a multiple eigenvalue, 
the spectral problem 
\eqref{eq:sp:pma-quasiperiodicsols} is degenerate.
Under the special reduction of $p_2=q_2=0$,
\eqref{eq:sp:pma-quasiperiodicsols} is equivalent to 
the AKNS spectral problem 
\cite{AblowitzKNS-SAM1974}, and thus 
it
is called a four-component AKNS spectral problem.

To derive the associated soliton hierarchy, 
we first solve the stationary zero curvature equation 
\eqref{eq:szce:pma-quasiperiodicsols}
corresponding to \eqref{eq:sp:pma-quasiperiodicsols}.
We suppose that a solution $W$ is given by 
\begin{equation}
W=
\left[\begin{array}{cc}
a&b \vspace{2mm}\\
c&d
\end{array}\right], \end{equation}
where $a$ is a scalar,
 $ b^T$ and $  c$ are two-dimensional columns, and $d$ is a $2 \times 2 $ matrix. 
Then the stationary zero curvature equation \eqref{eq:szce:pma-quasiperiodicsols} becomes
\begin{equation} 
a_x=pc-bq,\ b_x=-3\lambda b+pd-ap,\ 
c_x=3\lambda c+qa-dq,\ d_x=qb-cp.
\label{eq:equivalentformofV_x=[U,V]in4cAKNS:pma-quasiperiodicsols:pma-quasiperiodicsols}
\end{equation} 
We seek a formal series solution as
\begin{equation} 
W=
\left[\begin{array}{cc}
a&b \vspace{2mm}\\
c&d
\end{array}\right]=
\sum_{k=0}^\infty W_k\la^{-k},\ 
W_k=W_k(u)=
\left[\begin{array}{cc}
a^{[k]} &b^{[k]} \vspace{2mm}\\
c^{[k]}&d^{[k]}
\end{array}\right] ,\ k\ge 0,
\end{equation}
with $b^{[k]},c^{[k]}$ and $d^{[k]}$ being assumed to be 
\begin{equation}
b^{[k]}=(b^{[k]}_1,b^{[k]}_2 ),\
c^{[k]}=(c^{[k]}_1,c^{[k]}_2 )^T,\
d^{[k]}=(d^{[k]}_{ij})_{2 \times 2 },\ k\ge 0.
\end{equation}
Thus, the system (\ref{eq:equivalentformofV_x=[U,V]in4cAKNS:pma-quasiperiodicsols:pma-quasiperiodicsols}) 
equivalently leads to the following recursion relations:
\begin{subequations} \label{eq:rrin4cAKNS:pma-quasiperiodicsols}
\begin{gather}
b^{[0]}=0, \ c^{[0]}=0,\ a^{[0]}_x=0,\ d^{[0]}_x=0,\vspace{2mm}\\
b^{[k+1]}= \D \frac 1{3}( -b^{[k]}_x +pd^{[k]} -a^{[k]}p),\ k\ge 0,
\label{eq:rrforb^kof4cAKNS:pma-quasiperiodicsols}
\vspace{2mm}\\
c^{[k+1]}=\D  \frac 1{3}( c^{[k]}_x -qa^{[k]} +d^{[k]}q),\ k\ge 0,
\label{eq:rrforc^kof4cAKNS:pma-quasiperiodicsols}
\vspace{2mm}\\
a^{[k]}_x=pc^{[k]}-b^{[k]}q,\ d_x^{[k]}=qb^{[k]}-c^{[k]}p, \ k\ge 1 .
\label{eq:rrfora^kd^kof4cAKNS:pma-quasiperiodicsols}
\end{gather}
\end{subequations}

We choose the initial values as follows
\begin{equation}
a^{[0]}=-2 ,\  d^{[0]}=I_2,
\label{eq:initialvaluesof4cAKNS:pma-quasiperiodicsols}
\end{equation} 
where $I_2=\textrm{diag}(1,1)$,
and take constants of integration in \eqref{eq:rrfora^kd^kof4cAKNS:pma-quasiperiodicsols}
to be zero: 
\begin{equation} 
 W_k|_{u=0}=0,\  k\geq 1.
\label{eq:requirementofuniquenessin4cAKNS:pma-quasiperiodicsols}
\end{equation} 
Therefore, 
with $ 
a^{[0]}$ and $  d^{[0]}$
given by \eqref{eq:initialvaluesof4cAKNS:pma-quasiperiodicsols},
all matrices $W_k ,\ k\ge 1$, will be uniquely determined. 
For example, it follows from \eqref{eq:rrin4cAKNS:pma-quasiperiodicsols}
 that
\begin{subequations} 
\begin{gather}
b^{[1]}_i=p_i,\  c^{[1]}_i=q_i,\ a^{[1]}=0,\ 
d^{[1]}_{ij}=0;\notag \\
 b_i^{[2]}=-\frac{1}{3 }p_{i,x},\  c_i^{[2]}=\frac{1}{3 }q_{i,x},\ 
a^{[2]}=\frac{1}{3 }(  p_1q_1+p_2q_2),\ 
d^{[2]}_{ij}=-\frac{1}{3 }p_jq_i;
\notag \\
b_i^{[3]}=\frac{1}{9}[p_{i,xx}-2
(  p_1q_1+p_2q_2)p_i],\
c_i^{[3]}=\frac{1}{9}[q_{i,xx}-2(  p_1q_1+p_2q_2)q_i],
\notag\\
a^{[3]}=\frac{1}{9 }(  p_1q_{1,x}-
p_{1,x}q_{1}+
p_2q_{2,x}-
p_{2,x}q_{2}
),\ d^{[3]}_{ij}=
\frac{1}{9 }(  
p_{j,x}q_{i}-p_jq_{i,x} )
;\notag
\\
b_i^{[4]}=
-\frac 1{27}
[
p_{i,xxx}-
3(p_{1} q_{1}+p_2q_2)  p_{i,x}  -3(p_{1,x} q_{1}+p_{2,x}q_2)p_i
]
,
\notag
\\
c_i^{[4]}=
\frac 1{27}
[
q_{i,xxx}-
3(p_{1} q_{1}+p_2q_2)  q_{i,x}  -3(p_{1} q_{1,x}+p_{2}q_{2,x})q_i
]
,
\notag
\\
a^{[4]}=
-\frac 1{27}[
3(p_{1}  q_{1} +p_2 q_{2})^2
-p_{1}  q_{1,xx}  +
 p_{1,x}  
q_{1,x}  - p_{1,xx} q_{1}  
-p_{2}  q_{2,xx}  
+
p_{2,x} q_{2,x}
   -
 p_{2,xx} q_{2}]
,
\notag
\\
d_{ij}^{[4]}=
\frac 1{27}
[
3p_{j} (p_1q_1+ p_{2}  q_{2}) q_{i} 
-p_{
j,xx} q_i + p_{j,x}
  q_{i,x} 
 - p_jq_{i,xx}
  ]
;\notag
\end{gather}
\end{subequations}
where $1\le i,j\le 2$.
Based on \eqref{eq:rrfora^kd^kof4cAKNS:pma-quasiperiodicsols},
we can obtain,
from \eqref{eq:rrforb^kof4cAKNS:pma-quasiperiodicsols} and \eqref{eq:rrforc^kof4cAKNS:pma-quasiperiodicsols},
 a recursion relation for 
$b^{[k]}$ and $c^{[k]}$:
\begin{equation} 
 \left[ \begin{array}{c}c^{[k+1]}\vspace{2mm}
\\ b^{[k+1]T}\end{array}\right]
=\Psi  \left[ \begin{array}{c}c^{[k]}\vspace{2mm} \\ b^{[k]T}
\end{array}\right],\ k\ge 1,
\end{equation} 
where $\Psi $ is a $4\times 4$ matrix operator 
\begin{equation} 
\Psi =\frac{1}{3 }\ju{(\partial-\sum\limits_{i=1}^2 q_i
\partial^{-
1}p_i)I_2  -q\partial^{-1}p}{q\partial^{-1}q^T+(q\partial^{-1}q^T)^T}
{-p^T\partial^{-1}p-(p^T\partial^{-1}p)^T}
{(-\partial+\sum\limits_{i=1}^2 p_i\partial^{-1}q_i)I_2 +p^T
\partial^{-1}q^T} .
\end{equation} 

As usual, for all integers $r\ge 0$, 
we introduce
the following Lax matrices 
\be V^{[r]}=V^{[r]}(u,\lambda)=(V_{ij}^{[r]})_{3 \times 3}=
(\lambda ^rW)_+=\sum_{k=0}^r W_k\lambda ^{r-k} ,\ r\ge 0,
\label{eq:defofV_{r}:pma-quasiperiodicsols}
\ee
where the modification terms are taking as zero.
Note that we have  
\be V^{[r+1]}=\sum_{k=0}^{r+1}W_k\lambda ^{r-k+1}=
\lambda \sum_{k=0}^{r+1}W_k
\lambda ^{r-k}
=\lambda V^{[r]}+W_{r+1} ,\ r\ge 0.
\label{eq:recursionrelationofV_{r}:pma-quasiperiodicsols}
\ee
The compatibility conditions of
\eqref{eq:spofsh:pma-quasiperiodicsols}, i.e., the zero curvature equations
\eqref{eq:gCZCEs:pma-quasiperiodicsols}, 
generate 
the four-component AKNS soliton hierarchy
\be 
u_{t_r}=
\left [\begin{array}{l}p^T\vspace{2mm} \\ q
\end{array}\right] _{t_r}
=K_r=
\left[\begin{array}{c}-3b^{[r+1]T}\vspace{2mm} \\
3c^{[r+1]}
\end{array}\right]
,\ r\ge 0.
\label{eq:4cAKNSsh:pma-quasiperiodicsols}
\ee
The first two nonlinear
systems in this soliton hierarchy \eqref{eq:4cAKNSsh:pma-quasiperiodicsols}
read
\begin{subequations}
\begin{gather}
p_{i,t_2}=-\frac{1}{3 }[p_{i,xx}-2(p_1q_1+p_2q_2)p_i], \ 1\le i\le 2,\\
q_{i,t_2}=\frac{1}{3 }[q_{i,xx}-2(p_1q_1+p_2q_2)q_i],\ 1\le i\le 2,
\end{gather}
\label{eq:NLSof4cAKNS:pma-quasiperiodicsols}\end{subequations}
and 
\begin{subequations}
\begin{gather}
p_{i,t_3}=
\frac 1{9}
[
p_{i,xxx}-
3(p_{1} q_{1}+p_2q_2)  p_{i,x}  -3(p_{1,x} q_{1}+p_{2,x}q_2)p_i
],\ 1\le i\le 2,\\
q_{i,t_3}=
\frac 1{9}
[
q_{i,xxx}-
3(p_{1} q_{1}+p_2q_2)  q_{i,x}  -3(p_{1} q_{1,x}+p_{2}q_{2,x})q_i
], \ 1\le i\le 2,
\end{gather}
\label{eq:MKdVof4cAKNS:pma-quasiperiodicsols}\end{subequations}
which are the four-component versions of the AKNS 
systems of nonlinear Schr\"odinger equations and modified Korteweg-de Vries equations, respectively.
The four-component AKNS equations \eqref{eq:NLSof4cAKNS:pma-quasiperiodicsols} can be reduced to the Manokov system \cite{Manakov-SPJETP1974}, for which a decomposition into finite-dimensional integrable Hamiltonian systems was given in \cite{ChenZ-CAM2012}, whileas
the four-component AKNS equations \eqref{eq:MKdVof4cAKNS:pma-quasiperiodicsols} contain various mKdV equations, for which there exist different kinds of integrable decompositions (see, e.g., \cite{Ma-PA1995,YuZ-PLA2006}).  

We point out that the four-component AKNS soliton hierarchy \eqref{eq:4cAKNSsh:pma-quasiperiodicsols}
has a Hamiltonian structure \cite{MaZ-JNMP2002},
which can be generated through the trace identity \cite{Tu-JPA1989},
or more generally, the variational identity \cite{MaC-JPA2006}.
Actually, we have 
$$ \textrm{tr}(W\frac{\partial U}{\partial \lambda})
=-2a +\textrm{tr}(d)
=\sum_{k= 0}^\infty (-2a^{[k]}+
d_{11}^{[k]}+d_{22}^{[k]})\lambda ^{-k},$$
 and
 $$ \textrm{tr}(W\frac{\partial U}{\partial u})
=\left[\begin{array}{c}c\vspace{2mm} \\
b^{T}\end{array}\right]
=\sum_{k\ge 0}G_{k-1}\lambda^{-k}.$$
 Inserting these expressions 
into the trace identity and considering the case of $k=2$,
 we get $\gamma=0$ and thus we have
  \begin{equation} \frac{\delta \tilde {H}_{k}}{\delta u}=G_{k-1}, \
  \tilde { H}_k=\frac 1 k \int (2a^{[k+1]}-d_{11}^{[k+1]}-d_{22}^{[k+1]})\,dx,\ 
G_{k-1}= \left[\begin{array}{c}c^{[k]}\vspace{2mm} \\
b^{[k]T}\end{array}\right],\ k\ge 1
. \label{4cAKNS:pma-quasiperiodicsolshh} \end{equation} 
A 
bi-Hamiltonian structure of the four-component AKNS equations
(\ref{eq:4cAKNSsh:pma-quasiperiodicsols}) then follows:
\begin{equation}
u_{t_r}=K_r=JG_r=
J\frac{\delta \tilde {H}_{r+1}}{\delta u}=M\frac{\delta \tilde {H}_r}{\delta u},\ r\ge 1,
\label{eq:biHamiltonianstructureof4cAKNS:pma-quasiperiodicsols}
\end{equation}
where the Hamiltonian pair $(J,M=J\Psi )$ is given by 
\begin{subequations}
\begin{gather} 
J=\left[\begin{array}{cc}0&-3I_2\vspace{2mm}\\
3I_2&0
\end{array}\right] ,
\\ 
M=\ju 
{p^T\part ^{-1}p+(p^T\part ^{-1}p)^T }{(\part -\D 
\sum_{i=1}^2 p_i\part 
^{-1}q_i)I_2-p^T
\part ^{-1}q^T }{(\part -\D \sum_{i=1}^2 p_i
\part ^{-1}q_i)I_2-q\part ^{-1}p}{q\part ^{-1}q^T+(q\part ^{-1}q^T)^T }.
\end{gather}
\end{subequations} 
Adjoint symmetry constraints or equivalently symmetry constraints separate the four-component AKNS equations into two commuting finite-dimensional Liouville integrable Hamiltonian systems \cite{MaZ-JNMP2002}.

\section{Trigonal curves and Baker-Akhiezer functions}

\label{sec:BAfunctions:pma-quasiperiodicsols}

For each integer $n\ge 1$, let us take a linear combination of the Lax matrices
\be 
W^{[n]}=W^{[n]}(u,\lambda)=(W^{[n]}_{ij})_{3\times 3}=
\sum_{k=0}^n \alpha _k V^{[n-k]},
\ee
where the Lax matrices
$V^{[k]},\ 0\le k\le n$, are given by \eqref{eq:defofV_{r}:pma-quasiperiodicsols} and
$\alpha_k,\ 0\le k\le n,$ are arbitrary constants but $\alpha _0\ne 0$.
 Its corresponding characteristic polynomial 
 reads
\be 
{\cal F}_m(\lambda ,y)=
\textrm{det} (yI_3-W^{[n]}) =y^3 +yS_m(\lambda ) -T_m(\lambda) , \ee 
where $I_3=\textrm{diag}(1,1,1)$, $S_m$ and $T_m$ are two polynomials of $\lambda $ with degrees 
$\deg (S_m)=2n$ and $\deg (T_m)=3n$, defined by  
\be 
S_m=\sum_{1\le i< j\le 3} \left | \ba {cc} W_{ii}^{[n]} & W_{ij}^{[n]} \vspace{2mm}\\
W_{ji}^{[n]} & W_{jj}^{[n]} \ea \right |,
\label{eq:defofS_m:pma-quasiperiodicsols} 
\ee 
and 
\be 
T_m=\det W^{[n]}=\left |\ba {ccc} 
W_{11}^{[n]} & W_{12}^{[n]}  &W_{13}^{[n]}  
\vspace{2mm}\\
W_{21}^{[n]} & W_{22}^{[n]}  &W_{23}^{[n]}  
\vspace{2mm}\\
W_{31}^{[n]} & W_{32}^{[n]}  &W_{33}^{[n]}  
\ea \right |,
\ee
and $m=\max (\deg (S_m),\deg (T_m))=3n$.

Using the combined Lax matrix $W^{[n]}$, we introduce a trigonal curve 
 ${\cal K}_g$ of degree $m$ as follows:
\be {\cal K}_g =\{P=(\lambda ,y)\in \mathbb{C}^2\, |\,  \textrm{det} (yI_3-W^{[n]}) =y^3 +yS_m(\lambda ) -T_m(\lambda) =0\}.\ee
Note that the corresponding 
discriminant $\Delta=-27T_m^2-4S_m^3$, a polynomial of $\lambda $ of degree $4n-2$, is 
not zero at infinity,
 and thus, the curve has three non-branch points at infinity
\cite{Mumford-book19834}, which we denote by $ P_{\infty_i},\ 1\le i\le  3$. 
The curve ${\cal K}_g$ 
is 
compactified 
by adding those three points at infinity and its compactification 
is still denoted by ${\cal K}_g$ for the sake of convenience. 
The curve ${\cal K}_g$ is called to be nonsingular, if we have
$(\frac {\partial {\cal F}_m}{\partial \lambda },  
\frac {\partial {\cal F}_m}{\partial y })\ne 0$, 
while ${\cal F}_m(\lambda ,y)=0$. 
When ${\cal K}_g$ is nonsingular, it becomes a three-sheeted Riemann surface of arithmetical 
genus determined by the Riemann-Hurwitz formula:
\be g= \frac f 2 -k +1 = 2n-3,\ee
where $f=4n-2$ is the total multiplicity of 
its branch points and $k=3$ is the number of sheets.  
The compact Riemann surface ${\cal K}_g$ consists of 
points satisfying ${\cal F}_m(\lambda ,y) =0$ and the three points at infinity:
$\{P_{\infty_1},P_{\infty_2},P_{\infty_3}\}$.

For a fixed $\lambda \in \mathbb{C}$, we denote 
the three branches of $y(\lambda )$ satisfying ${\cal F}_m(\lambda ,y) =0$ by $y_i=y_i(\lambda )$, $1\le i\le 3$, and thus, we have 
\be 
(y-y_1(\lambda))(y-y_2(\lambda))(y-y_3(\lambda))=y^3+yS_m-T_m=0,
\ee 
from which we can easily get
\be
\left\{
\ba {l}
y_1+y_2+y_3=0,
\vspace{2mm}\\
y_1y_2+y_1y_3+y_2y_3=S_m,
\vspace{2mm}\\
y_1y_2y_3=T_m,
\vspace{2mm}\\
y_1^2+y_2^2+y_3^2=-2S_m,
\vspace{2mm}\\
y_1^3+y_2^3+y_3^3=3T_m,
\vspace{2mm}\\
(y_1+y_2)y_3^2+(y_2+y_3)y_1^2+(y_3+y_1)y_2^2=-3T_m,
\vspace{2mm}\\
y_1^2y_2^2+y_1^2y_3^2+y_2^2y_3^2=S^2_m,
\vspace{2mm}\\
(3y_1^2+S_m)(3y_2^2+S_m)(3y_3^2+S_m)=-\Delta,
\ea \right.
\label{eq:propertyfory_1y_2y_3:pma-quasiperiodicsols}
\ee
and further 
we have 
\be 
\sum_{i=1}^3 \frac 1 {3y_i^2+S_m}=0,\ \sum_{i=1}^3 \frac {y_i} {3y_i^2+S_m}=0.
\ee
The points $(\lambda ,y_1(\lambda))$, $(\lambda ,y_2(\lambda))$
and $(\lambda ,y_3(\lambda))$ are on the three different sheets of 
the Riemann surface ${\cal K}_g$.
The holomorphic map $* $, changing sheets, is defined by 
\be 
*:
{\cal K}_g\to {\cal K}_g , \ P=(\lambda ,y_i(\lambda ))\to P^*=(\lambda , y_{i+1\,  (\textrm{mod}\, 3)}(\lambda )), \ 1\le i\le 3,
\ee 
and
$P^{**}=(P^*)^*$, etc.
 Moreover, positive divisors on ${\cal K}_g$ of degree $k$
are denoted by 
\be \ba {l}
{\cal D}_{P_1,\cdots,P_{k}}: {\cal K}_g\to \mathbb{N}_0=\mathbb{N} \cup \{0\},
\vspace{2mm}\\ 
P\mapsto 
{\cal D}_{P_1,\cdots,P_k}(P)=
\left\{\ba {ll} 
l, & \textrm{if}\ P \ \textrm{occurs $l$ times in} \ 
  \{P_1,\cdots ,P_k\},
\vspace{2mm}\\
0, & \textrm{if}\ P\not\in \{P_1,\cdots ,P_k\}.
\ea \right.
\ea
\ee
Therefore, a divisor of a meromorphic function $f$  on ${\cal K}_g$ reads 
\be 
(f(P))={\cal D}_{P_1,\cdots,P_k}(P)-{\cal D}_{Q_1,\cdots,Q_l}(P),
\ee 
if $f$ has zeros $P_i$, $1\le i\le k$, and poles 
$Q_i$, $1\le i\le l$. The space of divisors on ${\cal K}_g$ 
is denoted by $\textrm{Div}({\cal K}_g)$.

We now introduce a vector of associated Baker-Akhiezer functions 
$\psi (P,x,x_0,t_r,t_{0,r})$ as follows:
\bea
&&
\psi_x (P,x,x_0,t_r,t_{0,r}) =U(u(x,t_r),\lambda (P))
\psi (P,x,x_0,t_r,t_{0,r}),
\label{eq:BAfunction_x:pma-quasiperiodicsols}
\\
&& \psi_{t_r}(P,x,x_0,t_r,t_{0,r})=
V^{[r]}(u(x,t_r),\lambda (P))
\psi (P,x,x_0,t_r,t_{0,r}),
\label{eq:BAfunction_t_r:pma-quasiperiodicsols}
\\
&& W^{[n]}(u(x,t_r),\lambda (P))
\psi 
(P,x,x_0,t_r,t_{0,r})
=y(P)\psi (P,x,x_0,t_r,t_{0,r}),\qquad
\label{eq:BAfunction_eigenprop:pma-quasiperiodicsols}
\\
&&
\psi_i(P,x_0,x_0,t_{0,r},t_{0,r})=1,
\ 1\le i\le 3,\label{eq:BAfunction_initialcond:pma-quasiperiodicsols}
\eea
where 
$x,t_r,x_0,t_{0,r},\lambda(P),y(P)\in \mathbb{C},$ and 
$P=(\lambda ,y) \in {\cal K}_g\backslash \{P_{\infty_1},P_{\infty_2},P_{\infty_3}\}$. 
The compatibility conditions of the equations  \eqref{eq:BAfunction_x:pma-quasiperiodicsols},
\eqref{eq:BAfunction_t_r:pma-quasiperiodicsols} and \eqref{eq:BAfunction_eigenprop:pma-quasiperiodicsols} engender 
that 
\bea &&
W_x^{[n]}=[U,W^{[n]}],\label{eq:Laxeqn_xforW^{[n]}:pma-quasiperiodicsols}
\\
&& 
W_{t_r}^{[n]}=[V^{[r]},W^{[n]}],\label{eq:Laxeqn_t_rforW^{[n]}:pma-quasiperiodicsols}
 \eea
besides the $r$-th zero curvature equation in \eqref{eq:gCZCEs:pma-quasiperiodicsols}.
Note that the matrix $yI_3-W^{[n]}$ also satisfies 
the Lax equations in \eqref{eq:Laxeqn_xforW^{[n]}:pma-quasiperiodicsols} and \eqref{eq:Laxeqn_t_rforW^{[n]}:pma-quasiperiodicsols}, and so, the characteristic polynomial 
${\cal F}_m(\lambda ,y)=\textrm{det} (yI_3-W^{[n]}) $ of the 
combined Lax
matrix $W^{[n]}$ is a constant, independent of the variables $x$ and $t_r$, when $u$ solves the $r$-th four-component AKNS equations 
\eqref{eq:4cAKNSsh:pma-quasiperiodicsols}.

Associated with the Baker-Akhiezer functions, we define a set of meromorphic functions
\be 
\phi_{ij}=\phi_{ij}(P,x,x_0,t_r,t_{0,r})=\frac {\psi_i(P,x,x_0,t_r,t_{0,r})}{\psi_j(P,x,x_0,t_r,t_{0,r}) }, \ 1\le i,j\le 3.
\label{eq:DefofMeromorphicFunctions:pma-quasiperiodicsols} 
\ee
Based on \eqref{eq:BAfunction_eigenprop:pma-quasiperiodicsols}, we can have 
\be
\phi_{ij}=
\frac {y W^{[n]}_{ik} + C^{[m]}_{ij}   }{y W^{[n]}_{jk}+A^{[m]}_{ij} }
= \frac {F^{[m]}_{ij}}{y^2 W^{[n]}_{ik}-y C _{ij}^{[m]}+D^{[m]}_{ij}}
=\frac {y^2 W^{[n]}_{jk}-y A _{ij}^{[m]}+B^{[m]}_{ij} }{E^{[m]}_{ij} }
,
\label{eq:PropertyofMeromorphicFunctions:pma-quasiperiodicsols}
\ee
with
\bea
&& 
A^{[m]}_{ij}=W_{ji}^{[n]}W_{ik}^{[n]}-W_{jk}^{[n]}W_{ii}^{[n]},
\label{eq:defofA^{[m]}_{ij}:pma-quasiperiodicsols}
\\
&&
B^{[m]}_{ij}=W_{jk}^{[n]}(W_{jj}^{[n]}W_{kk}^{[n]}-W_{jk}^{[n]}W_{kj}^{[n]})
+
W_{ji}^{[n]}(W_{jj}^{[n]}W_{ik}^{[n]}-W_{jk}^{[n]}W_{ij}^{[n]}) ,
\label{eq:defofB^{[m]}_{ij}:pma-quasiperiodicsols}
\\
&&
C^{[m]}_{ij}=A^{[m]}_{ji},\ D^{[m]}_{ij}=B^{[m]}_{ji}, 
\label{eq:defofCandD^{[m]}_{ij}:pma-quasiperiodicsols}
\\
&&
E^{[m]}_{ij}=(W_{jk}^{[n]})^2W_{ki}^{[n]}+W_{ji}^{[n]}W_{jk}^{[n]}(W_{ii}^{[n]}-W_{kk}^{[n]})-(W_{ji}^{[n]})^2W_{ik}^{[n]} ,
\label{eq:defofE^{[m]}_{ij}:pma-quasiperiodicsols}
\\
&&
 F^{[m]}_{ij}=E_{ji}^{[m]},
\label{eq:identity1ofEandF^{[m]}_{ij}:pma-quasiperiodicsols}
\eea
where $\{i,j,k\}=\{1,2,3\}$. 
By the notation $\{i,j,k\}=\{1,2,3\}$, we mean here and hereafter to take 
$1\le i,j,k\le 3$ arbitrarily but three different natural numbers.  
Obviously from \eqref{eq:defofA^{[m]}_{ij}:pma-quasiperiodicsols} and \eqref{eq:defofE^{[m]}_{ij}:pma-quasiperiodicsols},
we can obtain 
\bea && 
E^{[m]}_{ij}=-E^{[m]}_{kj},\qquad\qquad
\label{eq:identity2ofEandF^{[m]}_{ij}:pma-quasiperiodicsols}
\\
&& 
E^{[m]}_{ij}=W^{[n]}_{jk}A^{[m]}_{kj}-
W^{[n]}_{ji}A^{[m]}_{ij},
\qquad \qquad
\label{eq:relationofEandA^{[m]}_{ij}:pma-quasiperiodicsols}
\eea
where $\{i,j,k\}=\{1,2,3\}$.

From the expressions of the meromorphic functions $\phi_{ij}$, $1\le i,j\le 3$, in \eqref{eq:PropertyofMeromorphicFunctions:pma-quasiperiodicsols}, using $y^3=-yS_m+T_m$, we can also directly derive the following relations:
\bea
&& 
W_{ik}^{[n]}E_{ij}^{[m]}=-(W_{jk}^{[n]})^2S_m + W_{jk}^{[n]}B_{ij}^{[m]}
-(A_{ij}^{[m]})^2,
\label{eq:MainIdentity1:pma-quasiperiodicsols} 
\\
&& C_{ij}^{[m]}E_{ij}^{[m]}=(W_{jk}^{[n]})^2T_m+A_{ij}^{[m]}B_{ij}^{[m]},
\label{eq:MainIdentity2:pma-quasiperiodicsols} 
\\
&&
-(W_{ik}^{[n]})^2S_m-(C_{ij}^{[m]})^2+W_{ik}^{[n]}D_{ij}^{[m]} = W_{jk}^{[n]}F_{ij}^{[m]},
\label{eq:MainIdentity3:pma-quasiperiodicsols} 
\\
&&
(W_{ik}^{[n]})^2T_m+C_{ij}^{[m]}D_{ij}^{[m]}
=A_{ij}^{[m]}F_{ij}^{[m]},
\label{eq:MainIdentity4:pma-quasiperiodicsols} 
\eea
where $\{i,j,k\}=\{1,2,3\}$,
and 
\bea
&&
-W_{ik}^{[n]}W_{jk}^{[n]}S_m + W_{ik}^{[n]}B_{ij}^{[m]}
+W_{jk}^{[n]}D_{ij}^{[m]}
+A_{ij}^{[m]}C_{ij}^{[m]}
=0
,\\
&& 
W_{ik}^{[n]}W_{jk}^{[n]}T_m + W_{ik}^{[n]}A_{ij}^{[m]}S_m
+W_{jk}^{[n]}C_{ij}^{[m]}S_m
-
B_{ij}^{[m]}C_{ij}^{[m]}
-A_{ij}^{[m]}D_{ij}^{[m]}
=0
,
\\
&&
W_{ik}^{[n]}A_{ij}^{[m]}T_m + W_{jk}^{[n]}C_{ij}^{[m]}T_m
-B_{ij}^{[m]}D_{ij}^{[m]}
+E_{ij}^{[m]}F_{ij}^{[m]}
=0,
\eea
where $\{i,j,k\}=\{1,2,3\}$.
Actually, 
 owing to \eqref{eq:defofCandD^{[m]}_{ij}:pma-quasiperiodicsols}
and \eqref{eq:identity1ofEandF^{[m]}_{ij}:pma-quasiperiodicsols}, 
\eqref{eq:MainIdentity3:pma-quasiperiodicsols} and 
\eqref{eq:MainIdentity4:pma-quasiperiodicsols} are also consequences of \eqref{eq:MainIdentity1:pma-quasiperiodicsols} and \eqref{eq:MainIdentity2:pma-quasiperiodicsols}, respectively.

In what follows, we first derive two derivative formulas 
 with respect to $x$ and $t_r$ for the meromorphic functions 
$\phi_{ij}$, $1\le i,j\le 3$.

\begin{lem}Suppose that 
\eqref{eq:BAfunction_x:pma-quasiperiodicsols} and
\eqref{eq:BAfunction_t_r:pma-quasiperiodicsols} 
hold. Then 
the meromorphic functions 
$\phi_{ij}$, $1\le i,j\le 3$, defined by \eqref{eq:DefofMeromorphicFunctions:pma-quasiperiodicsols}, satisfy the following Riccati type equations: 
\bea &&
\phi_{ij,x}=(U_{ii}-U_{jj})\phi_{ij}
+U_{ij}+U_{ik}\phi_{kj}-U_{ji}\phi_{ij}^2-U_{jk}\phi_{ij}\phi_{kj},
\label{eq:phi_{ij}_x:pma-quasiperiodicsols}
\\
&&
\phi_{ij,t_r}=(V^{[r]}_{ii}-V^{[r]}_{jj})\phi_{ij}
+V^{[r]}_{ij}+V^{[r]}_{ik}\phi_{kj}-V^{[r]}_{ji}\phi_{ij}^2-V^{[r]}_{jk}\phi_{ij}\phi_{kj},
\label{eq:phi_{ij}_t_m:pma-quasiperiodicsols}
\eea
where $\{i,j,k\}=\{1,2,3\}$.
\end{lem}

\noindent {\bf {\it Proof}}:
We prove the $x$-derivative part. The proof of the $t_r$-derivative part is 
similar.
Observing \eqref{eq:BAfunction_x:pma-quasiperiodicsols}, we have 
\[
\ba {l}
\D \frac {\phi_{ij,x}}{\phi_{ij}} = \Bigr(\ln \frac {\psi_i}{\psi_j} \Bigl) _x
=
\D \frac {\psi_{i,x}}{\psi_{i}}-
\frac {\psi_{j,x}}{\psi_{j}}
\vspace{2mm}\\
=
\D \frac {\sum_{k=1}^3 U_{ik}\psi_k}{\psi_i}-
 \D \frac {\sum_{k=1}^3 U_{jk}\psi_k}{\psi_j}
\vspace{2mm}\\
=
\D \sum_{k=1}^3 (U_{ik}\phi_{ki}- 
U_{jk}\phi_{kj}).
\ea
 \]
The $x$-derivative part \eqref{eq:phi_{ij}_x:pma-quasiperiodicsols} follows.
\hfill $\Box$

Secondly, we directly verify the following relations between $B_{ij}^{[m]}$, $D_{ij}^{[m]}$, $E_{ij}^{[m]}$ and $F_{ij}^{[m]}$. 
  
\begin{lem}
\label{lem:propertiesforEFandBD_{ij}^{[m]}:pma-quasiperiodicsols}
Let $B_{ij}^{[m]}$, $D_{ij}^{[m]}$,
$E_{ij}^{[m]}$ and $F_{ij}^{[m]}$, $1\le i,j\le 3$, be defined by 
\eqref{eq:defofB^{[m]}_{ij}:pma-quasiperiodicsols},
\eqref{eq:defofCandD^{[m]}_{ij}:pma-quasiperiodicsols},  \eqref{eq:defofE^{[m]}_{ij}:pma-quasiperiodicsols}
and \eqref{eq:identity1ofEandF^{[m]}_{ij}:pma-quasiperiodicsols}, respectively. Then 
\bea &&
W_{jj}^{[n]}E_{ij}^{[m]}+ W_{ji}^{[n]}B_{ij}^{[m]}-W_{jk}^{[n]}B_{kj}^{[m]}=0,
\label{eq:propertiesforE_{ij}^{[m]}B_{ij}^{[m]}:pma-quasiperiodicsols}
\\
&& 
W_{ii}^{[n]}F_{ij}^{[m]}+ W_{ij}^{[n]}D_{ij}^{[m]}-W_{ik}^{[n]}D_{ik}^{[m]}=0,
\label{eq:propertiesforF_{ij}^{[m]}D_{ij}^{[m]}:pma-quasiperiodicsols}
\eea
where $\{i,j,k\}=\{1,2,3\}$.
\end{lem}

\noindent {\it Proof}:
From the definitions of $B_{ij}^{[m]}$ and $E_{ij}^{[m]}$
in \eqref{eq:defofB^{[m]}_{ij}:pma-quasiperiodicsols} and \eqref{eq:defofE^{[m]}_{ij}:pma-quasiperiodicsols}, a direct computation verifies the relation in \eqref{eq:propertiesforE_{ij}^{[m]}B_{ij}^{[m]}:pma-quasiperiodicsols}.

Further using \eqref{eq:defofCandD^{[m]}_{ij}:pma-quasiperiodicsols} and \eqref{eq:identity1ofEandF^{[m]}_{ij}:pma-quasiperiodicsols},
from \eqref{eq:propertiesforE_{ij}^{[m]}B_{ij}^{[m]}:pma-quasiperiodicsols}, 
we immediately get the relation in  \eqref{eq:propertiesforF_{ij}^{[m]}D_{ij}^{[m]}:pma-quasiperiodicsols}.
\hfill $\Box$

Now, we consider how to compute
derivatives of $E_{ij}^{[m]}$ and $F_{ij}^{[m]}$.
Thanks to $\textrm{tr}(W^{[n]})=0$, we can directly prove the following statements.

\begin{thm}
\label{thm:E_{21}F_{21}F_{31}^{[m]}_z:pma-quasiperiodicsols}
Let $S_m$, 
$B_{ij}^{[m]}$, $D_{ij}^{[m]}$,
$E_{ij}^{[m]}$ and $F_{ij}^{[m]}$, $1\le i,j\le 3$, be defined by 
\eqref{eq:defofS_m:pma-quasiperiodicsols}, 
\eqref{eq:defofB^{[m]}_{ij}:pma-quasiperiodicsols},
\eqref{eq:defofCandD^{[m]}_{ij}:pma-quasiperiodicsols},  \eqref{eq:defofE^{[m]}_{ij}:pma-quasiperiodicsols}
and \eqref{eq:identity1ofEandF^{[m]}_{ij}:pma-quasiperiodicsols}, respectively. 
If $W^{[n]}_z=[V,W^{[n]}]$, where $V=(V_{ij})_{3\times 3}$, then we have
\bea &&
E^{[m]}_{ij,z}=(2 V_{jj}-V_{ii}-V_{kk}) E^{[m]}_{ij}
-V_{ji}(2W^{[n]}_{jk}S_m -3 B^{[m]}_{ij})
+V_{jk}(2W^{[n]}_{ji}S_m-3 B^{[m]}_{kj}),
\label{eq:E_{ij}^{[m]}_z:pma-quasiperiodicsols} 
\\
&&
F^{[m]}_{ij,z}=(2 V_{ii}-V_{jj}-V_{kk}) F^{[m]}_{ij}
-V_{ij}(2W^{[n]}_{ik}S_m -3 D^{[m]}_{ij})
+V_{ik}(2W^{[n]}_{ij}S_m-3 D^{[m]}_{ik}),\qquad \quad
\label{eq:F_{ij}^{[m]}_z:pma-quasiperiodicsols} 
\eea
where $\{i,j,k\}=\{1,2,3\}$.
\end{thm}

Applying this theorem, we can easily obtain the following relations between two derivatives of
$E_{ij}^{[m]}$ and $F_{ij}^{[m]}$.

\begin{thm}
\label{thm:E_{ij}^{[m]}_z_1andz_2:pma-quasiperiodicsols}
Let $E_{ij}^{[m]}$ and $F_{ij}^{[m]}$ be defined by 
\eqref{eq:defofE^{[m]}_{ij}:pma-quasiperiodicsols} and \eqref{eq:identity1ofEandF^{[m]}_{ij}:pma-quasiperiodicsols}, respectively. 
If $W_{z_k}^{[n]}=[V^{(k)},W^{[n]}]$, where $V^{(k)}=(V^{(k)}_{ij})_{3\times 3}$, $1\le k\le 2$,
then we have
\bea 
&&
 (V_{ji}^{(1)}W_{jk}^{[n]}- V_{jk}^{(1)}W_{ji}^{[n]})
E_{ij,z_2}^{[m]}-
(V_{ji}^{(2)}W_{jk}^{[n]}- V_{jk}^{(2)}W_{ji}^{[n]})
E_{ij,z_1}^{[m]}
\nonumber
\\
&&
=E_{ij}^{[m]}\bigl[
(2V_{jj}^{(2)}-V_{ii}^{(2)}-V_{kk}^{(2)}) ( V_{ji}^{(1)}W_{jk}^{[n]}- V_{jk}^{(1)}W_{ji}^{[n]})
\bigr.
\nonumber
\\
&&
\quad -\bigl.
(2V_{jj}^{(1)}-V_{ii}^{(1)}-V_{kk}^{(1)}) ( V_{ji}^{(2)}W_{jk}^{[n]}- V_{jk}^{(2)}W_{ji}^{[n]})
\bigr.
\nonumber
\\
&&
\quad 
\bigl.
+3(V_{jk}^{(1)}V_{ji}^{(2)}-V_{jk}^{(2)}V_{ji}^{(1)})W_{jj}^{[n]}
\bigr],
\label{eq:relationofE_{ij}^{[m]}_z_1andz_2:pma-quasiperiodicsols}
\\
&&
 (V_{ij}^{(1)}W_{ik}^{[n]}- V_{ik}^{(1)}W_{ij}^{[n]})
F_{ij,z_2}^{[m]}-
(V_{ij}^{(2)}W_{ik}^{[n]}- V_{ik}^{(2)}W_{ij}^{[n]})
F_{ij,z_1}^{[m]}
\nonumber
\\
&&
=F_{ij}^{[m]}\bigl[
(2V_{ii}^{(2)}-V_{jj}^{(2)}-V_{kk}^{(2)}) ( V_{ij}^{(1)}W_{ik}^{[n]}- V_{ik}^{(1)}W_{ij}^{[n]})
\bigr.
\nonumber
\\
&&
\quad -\bigl.
(2V_{ii}^{(1)}-V_{jj}^{(1)}-V_{kk}^{(1)}) ( V_{ij}^{(2)}W_{ik}^{[n]}- V_{ik}^{(2)}W_{ij}^{[n]})
\bigr.
\nonumber
\\
&&
\quad 
\bigl.
+3(V_{ik}^{(1)}V_{ij}^{(2)}-V_{ik}^{(2)}V_{ij}^{(1)})W_{ii}^{[n]}
\bigr],\label{eq:relationofF_{ij}^{[m]}_z_1andz_2:pma-quasiperiodicsols}
\eea
where $\{i,j,k\}=\{1,2,3\}$.
\end{thm}

Note that 
 \eqref{eq:relationofE_{ij}^{[m]}_z_1andz_2:pma-quasiperiodicsols}
and \eqref{eq:relationofF_{ij}^{[m]}_z_1andz_2:pma-quasiperiodicsols}
tell
that the weighted differences 
between two derivatives  
are multiples of 
$E_{ij}^{[m]}$ and $F_{ij}^{[m]}$, respectively. 

\begin{thm}
\label{thm:propertiesforphi_{ij}phi_{ij}^*phi_{ij}^**:pma-quasiperiodicsols}
Let $P=(\lambda , y(P))\in {\cal K}_g\backslash \{P_{\infty_1},P_{\infty_2},P_{\infty_3}\}$ and
\eqref{eq:BAfunction_eigenprop:pma-quasiperiodicsols}
hold. If $W^{[n]}_z=[V,W^{[n]}]$, where $V=(V_{ij})_{3\times 3}$, 
then the meromorphic functions $\phi_{ij}$, $1\le i,j\le 3$, 
defined by \eqref{eq:PropertyofMeromorphicFunctions:pma-quasiperiodicsols},
satisfy 
\bea 
&&
\phi_{ij}(P)+\phi_{ij}(P^*)+\phi_{ij}(P^{**})=
\frac {3B_{ij}^{[m]}-2 W_{jk}^{[n]}S_m}{
E_{ij}^{[m]}}
\nonumber \\
&&  \quad =
\frac {1}{V_{ji}W_{jk}^{[n]}-V_{jk}W_{ji}^{[n]}}
\bigl\{
W_{jk}^{[n]}
\bigl[
\frac {E_{ij,z}^{[m]}}{E_{ij}^{[m]}} -(2V_{jj}-V_{ii}-V_{kk})
\bigr]
+3V_{jk}W_{jj}^{[n]} \bigr\}
,
\label{eq:property1forphi_{ij}phi_{ij}^*phi_{ij}^**:pma-quasiperiodicsols}
\\
&&
\phi_{ij}(P)\phi_{ij}(P^*)\phi_{ij}(P^{**})=
\frac {F_{ij}^{[m]} }{E_{kj}^{[m]}}
,
\label{eq:property2forphi_{ij}phi_{ij}^*phi_{ij}^**:pma-quasiperiodicsols}
\\
&&
W_{ij}^{[n]}[\phi_{ji}(P)+\phi_{ji}(P^*)+\phi_{ji}(P^{**})
]
+W_{ik}^{[n]}
[\phi_{ki}(P)+\phi_{ki}(P^*)+\phi_{ki}(P^{**})]=-3W_{ii}^{[n]},\qquad\quad
\label{eq:property3forphi_{ij}phi_{ij}^*phi_{ij}^**:pma-quasiperiodicsols}
\\
&&
V_{ij}[\phi_{ji}(P)+\phi_{ji}(P^*)+\phi_{ji}(P^{**})
]
+V_{ik}
[\phi_{ki}(P)+\phi_{ki}(P^*)+\phi_{ki}(P^{**})] \qquad \quad
\nonumber 
\\ &&
\quad =
\frac {E_{ji,z}^{[m]}}{E_{ji}^{[m]}}-(2V_{ii}-V_{jj}-V_{kk})
,
\label{eq:property4forphi_{ij}phi_{ij}^*phi_{ij}^**:pma-quasiperiodicsols}
\eea
where $\{i,j,k\}=\{1,2,3\}$.
\end{thm}

\noindent {\it Proof}:
First, we start with the last equality in 
\eqref{eq:PropertyofMeromorphicFunctions:pma-quasiperiodicsols}, and make use of 
\eqref{eq:propertyfory_1y_2y_3:pma-quasiperiodicsols}. Then we have 
\[
\ba {l}
\D \quad \phi_{ij}(P)+\phi_{ij}(P^*)+\phi_{ij}(P^{**})
\vspace{2mm}\\
\D = \frac {(y_1^2+y_2^2+y_3^2)W_{jk}^{[n]}-(y_1+y_2+y_3)A_{ij}^{[m]}+3 B_{ij}^{[m]}}{E_{ij}^{[m]}}
\vspace{2mm}\\
\D =
\frac {3B_{ij}^{[m]}-2 W_{jk}^{[n]}S_m}{
E_{ij}^{[m]}},
\ea 
\] 
which is exactly the first equality in \eqref{eq:property1forphi_{ij}phi_{ij}^*phi_{ij}^**:pma-quasiperiodicsols}.
To prove the second equality in \eqref{eq:property1forphi_{ij}phi_{ij}^*phi_{ij}^**:pma-quasiperiodicsols},
we first note that 
from 
\eqref{eq:E_{ij}^{[m]}_z:pma-quasiperiodicsols},
we have 
\[2S_m=\frac {
E_{ij,z}^{[m]}-(2V_{jj}-V_{ii}-V_{kk})E_{ij}^{[m]}+
3(V_{jk}B_{kj}^{[m]}-V_{ji}B_{ij}^{[m]})
}
{V_{ji}W_{jk}^{[n]}-V_{jk}W_{ji}^{[n]}} .\] 
Then 
making use of 
\eqref{eq:propertiesforE_{ij}^{[m]}B_{ij}^{[m]}:pma-quasiperiodicsols}, we can directly
verify the second equality in \eqref{eq:property1forphi_{ij}phi_{ij}^*phi_{ij}^**:pma-quasiperiodicsols}, starting from 
the first equality in \eqref{eq:property1forphi_{ij}phi_{ij}^*phi_{ij}^**:pma-quasiperiodicsols}.

Secondly,
we use \eqref{eq:propertyfory_1y_2y_3:pma-quasiperiodicsols} and 
  the first equality in 
\eqref{eq:PropertyofMeromorphicFunctions:pma-quasiperiodicsols} to get
\[
\ba {l}
\D \quad \phi_{ij}(P)\phi_{ij}(P^*)\phi_{ij}(P^{**})
\vspace{2mm}\\
\D =\frac {y_1y_2y_3 (W_{ik}^{[m]} )^3 + (y_1y_2+y_1y_3+y_2y_3) (W_{ik}^{[m]})^2 
C_{ij}^{[m]}+(y_1+y_2+y_3)W_{ik}^{[m]} (C_{ij}^{[m]})^2+(C_{ij}^{[m]})^3
}{
y_1y_2y_3 (W_{jk}^{[m]} )^3 + (y_1y_2+y_1y_3+y_2y_3) (W_{jk}^{[m]})^2 
A_{ij}^{[m]}+(y_1+y_2+y_3)W_{jk}^{[m]} (A_{ij}^{[m]})^2+(A_{ij}^{[m]})^3
}
\vspace{2mm}\\
\D = \frac {T_m(W_{ik}^{[m]} )^3 +S_m (W_{ik}^{[m]} )^2 C_{ij}^{[m]}+ (C_{ij}^{[m]} )^3 }{
T_m(W_{jk}^{[m]} )^3 +S_m (W_{jk}^{[m]} )^2 A_{ij}^{[m]}+ (A_{ij}^{[m]} )^3 
}.
\ea \]
Then, based on the properties in 
\eqref{eq:defofCandD^{[m]}_{ij}:pma-quasiperiodicsols},
\eqref{eq:identity1ofEandF^{[m]}_{ij}:pma-quasiperiodicsols}
and \eqref{eq:identity2ofEandF^{[m]}_{ij}:pma-quasiperiodicsols},
a direct application of 
\eqref{eq:MainIdentity1:pma-quasiperiodicsols} and \eqref{eq:MainIdentity4:pma-quasiperiodicsols} 
yields the equality \eqref{eq:property2forphi_{ij}phi_{ij}^*phi_{ij}^**:pma-quasiperiodicsols}.

Thirdly, using \eqref{eq:BAfunction_eigenprop:pma-quasiperiodicsols} in the definition of the Baker-Akhiezer functions, we have 
\[
\sum_{j=1}^3 W_{ij}^{[n]}\phi_{ji}(P)=y_1,\ 
\sum_{j=1}^3 W_{ij}^{[n]}\phi_{ji}(P^*)=y_2,\ 
\sum_{j=1}^3 W_{ij}^{[n]}\phi_{ji}(P^{**})=y_3,
\]
and then, based on 
\eqref{eq:propertyfory_1y_2y_3:pma-quasiperiodicsols},
summing them up generates 
the equality \eqref{eq:property3forphi_{ij}phi_{ij}^*phi_{ij}^**:pma-quasiperiodicsols}.

Finally, note that the derivative formula
\eqref{eq:E_{ij}^{[m]}_z:pma-quasiperiodicsols} 
guarantees 
\[
\frac {E_{ji,z}^{[m]}}{E_{ji}^{[m]}}-(2V_{ii}-V_{jj}-V_{kk})
=
\frac {E_{ki,z}^{[m]}}{E_{ki}^{[m]}}-(2V_{ii}-V_{kk}-V_{jj}),
\]
where $\{i,j,k\}=\{1,2,3\}$.
Then, making use of 
the second equality in \eqref{eq:property1forphi_{ij}phi_{ij}^*phi_{ij}^**:pma-quasiperiodicsols}, we can 
arrive at
  the equality \eqref{eq:property4forphi_{ij}phi_{ij}^*phi_{ij}^**:pma-quasiperiodicsols} by a direct computation. This completes the proof of the theorem. \hfill $\Box$

When 
\eqref{eq:BAfunction_x:pma-quasiperiodicsols},
\eqref{eq:BAfunction_t_r:pma-quasiperiodicsols} 
and \eqref{eq:BAfunction_eigenprop:pma-quasiperiodicsols}
in the definition of the Baker-Akhiezer functions
hold, 
we have 
the two Lax equations in 
\eqref{eq:Laxeqn_xforW^{[n]}:pma-quasiperiodicsols} and \eqref{eq:Laxeqn_t_rforW^{[n]}:pma-quasiperiodicsols}.
Thus, upon noting $\textrm{tr}(U)=\textrm{tr} (V^{[r]})=0$,
Theorem \ref{thm:propertiesforphi_{ij}phi_{ij}^*phi_{ij}^**:pma-quasiperiodicsols}
with $V=U$ and $V^{[r]}$ yields that 
\bea 
&&
\quad \phi_{ij}(P)+\phi_{ij}(P^*)+\phi_{ij}(P^{**})
\nonumber 
\\
&& 
=
\frac {1}{U_{ji}W_{jk}^{[n]}-U_{jk}W_{ji}^{[n]}}
\bigl[
W_{jk}^{[n]}
\bigl(
\frac {E_{ij,x}^{[m]}}{E_{ij}^{[m]}} -3U_{jj}\bigr)
+3U_{jk}W_{jj}^{[n]} \bigr]
,
\\
&&
\quad 
\phi_{ij}(P)+\phi_{ij}(P^*)+\phi_{ij}(P^{**})
\nonumber 
\vspace{2mm}
\\
&&
=
\frac {1}{V_{ji}^{[r]}W_{jk}^{[n]}-V_{jk}^{[r]}W_{ji}^{[n]}}
\bigl[
W_{jk}^{[n]}
\bigl(
\frac {E_{ij,t_r}^{[m]}}{E_{ij}^{[m]}} -3V_{jj}^{[r]}\bigr)
+3V_{jk}^{[r]}W_{jj}^{[n]} \bigr]
,
\eea
and 
\bea 
&& 
U_{ij}[\phi_{ji}(P)+\phi_{ji}(P^*)+\phi_{ji}(P^{**})
]
+U_{ik}
[\phi_{ki}(P)+\phi_{ki}(P^*)+\phi_{ki}(P^{**})]
\nonumber 
\\ &&
=\frac {E_{ji,x}^{[m]}}{E_{ji}^{[m]}}-3U_{ii},
\label{eq:SpecialCase1property4forphi_{ij}phi_{ij}^*phi_{ij}^**:pma-quasiperiodicsols}
\\
&&
V_{ij}^{[r]}[\phi_{ji}(P)+\phi_{ji}(P^*)+\phi_{ji}(P^{**})
]
+V_{ik}^{[r]}
[\phi_{ki}(P)+\phi_{ki}(P^*)+\phi_{ki}(P^{**})]\qquad\quad
\nonumber 
\\
&& =
\frac {E_{ji,t_r}^{[m]}}{E_{ji}^{[m]}}
-3V_{ii}^{[r]},\qquad 
\label{eq:SpecialCase2property4forphi_{ij}phi_{ij}^*phi_{ij}^**:pma-quasiperiodicsols}
\eea
where $\{i,k,j\}=\{1,2,3\}$ (see \cite{GengWH-JNS2013} for the Kaup-Kupershmidt case
and \cite{GengZD-AM2014}
for the coupled KdV case).

In view of the relations in 
\eqref{eq:identity1ofEandF^{[m]}_{ij}:pma-quasiperiodicsols} and \eqref{eq:identity2ofEandF^{[m]}_{ij}:pma-quasiperiodicsols},
we only need to explore properties of the three sums 
$E^{[m]}_{21},F^{[m]}_{21}$ and $F^{[m]}_{31}$, to determine 
dynamics of zeros and poles of 
the meromorphic functions
$\phi_{ij}$, $1\le i,j\le 3$. For all other sums, we can generate 
similar results. For example, the relations 
\[ E_{32}^{[m]} = -E_{12}^{[m]}= -F_{21}^{[m]},\ 
F_{13}^{[m]}=E_{31}^{[m]}=-E_{21}^{[m]}
, \]
 permit one to draw analogies for $E_{32}^{[m]}$ and $F_{13}^{[m]}$.

Taking $V=U$ and $V^{[r]}$, and noting $\textrm{tr}(U)=\textrm{tr}(V^{[r]})=0$, directly from Theorem \ref{thm:E_{21}F_{21}F_{31}^{[m]}_z:pma-quasiperiodicsols}, 
we can obtain the following derivative formulas in the four-component AKNS case.

\begin{thm}
Let 
$E_{21}^{[m]},F_{21}^{[m]}$ and 
$F_{31}^{[m]}$ be defined by 
\eqref{eq:defofE^{[m]}_{ij}:pma-quasiperiodicsols} and \eqref{eq:identity1ofEandF^{[m]}_{ij}:pma-quasiperiodicsols}, and 
$(\lambda ,x,t_r)\in \mathbb{C}^3$. Suppose that 
\eqref{eq:BAfunction_x:pma-quasiperiodicsols},
\eqref{eq:BAfunction_t_r:pma-quasiperiodicsols} and 
\eqref{eq:BAfunction_eigenprop:pma-quasiperiodicsols}
hold. Then we have
\bea 
&& 
E_{21,x}^{[m]}=-6\lambda E_{21}^{[m]} -p_1 (2 W^{[n]}_{13} S_m -3 B^{[m]}_{21})+p_2(
2 W^{[n]}_{12} S_m -3 B^{[m]}_{31}
),\qquad
\label{eq:formula1forderivativesofEFFwrtx:pma-quasiperiodicsols}
 \\
&&
F_{21,x}^{[m]}=3\lambda F_{21}^{[m]} -q_1 (2 W^{[n]}_{23} S_m -3 D^{[m]}_{21})
,
\label{eq:formula2forderivativesofEFFwrtx:pma-quasiperiodicsols}
 \\
&&
F_{31,x}^{[m]}=3 \lambda E_{31}^{[m]} -q_2 (2 W^{[n]}_{32} S_m -3 D^{[m]}_{31})
,
\label{eq:formula3forderivativesofEFFwrtx:pma-quasiperiodicsols}
\eea
and 
\bea 
&& 
E_{21,t_r}^{[m]}=3 V^{[r]}_{11} E_{21}^{[m]} -V_{12}^{[r]} (2 W^{[n]}_{13} S_m -3 B^{[m]}_{21})+V_{13}^{[r]}(
2 W^{[n]}_{12} S_m -3 B^{[m]}_{31}
),\qquad
\label{eq:formula1forderivativesofEFFwrtt_r:pma-quasiperiodicsols}
 \\
&&
F_{21,t_r}^{[m]}=3 V_{22}^{[r]} F_{21}^{[m]} -V_{21}^{[r]} (2 W^{[n]}_{23} S_m -3 D^{[m]}_{21})+V_{23}^{[r]} (2 W^{[n]}_{21} S_m -3 D^{[m]}_{23})
,
\label{eq:formula2forderivativesofEFFwrtt_r:pma-quasiperiodicsols}
 \\
&&
F_{31,t_r}^{[m]}=3 V_{33}^{[r]} F_{31}^{[m]} - V_{31}^{[r]} (2 W^{[n]}_{32} S_m -3 D^{[m]}_{31})+V^{[r]}_{32}(2 W^{[n]}_{31} S_m -3 D^{[m]}_{32})
.
\label{eq:formula3forderivativesofEFFwrtt_r:pma-quasiperiodicsols}
\eea
\end{thm}

We can further present the derivatives of $E_{21}^{[m]}$, $F_{21}^{[m]}$ and $F_{31}^{[m]}$ with respect to $t_r$ 
in terms of $E_{21}^{[m]}$, $F_{21}^{[m]}$ and $F_{31}^{[m]}$, and their derivatives with respect to $x$. 

\begin{thm}
Let 
$E_{21}^{[m]},F_{21}^{[m]}$ and 
$F_{31}^{[m]}$ be defined by 
\eqref{eq:defofE^{[m]}_{ij}:pma-quasiperiodicsols} and \eqref{eq:identity1ofEandF^{[m]}_{ij}:pma-quasiperiodicsols}, and 
$(\lambda ,x,t_r)\in \mathbb{C}^3$. Suppose that 
\eqref{eq:BAfunction_x:pma-quasiperiodicsols},
\eqref{eq:BAfunction_t_r:pma-quasiperiodicsols} and 
\eqref{eq:BAfunction_eigenprop:pma-quasiperiodicsols}
hold. Then we have
\bea &&
E_{21,t_r}^{[m]}=
E_{21,x}^{[m]}\frac {W^{[n]}_{13}V^{[r]}_{12}- W^{[n]}_{12}V^{[r]}_{13}
 }{p_1 W^{[n]}_{13} -p_2 W^{[n]}_{12} } \nonumber 
\\ && \ \ \ +
E_{21}^{[m]}\Bigl[3 \Bigl( V^{[r]}_{11} -  \frac {p_1 V^{[r]}_{13}-p_2V^{[r]}_{12}
}{p_1 W^{[n]}_{13}-p_2 W^{[n]}_{12} } W^{[n]}_{11}  \Bigr) 
+6\lambda \frac {W^{[n]}_{13}V^{[r]}_{12}- W^{[n]}_{12}V^{[r]}_{13}
 }{p_1 W^{[n]}_{13} -p_2 W^{[n]}_{12} } 
\Bigr],\qquad
\label{eq:E_{21}^{[m]}_t_randx:pma-quasiperiodicsols}
\\
&& 
F_{21,t_r}^{[m]}=
F_{21,x}^{[m]}\frac {W^{[n]}_{23}V^{[r]}_{21}- W^{[n]}_{21}V^{[r]}_{23}
 }{q_1 W^{[n]}_{23}  } \nonumber 
\\ && \ \ \ + 
F_{21}^{[m]}\Bigl[3 \Bigl( V^{[r]}_{22} -  \frac { W^{[n]}_{22} 
}{ W^{[n]}_{23} } V^{[r]}_{23} \Bigr) 
+3\lambda \frac {W^{[n]}_{21}V^{[r]}_{23}- W^{[n]}_{23}V^{[r]}_{21}
 }{q_1 W^{[n]}_{23} } 
\Bigr],
\label{eq:F_{21}^{[m]}_t_randx:pma-quasiperiodicsols}
\\
&& 
F_{31,t_r}^{[m]}=
F_{31,x}^{[m]}\frac {W^{[n]}_{32}V^{[r]}_{31}- W^{[n]}_{31}V^{[r]}_{32}
 }{q_2 W^{[n]}_{32}  } \nonumber 
\\ && \ \ \ + 
F_{31}^{[m]}\Bigl[3 \Bigl( V^{[r]}_{33} -  \frac { W^{[n]}_{33} 
}{ W^{[n]}_{32} } V^{[r]}_{32} \Bigr) 
+3\lambda \frac {W^{[n]}_{31}V^{[r]}_{32}- W^{[n]}_{32}V^{[r]}_{31}
 }{q_2 W^{[n]}_{32} } 
\Bigr].
\label{eq:F_{31}^{[m]}_t_randx:pma-quasiperiodicsols}
\eea
\end{thm}

\noindent {\bf Proof:}
Note that \eqref{eq:BAfunction_x:pma-quasiperiodicsols},
\eqref{eq:BAfunction_t_r:pma-quasiperiodicsols} 
and \eqref{eq:BAfunction_eigenprop:pma-quasiperiodicsols}
imply the Lax equations 
\eqref{eq:Laxeqn_xforW^{[n]}:pma-quasiperiodicsols} and \eqref{eq:Laxeqn_t_rforW^{[n]}:pma-quasiperiodicsols}.
Upon taking
$V^{(1)}=U$ and $z=x$, and $V^{(2)}=V^{[r]}$ and $z=t_r$, 
 Theorem \ref{thm:E_{ij}^{[m]}_z_1andz_2:pma-quasiperiodicsols}
 immediately leads to 
the three derivative relations 
in 
\eqref{eq:E_{21}^{[m]}_t_randx:pma-quasiperiodicsols},
\eqref{eq:F_{21}^{[m]}_t_randx:pma-quasiperiodicsols} and 
\eqref{eq:F_{31}^{[m]}_t_randx:pma-quasiperiodicsols}.
The proof is finished.
\hfill $\Box$ 

Directly applying the following three equalities
\bea &&
W^{[n]}_{11}E_{21}^{[m]} + W^{[n]}_{12}B_{21}^{[m]} 
-W^{[n]}_{13}B_{31}^{[m]} =0,
\label{eq:AdditionalIdentity1:pma-quasiperiodicsols} 
\\
&&
W^{[n]}_{21}D_{21}^{[m]} + W^{[n]}_{22}F_{21}^{[m]} 
-W^{[n]}_{23}D_{23}^{[m]} =0,
\label{eq:AdditionalIdentity2:pma-quasiperiodicsols} 
\\
&&
W^{[n]}_{31}D_{31}^{[m]} + W^{[n]}_{33}F_{31}^{[m]} 
-W^{[n]}_{32}D_{32}^{[m]} =0,
\label{eq:AdditionalIdentity3:pma-quasiperiodicsols} 
\eea
which are consequences of \eqref{eq:propertiesforE_{ij}^{[m]}B_{ij}^{[m]}:pma-quasiperiodicsols} and 
\eqref{eq:propertiesforF_{ij}^{[m]}D_{ij}^{[m]}:pma-quasiperiodicsols},
we can represent 
all terms on the right-hand side of each equation in 
\eqref{eq:formula1forderivativesofEFFwrtt_r:pma-quasiperiodicsols},
\eqref{eq:formula2forderivativesofEFFwrtt_r:pma-quasiperiodicsols} and
\eqref{eq:formula3forderivativesofEFFwrtt_r:pma-quasiperiodicsols},
 in terms of $E_{21}^{[m]}$, $F_{21}^{[m]}$ and $F_{31}^{[m]}$ and their derivatives with respect to $x$ in
\eqref{eq:formula1forderivativesofEFFwrtx:pma-quasiperiodicsols}, \eqref{eq:formula2forderivativesofEFFwrtx:pma-quasiperiodicsols} and \eqref{eq:formula3forderivativesofEFFwrtx:pma-quasiperiodicsols},  
which also presents
the three derivative relations 
in 
\eqref{eq:E_{21}^{[m]}_t_randx:pma-quasiperiodicsols},
\eqref{eq:F_{21}^{[m]}_t_randx:pma-quasiperiodicsols} and 
\eqref{eq:F_{31}^{[m]}_t_randx:pma-quasiperiodicsols}, precisely.

\section{Characteristic variables and Dubrovin type equations}

\label{sec:Dubrovineqns:pma-quasiperiodicsols}

It is direct to see that the degrees of $E_{21}^{[m]},F_{21}^{[m]}$ and  
$F_{31}^{[m]}$ are $g$, $g+1$ and $g+1$, respectively. Thus,
 we can assume that 
\bea && 
E_{21}^{[m]}(x,t_r)=e_{21}^{[m]}(\alpha ,u) \prod_{j=1}^{g} (\lambda - \mu_j(x,t_r)),
\label{eq:defofmu_j:pma-quasiperiodicsols}
\\
&&
F_{21}^{[m]}(x,t_r)=f_{21}^{[m]}(\alpha ,u) \prod_{j=0}^{g} (\lambda - \nu_j(x,t_r)),
\label{eq:defofnu_j:pma-quasiperiodicsols}
\\
&&
 F_{31}^{[m]}(x,t_r)=f_{31}^{[m]}(\alpha ,u) \prod_{j=0}^{g} (\lambda - \xi_j(x,t_r) ),
\label{eq:defofxi_j:pma-quasiperiodicsols}
\eea
where $e_{21}^{[m]},f_{21}^{[m]}$ and $f_{31}^{[m]}$ are three non-zero functions depending on $\alpha =(\alpha _0,\alpha _1,\cdots, \alpha_n)$ and $u=(p_1,p_2,q_1,q_2)^T$. We call those roots characteristic variables associated with 
the Baker-Akhiezer functions.

In light of \eqref{eq:PropertyofMeromorphicFunctions:pma-quasiperiodicsols}, 
we can introduce the following three sets of particular points in ${\cal K}_g$: 
\bea && 
\hat {\mu}_j(x,t_r)=(\mu_j (x,t_r), y(\mu_j(x,t_r))) 
=\Bigl(\mu_j (x,t_r), -\Bigl.\frac {A_{21}^{[m]}(x,t_r)}{W_{13}^{[n]}(x,t_r)}
\Bigr|_{\lambda =\mu_j(x,t_r)}\Bigr)
\nonumber 
\\
&& \qquad \qquad \qquad \qquad \qquad \qquad \qquad \ \,
=\Bigl(\mu_j (x,t_r), -\Bigl.\frac {A_{31}^{[m]}(x,t_r)}{W_{12}^{[n]}(x,t_r)}
\Bigr|_{\lambda =\mu_j(x,t_r)}
\Bigr)
,\ 1\le j\le g,
\\
&&
\hat {\nu}_j(x,t_r)
=(\nu_j (x,t_r), y(\nu_j(x,t_r))) 
=\Bigl(\nu_j (x,t_r), -\Bigl.\frac {C_{21}^{[m]}(x,t_r)}{W_{23}^{[n]}(x,t_r)}
\Bigr|_{\lambda =\nu_j(x,t_r)}
\Bigr)
, \ 0\le j\le g,
\\ 
&&
\hat {\xi}_j(x,t_r)
=(\xi_j (x,t_r), y(\xi_j(x,t_r))) 
=\Bigl(\xi_j (x,t_r), -\Bigl.\frac {C_{31}^{[m]}(x,t_r)}{W_{32}^{[n]}(x,t_r)}
\Bigr|_{\lambda =\xi_j(x,t_r)}
\Bigr)
, \ 0\le j\le g,\qquad \quad\ 
\eea
where $(x,t_r)\in \mathbb{C}^2$.
To determine zeros and poles of the Baker-Akhiezer functions $\psi_i$, $1\le i\le 3$, we set
\be 
J^{(i)}_r=U_{i1}\phi_{1i}+U_{i2}\phi_{2i}
+U_{i3}\phi_{3i}, \ I_r^{(i)}=V_{i1}^{[r]}\phi_{1i}
+V_{i2}^{[r]}\phi_{2i}
+V_{i3}^{[r]}\phi_{3i}, 
\ 1\le i\le 3.
\ee
Note that 
\eqref{eq:BAfunction_x:pma-quasiperiodicsols} and 
\eqref{eq:BAfunction_t_r:pma-quasiperiodicsols} give
\be 
\frac {\psi_{i,x}(P,x,x_0,t_r,t_{0,r})}{\psi_i(P,x,x_0,t_r,t_{0,r})}
=J_r^{(i)}(P,x,t_r),\ 
1\le i\le 3,
\label{eq:defofJ_r^{(i)}:pma-quasiperiodicsols}
\ee and 
\be 
\frac {\psi_{i,t_r}(P,x,x_0,t_r,t_{0,r})}{\psi_i(P,x,x_0,t_r,t_{0,r})}
=I_r^{(i)}(P,x,t_r),
 \ 1\le i\le 3,
\label{eq:defofI_r^{(i)}:pma-quasiperiodicsols}
\ee
respectively. 
It follows that the basic conservation laws associated with Lax pairs hold, i.e., 
\be (I_r^{(i)})_x=(\frac {\psi_{i,t_r}}{\psi_i})_x
=(\frac {\psi_{i,x}}{\psi_i})_{t_r}=
(J_r^{(i)})_{t_r},\ 1\le i\le 3,
\label{eq:compatibilityconditions:pma-quasiperiodicsols}
 \ee
from which we can also generate infinitely many conservation laws by observing Laurent series of the conserved quantities 
$J_r^{(i)}$, $1\le i\le 3$,
and the conserved fluxes 
$I_r^{(i)}$, $1\le i\le 3$,
at $\lambda =\infty$ (or $ \zeta =\lambda ^{-1}=0$). 
Furthermore,
\eqref{eq:defofJ_r^{(i)}:pma-quasiperiodicsols} and 
\eqref{eq:defofI_r^{(i)}:pma-quasiperiodicsols}
imply the expressions for the Baker-Akhiezer functions $\psi_i$, $1\le i\le 3$,
\be 
\psi_i(P,x,x_0,t_r,t_{0,r})=
\textrm{exp}\Bigl(  \int_{x_0}^x J^{(i)}_r(P,x',t_r) \, dx'+\int_{t_{0,r}}^{t_r}I^{(i)}_r(P,x_0,t')\, dt' \Bigr),
 \ 1\le i\le 3,
\label{eq:expressionsforpsi_i:pma-quasiperiodicsols}
\ee
upon taking advantage of the basic 
conservation laws in \eqref{eq:compatibilityconditions:pma-quasiperiodicsols}.

Let us first determine general dynamics of zeros of $E_{21}^{[m]},F_{21}^{[m]}$ and $F_{31}^{[m]}$.

\begin{thm}
\label{thm:generalDubrovintypeequations:pma-quasiperiodicsols}
Let $W_z^{[n]}=[V,W^{[n]}]$, where $V=(V_{ij})_{3\times 3}$. If
$\mu_i\ne \mu_j,\ \nu_i\ne \nu_j$ and $ \xi_i\ne \xi _j$ for $i\ne j$, then
the zeros of $E_{21}^{[m]},F_{21}^{[m]}$ and $F_{31}^{[m]}$ satisfy 
the Dubrovin type equations
\bea 
&&\mu_{j,z}= 
-\frac {[(V_{12}W_{13}^{[n]}- V_{13}W_{12}^{[n]})
(3y^2+S_m)]\bigl.\bigr|_{\lambda =\mu_j}
}{e_{21}^{[m]}\prod_{k=1,\,k\ne j}^g (\mu _j-\mu _k) },\ 1\le j\le g,
\label{eq:mu_j_z:pma-quasiperiodicsols}
\\ 
&&
\nu_{j,z}=
-\frac {[(V_{21}W_{23}^{[n]}- V_{23}W_{21}^{[n]})
(3y^2+S_m)]\bigl.\bigr|_{\lambda =\nu_j}
}{f_{21}^{[m]}\prod_{k=0,\,k\ne j}^g (\nu _j-\nu _k) }
, \ 0\le j\le g, 
\label{eq:nu_j_z:pma-quasiperiodicsols}
\\ 
&&
\xi_{j,z}= 
-\frac {[(V_{31}W_{32}^{[n]}- V_{32}W_{31}^{[n]})
(3y^2+S_m)]\bigl.\bigr|_{\lambda =\xi_j}
}{f_{31}^{[m]}\prod_{k=0,\,k\ne j}^g (\xi _j-\xi _k) }
, \ 0\le j\le g.
\label{eq:xi_j_z:pma-quasiperiodicsols}
\eea
\end{thm}

\noindent {\it Proof}: 
We first prove the Dubrovin type equation
 \eqref{eq:mu_j_z:pma-quasiperiodicsols}. 
Using 
\eqref{eq:MainIdentity1:pma-quasiperiodicsols} and \eqref{eq:AdditionalIdentity1:pma-quasiperiodicsols},
we have 
\[
\ba {l}
\D (y^2+S_m)|_{\lambda =\mu_j}= \Bigl[
\Bigl(-\frac {A_{21} ^{[m]}} {W_{13}^{[n]}}\Bigr)^2 +S_m
\Bigr]\Bigl.\Bigr|_{\lambda =\mu_j}
\vspace{2mm}\\
\D =\Bigl.\frac {(A_{21}^{[m]})^2+(W_{13}^{[n]})^2S_m}{
(W_{13}^{[n]})^2
} \Bigr|_{\lambda =\mu_j}
=\Bigl.\frac {
W_{13}^{[n]}B_{21}^{[m]}
}{(W_{13}^{[n]})^2} \Bigr|_{\lambda =\mu_j}
\vspace{2mm}\\
\D =\Bigl.\frac {B_{21}^{[m]} }{W_{13}^{[n]}} \Bigr|_{\lambda =\mu_j}
=\Bigl.\frac {B_{31}^{[m]} }{W_{12}^{[n]}} \Bigr|_{\lambda =\mu_j}, \ 1\le j\le g.
\ea
\]
Following these two expressions for 
$B_{21}^{[m]}$ and $B_{31}^{[m]}$, we have
\[
(V_{12}B_{21}^{[m]}-V_{13}B_{31}^{[m]})\bigl.\bigr|_{\lambda =\mu_j}=
[
(V_{12}W_{13}^{[n]}-V_{13}W_{12}^{[n]})
(y^2+S_m)
]\bigl.\bigr|_{\lambda =\mu_j}, \ 1\le j\le g,
\]
and 
thus, applying the derivative formula \eqref{eq:E_{ij}^{[m]}_z:pma-quasiperiodicsols}, we can get
\be 
E^{[m]}_{21,z}\bigl.\bigr|_{\lambda =\mu_j} = 
[(V_{12}W_{13}^{[n]}- V_{13}W_{12}^{[n]})
(3y^2+S_m)]\bigl.\bigr|_{\lambda =\mu_j}
,\ 1\le j\le g.
\ee
Now, according to \eqref{eq:defofmu_j:pma-quasiperiodicsols},
this leads to 
the Dubrovin type equation \eqref{eq:mu_j_z:pma-quasiperiodicsols}
for $\mu_j$, $1\le j\le g$.

We secondly verify the Dubrovin type equation
 \eqref{eq:nu_j_z:pma-quasiperiodicsols}. 
Now using 
\eqref{eq:MainIdentity3:pma-quasiperiodicsols} and \eqref{eq:AdditionalIdentity2:pma-quasiperiodicsols},
we can compute that
\[
\ba {l}
\D (y^2+S_m)|_{\lambda =\nu_j}= \Bigl[
\Bigl(-\frac {C_{21} ^{[m]}} {W_{23}^{[n]}}\Bigr)^2 +S_m
\Bigr]\Bigl.\Bigr|_{\lambda =\nu_j}
\vspace{2mm}\\
\D =\Bigl.\frac {(C_{21}^{[m]})^2+(W_{23}^{[n]})^2S_m}{
(W_{23}^{[n]})^2
} \Bigr|_{\lambda =\nu_j}=
\Bigl.\frac {W_{23}^{[n]}D_{21}^{[m]}}{
(W_{23}^{[n]})^2
} \Bigr|_{\lambda =\nu_j}
\vspace{2mm}\\
\D =\Bigl.\frac {D_{21}^{[m]} }{W_{23}^{[n]}} \Bigr|_{\lambda =\nu_j}
=\Bigl.\frac {D_{23}^{[m]} }{W_{21}^{[n]}} \Bigr|_{\lambda =\nu_j}, \ 0\le j\le g.
\ea
\]
Based on these two expressions for 
$D_{21}^{[m]}$ and $D_{23}^{[m]}$, we get
\[
(V_{21}D_{21}^{[m]}-V_{23}D_{23}^{[m]})\bigl.\bigr|_{\lambda =\nu_j}=
[
(V_{21}W_{23}^{[n]}-V_{23}W_{21}^{[n]})
(y^2+S_m)
]\bigl.\bigr|_{\lambda =\nu_j}, \ 0\le j\le g,
\]
and 
thus, from the derivative formula \eqref{eq:F_{ij}^{[m]}_z:pma-quasiperiodicsols}, we can have 
\be 
F^{[m]}_{21,z}\bigl.\bigr|_{\lambda =\nu_j} = 
[(V_{21}W_{23}^{[n]}- V_{23}W_{21}^{[n]})
(3y^2+S_m)]\bigl.\bigr|_{\lambda =\nu_j}
,\ 0\le j\le g.
\ee
Then based on \eqref{eq:defofnu_j:pma-quasiperiodicsols},
this yields
the Dubrovin type equation \eqref{eq:nu_j_z:pma-quasiperiodicsols}
 for $\nu_j$, $0\le j\le g$.

We thirdly prove the Dubrovin type equation 
\eqref{eq:xi_j_z:pma-quasiperiodicsols}. 
Similarly using 
\eqref{eq:MainIdentity3:pma-quasiperiodicsols} and \eqref{eq:AdditionalIdentity3:pma-quasiperiodicsols},
we have
\[
\ba {l}
\D (y^2+S_m)|_{\lambda =\xi_j}= \Bigl[
\Bigl(-\frac {C_{31} ^{[m]}} {W_{32}^{[n]}}\Bigr)^2 +S_m
\Bigr]\Bigl.\Bigr|_{\lambda =\xi_j}
\vspace{2mm}\\
\D =\Bigl.\frac {(C_{31}^{[m]})^2+(W_{32}^{[n]})^2S_m}{
(W_{32}^{[n]})^2
} \Bigr|_{\lambda =\xi_j}
=
\Bigl.\frac {W_{32}^{[n]}D_{31}^{[m]}}{
(W_{32}^{[n]})^2
} \Bigr|_{\lambda =\xi_j}
\vspace{2mm}\\
\D =\Bigl.\frac {D_{31}^{[m]} }{W_{32}^{[n]}} \Bigr|_{\lambda =\xi_j}
=\Bigl.\frac {D_{32}^{[m]} }{W_{31}^{[n]}} \Bigr|_{\lambda =\xi_j}, \ 0\le j\le g.
\ea
\]
From these two expressions for 
$D_{31}^{[m]}$ and $D_{32}^{[m]}$, we obtain
\[
(V_{31}D_{31}^{[m]}-V_{32}D_{32}^{[m]})\bigl.\bigr|_{\lambda =\xi_j}=
[
(V_{31}W_{32}^{[n]}-V_{32}W_{31}^{[n]})
(y^2+S_m)
]\bigl.\bigr|_{\lambda =\xi_j}, \ 0\le j\le g,
\]
and 
then, applying
the derivative formula \eqref{eq:F_{ij}^{[m]}_z:pma-quasiperiodicsols}, we can get
\be 
F^{[m]}_{31,z}\bigl.\bigr|_{\lambda =\xi_j} = 
[(V_{31}W_{32}^{[n]}- V_{32}W_{31}^{[n]})
(3y^2+S_m)]\bigl.\bigr|_{\lambda =\xi_j}
,\ 0\le j\le g.
\ee
Finally, according to \eqref{eq:defofxi_j:pma-quasiperiodicsols},
this equality generates 
the Dubrovin type equation
\eqref{eq:xi_j_z:pma-quasiperiodicsols}
 for $\xi_j$, $0\le j\le g$.
The proof is finished.
\hfill $\Box$

In order to determine zeros and poles of the Baker-Akhiezer functions 
$\psi_i$, $1\le i\le 3$, we verify the following statements.

\begin{thm}
\label{thm:generalzerosofBakerAkhiezerfunctions:pma-quasiperiodicsols}
Let $W_z^{[n]}=[V,W^{[n]}]$, where $V=(V_{ij})_{3\times 3}$ with $\textrm{tr}(V)=0$.
If $\mu_i\ne \mu_j,\ \nu_i\ne \nu_j$ and $ \xi_i\ne \xi _j$ for $i\ne j$,
then 
we have 
\bea
&&
V_{11}+V_{12}\phi_{21}+V_{13}\phi_{31}
\mathop{=}\limits_{\lambda \to \mu_j}
\partial _z \ln (\lambda -\mu_j) +\textrm{O}(1),\ 1\le j\le g,\\
&&
V_{21}\phi_{12}+V_{22}+V_{23}\phi_{32}
\mathop{=}\limits_{\lambda \to \nu_j}
\partial _z \ln (\lambda -\nu_j) +\textrm{O}(1),\ 
0\le j\le g,
\\&&
V_{31}\phi_{13}+V_{22}\phi_{23}+V_{33}
\mathop{=}\limits_{\lambda \to \xi_j}
\partial _z \ln (\lambda -\xi_j) +\textrm{O}(1),\ 0\le j\le g.
\eea
\end{thm}

\noindent {\it Proof}:
We only prove the first statement. The proofs for the other two statements are similar. 

Using
\eqref{eq:identity2ofEandF^{[m]}_{ij}:pma-quasiperiodicsols} and noting $\textrm{tr}(V)=0$, 
we can compute that
\[
\ba {l}
\D \quad \ \, 
V_{11} +V_{12} \phi_{21}+V_{13}\phi_{31}
\vspace{2mm}
\\
\D 
\quad \ \, =
V_{11} +V_{12}
\frac {
y^2 W_{13}^{[n]}-yA^{[m]}_{21}+B^{[m]}_{21}
}
{
E_{21}^{[m]}
}
-V_{13} 
\frac {y^2 W_{12}^{[n]}-yA^{[m]}_{31}+B^{[m]}_{31}}{E_{21}^{[m]}}
\qquad
\vspace{2mm}
\\
\D \quad \ \, 
=  
\frac 13 \frac {E_{21,z}^{[m]}}{E_{21}^{[m]}} -\frac 23 
\frac {(V_{13}W_{12}^{[n]}- V_{12}W_{13}^{[n]} )S_m}{E_{21}^{[m]}}
\vspace{2mm}
\\
\D \quad \ \, 
\quad +\frac {
y^2(V_{12}W_{13}^{[n]}-V_{13}W_{12}^{[n]})-y(
V_{12}A_{21}^{[m]}-V_{13}A_{31}^{[m]})
}{E_{21}^{[m]}}
\vspace{2mm}
\\
\D \quad \ \, 
=  
\frac 13 \frac {E_{21,z}^{[m]}}{E_{21}^{[m]}}+\frac 23 
\frac {(V_{12}W_{13}^{[n]}- V_{13}W_{12}^{[n]})(3y^2+S_m )}{E_{21}^{[m]}}
\vspace{2mm}
\\
\D \quad \ \, 
\quad -\frac {
V_{12}W_{13}^{[n]}y(y+\frac 
{A_{21}^{[m]} }{W_{13}^{[n]} } )
-V_{13}W_{12}^{[n]}y(y+\frac {
A_{31}^{[m]}
}{
W_{21}^{[n]}
})
}{E_{21}^{[m]}}
\vspace{2mm}
\\
\D 
\quad
\mathop{=}\limits_{\lambda \to \mu _j }
-\frac {\mu _{j,z}}{\lambda - \mu _j }+\textrm{O}(1) 
\vspace{2mm}
\\
\D 
\quad
\mathop{=}\limits_{\lambda \to \mu _j }
\partial _z \ln  (\lambda - \mu _j )+\textrm{O}(1) ,
\ea
\]
where we have used 
the derivative formula
\eqref{eq:E_{ij}^{[m]}_z:pma-quasiperiodicsols} 
and the Dubrovin type equation 
\eqref{eq:mu_j_z:pma-quasiperiodicsols}.
The proof is finished.
\hfill $\Box $ 

Taking $V=U$ and $V^{[r]}$ and noting $\textrm{tr}(U)=\textrm{tr}(V^{[r]})=0$, we can have the following two conclusions from 
Theorem \ref{thm:generalDubrovintypeequations:pma-quasiperiodicsols} and Theorem
\ref{thm:generalzerosofBakerAkhiezerfunctions:pma-quasiperiodicsols}.

\begin{thm}
\label{thm:Dubrovintypeequations:pma-quasiperiodicsols}
Let $u=(p_1,p_2,q_1,q_2)^T$ solve the $r$-th four-component AKNS equations 
\eqref{eq:4cAKNSsh:pma-quasiperiodicsols}, and 
$\Omega_\mu$ be an open and connected set of $ \mathbb{C}^2 $.
If \be
\mu_i(x,t_r)\ne \mu_j(x,t_r),\ \nu_i(x,t_r)\ne \nu_j(x,t_r),\ \xi_i(x,t_r)\ne \xi_j(x,t_r)
\label{eq:distinctconditionsformu_jnu_jxi_j:pma-quasiperiodicsols}
\ee
 for $i\ne j $ and $(x,t_r)\in \Omega_\mu$,
then
the zeros of $E_{21}^{[m]},F_{21}^{[m]}$ and $F_{31}^{[m]}$ satisfy 
the Dubrovin type equations:
\bea 
&&\mu_{j,x}(x,t_r)= 
\frac {\bigl[(p_2(x,t_r)W_{12}^{[n]}-p_1(x,t_r)W_{13}^{[n]})(3y^2+S_m)\bigr]\bigl.\bigr|_{\lambda =\mu_j(x,t_r)}}
{e_{21}^{[m]}\prod_{k=1,\,k\ne j}^g (\mu_j(x,t_r)-\mu_k(x,t_r))}
,\ 
1\le j\le g,\qquad \quad 
\label{eq:mu_jwrtx:pma-quasiperiodicsols}
\\ 
&&
\nu_{j,x}(x,t_r)= 
-\frac { \bigl[q_1(x,t_r)W_{23}^{[n]}(3y^2+S_m)\bigr]\bigl.\bigr|_{\lambda =\nu_j(x,t_r)}}
{f_{21}^{[m]}\prod_{k=0,\,k\ne j}^g (\nu_j(x,t_r)-\nu_k(x,t_r))}
, \ 0\le j\le g, \label{eq:nu_jwrtx:pma-quasiperiodicsols}
\\ 
&&
\xi_{j,x}(x,t_r)= 
-\frac {\bigl[q_2(x,t_r)W_{32}^{[n]}(3y^2+S_m)\bigr]\bigl.\bigr|_{\lambda =\xi_j(x,t_r)}}
{f_{31}^{[m]}\prod_{k=0,\,k\ne j}^g (\xi_j(x,t_r)-\xi_k(x,t_r))}
, \ 0\le j\le g,\label{eq:xi_jwrtx:pma-quasiperiodicsols}
\eea
and 
\bea 
&&\mu_{j,t_r}
(x,t_r)= 
\frac {\bigl[(V_{13}^{[r]}W_{12}^{[n]}-V_{12}^{[r]}W_{13}^{[n]})(3y^2+S_m)\bigr]\bigl.\bigr|_{\lambda =\mu_j(x,t_r)}}
{e_{21}^{[m]}\prod_{k=1,\,k\ne j}^g (\mu_j(x,t_r)-\mu_k(x,t_r))}
,\ 
1\le j\le g,\qquad \quad \label{eq:mu_jwrtt_r:pma-quasiperiodicsols}
\\ 
&&
\nu_{j,t_r}
(x,t_r)= 
\frac {\bigl[(V_{23}^{[r]}W_{21}^{[n]}-V_{21}^{[r]}W_{23}^{[n]})(3y^2+S_m)\bigr]\bigl.\bigr|_{\lambda =\nu_j(x,t_r)}}
{f_{21}^{[m]}\prod_{k=0,\,k\ne j}^g (\nu_j(x,t_r)-\nu_k(x,t_r))}
, \ 0\le j\le g, \label{eq:nu_jwrtt_r:pma-quasiperiodicsols}
\\ 
&&
\xi_{j,t_r}
(x,t_r)= 
\frac {\bigl[(V_{32}^{[r]}W_{31}^{[n]}-V_{31}^{[r]}W_{32}^{[n]})(3y^2+S_m)\bigr]\bigl.\bigr|_{\lambda =\xi_j(x,t_r)}}
{f_{31}^{[m]}\prod_{k=0,\,k\ne j}^g (\xi_j(x,t_r)-\xi_k(x,t_r))}
, \ 0\le j\le g.\label{eq:xi_jwrtt_r:pma-quasiperiodicsols}
\eea
\end{thm}

\noindent {\it Proof}:
Note that now we have the Lax equations
\eqref{eq:Laxeqn_xforW^{[n]}:pma-quasiperiodicsols} and \eqref{eq:Laxeqn_t_rforW^{[n]}:pma-quasiperiodicsols}.
Two immediate applications of Theorem \ref{thm:generalDubrovintypeequations:pma-quasiperiodicsols} to 
the case of 
$V=U$ and $z=x$ and the case of $V=V^{[r]}$ and $z=t_r$
yield the Dubrovin type dynamical equations in 
\eqref{eq:mu_jwrtx:pma-quasiperiodicsols}, \eqref{eq:nu_jwrtx:pma-quasiperiodicsols} and 
\eqref{eq:xi_jwrtx:pma-quasiperiodicsols},
and \eqref{eq:mu_jwrtt_r:pma-quasiperiodicsols}, \eqref{eq:nu_jwrtt_r:pma-quasiperiodicsols} and 
\eqref{eq:xi_jwrtt_r:pma-quasiperiodicsols}, respectively.  This completes the proof of the theorem. 
\hfill $\Box$

\begin{thm}
\label{thm:zerosandpolesofpsi_i:pma-quasiperiodicsols}
Let $P=(\lambda ,y)\in {\cal K}_g\backslash \{P_{\infty_1},P_{\infty_2},P_{\infty_3}\}$, $(x,x_0,t_r,t_{0,r})\in \mathbb{C}^4$, and 
$\Omega_\mu$ be an open and connected set of $ \mathbb{C}^2 $.
Suppose that 
$u=(p_1,p_2,q_1,q_2)^T$ solves the $r$-th four-component AKNS equations \eqref{eq:4cAKNSsh:pma-quasiperiodicsols}. 
If 
the conditions in \eqref{eq:distinctconditionsformu_jnu_jxi_j:pma-quasiperiodicsols} hold 
 for $i\ne j $ and $(x,t_r)\in \Omega_\mu$, and 
\be
\hat{\mu}_j(x,t_r)\ne \hat{\mu}_j(x_0,t_{0,r}),\ 
\hat{\nu}_j(x,t_r)\ne 
\hat{\nu}_j(x_0,t_{0,r})
,\ \hat{\xi}_j(x,t_r)\ne \hat{\xi}_j(x_0,t_{0,r})
\label{eq:distinctconditionsformu_jnu_jxi_jat0:pma-quasiperiodicsols}
\ee
 for every $j$,
then 

\noindent (a) $\psi_1(P,x,x_0,t_r,t_{0,r})$ on ${\cal K}_g\backslash \{P_{\infty_1},P_{\infty_2},P_{\infty_3}\}$ has $g$ zeros, $\hat {\mu}_1(x,t_r),\cdots ,\hat{\mu}_g(x,t_r)$, and $g$ poles, 
$\hat {\mu}_1(x_0,t_{0,r}),\cdots ,\hat{\mu}_g(x_0,t_{0,r})$;
 
\noindent (b) $\psi_2(P,x,x_0,t_r,t_{0,r})$ on ${\cal K}_g\backslash \{P_{\infty_1},P_{\infty_2},P_{\infty_3}\}$ has $g+1$ zeros, $\hat {\nu}_0(x,t_r),\cdots ,\hat{\nu}_g(x,t_r)$, and $g+1$ poles, 
$\hat {\nu}_0(x_0,t_{0,r}),\cdots ,\hat{\nu}_g(x_0,t_{0,r})$;

\noindent (c) $\psi_3(P,x,x_0,t_r,t_{0,r})$ on ${\cal K}_g\backslash \{P_{\infty_1},P_{\infty_2},P_{\infty_3}\}$ has $g+1$ zeros, $\hat {\xi}_0(x,t_r),\cdots ,\hat{\xi}_g(x,t_r)$, and $g+1$ poles, 
$\hat {\xi}_0(x_0,t_{0,r}),\cdots ,\hat{\xi}_g(x_0,t_{0,r})$.

\end{thm}

\noindent {\it Proof}:
We only prove the statement (a), and the proofs for the other two statements can be given similarly. 

Noting that $\textrm{tr}(U)=\textrm{tr}(V^{[r]})=0$ and
considering two cases 
of Theorem \ref{thm:generalzerosofBakerAkhiezerfunctions:pma-quasiperiodicsols}
with  $V=U$ and $V=V^{[r]}$,
we have 
\[\ba{l}
J^{(1)}_r 
\mathop{=}\limits_{\lambda \to \mu _j }
\partial _x \ln  (\lambda - \mu _j )+\textrm{O}(1) ,
\vspace{2mm}\\
I^{(1)}_r
\mathop{=}\limits_{\lambda \to \mu _j }
\partial _{t_r} \ln  (\lambda - \mu _j )+\textrm{O}(1) ,
\ea\]
where $1\le j\le g$.
Consequently, for each $1\le j\le g$,
we can compute that 
\[
\ba {l}
\quad \psi_{1}(P,x,x_0,t_r,t_{0,r})
\vspace{2mm}\\
= \textrm{exp}(\int_{x_0}^x J_r^{(1)}(P,x',t_r)\, dx'
+\int_{t_{0,r}}^{t_r}I_{r}^{(1)}(P,x_0,t')\, dt'
\vspace{2mm}\\
=\frac {\lambda -\mu_j(x,t_r)} {\lambda -\mu_j (x_0,t_r)}
\frac {\lambda -\mu_j(x_0,t_r)}{\lambda -\mu_j (x_0,t_{0,r})}
\textrm{O}(1)
\vspace{2mm}\\
=\frac {\lambda -\mu_j(x,t_r)}{\lambda -\mu_j (x_0,t_{0,r})}
\textrm{O}(1)
\vspace{2mm}\\
=\left \{\ba {ll}
(\lambda -\mu_j(x,t_r))
\textrm{O}(1) & \textrm{for}\ P \ \textrm{near}\ \hat {\mu}_j(x,t_r)\ne \hat {\mu }_j(x_0,t_{0,r}),
\vspace{2mm}\\
 \textrm{O}(1) & \textrm{for}\ P \ \textrm{near}\ \hat {\mu}_j(x,t_r)= \hat {\mu }_j(x_0,t_{0,r}),
\vspace{2mm}\\
(\lambda -\mu_j(x_0,t_{0,r}))^{-1}
\textrm{O}(1) & \textrm{for}\ P \ \textrm{near}\ \hat {\mu}_j(x_0,t_{0,r})\ne \hat {\mu }_j(x,t_{r}),
\ea  \right.
\ea
\]
where $\textrm{O}(1) \ne 0$. Under the conditions in 
\eqref{eq:distinctconditionsformu_jnu_jxi_jat0:pma-quasiperiodicsols},
this leads to the statement (a), which completes the proof. 
\hfill $\Box$

This theorem determines zeros and poles of the Baker-Akhiezer functions 
$\psi_i$, $1\le i\le 3$, in ${\cal K}_g\backslash \{P_{\infty_1},P_{\infty_2},P_{\infty_3}\}$.

\section{Asymptotic behaviors}

\label{sec:AsymptoticBehaviors:pma-quasiperiodicsols}

In order to generate 
algebro-geometric solutions in terms of the Riemann theta functions, 
we need to explore asymptotic properties of the Baker-Akhiezer functions $\psi_i$, $1\le i\le 3$.

\subsection{Asymptotics of the first Baker-Akhiezer function}

We first start with determining asymptotic properties of the meromorphic functions 
$\phi_{21}$ and $\phi_{31}$ at the points at infinity.

\begin{lem}
\label{lem:asymptoticbehavioursofphi_{21}andphi_{31}:pma-quasiperiodicsols}
Let $u=(p_1,p_2,q_1,q_2)^T$ satisfy 
the $r$-th 
four-component
AKNS equations \eqref{eq:4cAKNSsh:pma-quasiperiodicsols} and $\zeta=\lambda^{-1}$.
Suppose that $P\in {\cal K}_g\backslash \{P_{\infty_1},P_{\infty_2},P_{\infty_3}\}$ and $(x,t_r)\in \mathbb {C}^2$. Then 
\be 
\phi_{21}(P,x,t_r)
\mathop{=}\limits_{\zeta\to  0}\left\{
                                                     \begin{array}{ll}
\frac 3 {p_1} \zeta^{-1} +\frac {p_{1,x}-p_1p_2\chi_{1,0} }{p_1^2}+\kappa_{1,1}\zeta
+\textrm{O}(\zeta^{2}),& \textrm{as}\ P\to  P_{\infty_1},
\vspace{2mm}
\\
 \kappa_{2,0}+ \kappa _{2,1}\zeta  
+\textrm{O}(\zeta^{2}),& \textrm{as}\ P\to  P_{\infty_2},
\vspace{2mm}\\
-\frac {q_1} 3 \zeta -
\frac {q_{1,x}}{9}\zeta ^2 
-
\frac 
{q_{1,xx}-p_1q_1^2-p_2q_1q_2}{27} 
\zeta ^3
+\textrm{O}(\zeta^{4}),& \textrm{as}\ P\to  P_{\infty_3},
                                                     \end{array}
                                                 \right.
\label{eq:asymptoticbehavioursofphi_{21}:pma-quasiperiodicsols}
\ee
and 
\be 
\phi_{31}(P,x,t_r)
\mathop{=}\limits_{\zeta\to  0}\left\{
                                                     \begin{array}{ll}
\chi_{1,0} +\chi_{1,1}\zeta 
+\textrm{O}(\zeta^{2}),& \textrm{as}\ P\to  P_{\infty_1},
\vspace{2mm}
\\
 \frac 3 {p_2} \zeta^{-1}+ \frac {p_{2,x}-p_1p_2\kappa_{2,0}}{p_2^2}+\chi_{2,1}\zeta   
+\textrm{O}(\zeta^{2}),& \textrm{as}\ P\to  P_{\infty_2},
\vspace{2mm}\\
-\frac {q_2} {3} \zeta 
- \frac {q_{2,x}}9
\zeta^2
-
\frac {q_{2,xx}-p_2q_2^2 -p_1q_1q_2}{27}
 \zeta ^3
+ 
+\textrm{O}(\zeta^{4}),& \textrm{as}\ P\to  P_{\infty_3},
                                                     \end{array}
                                                 \right.
\label{eq:asymptoticbehavioursofphi_{31}:pma-quasiperiodicsols}
\ee
where 
\[ 
\left\{
\ba {l}
(p_1\chi_{1,0})_x=p_1q_2,\vspace{2mm}\\
(p_1\chi_{1,1})_x= -\frac {\chi_{1,0}} {3p_1}
(p_1^2 p_{2,x}\chi_{1,0} - p_1p_{1,x}p_2  +p_1^3 q_1+p_1^2 p_2q_2-p_1p_{1,xx}+p_{1,x}^2),
\vspace{2mm}\\
 \kappa_{1,1}=\frac 1{3p_1^3}
(p_1^2 p_{2,x}\chi_{1,0}-p_1p_{1,x}p_2 \chi_{1,0} 
-3 p_1^2 p_2 \chi_{1,1}+p_1^3 q_1+p_1^2 p_2 q_2 -p_1 p_{1,xx}+p_{1,x}^2
),
\ea \right.
\]
and 
\[
\left\{\ba {l}
(p_2\kappa_{2,0})_x= p_2q_1,
\vspace{2mm}\\ 
(p_2\kappa_{2,1})_x
=\frac {\kappa_{2,0}} {3p_2}
(p_1p_2 p_{2,x}\kappa_{2,0} - p_{1,x}p_2^2 -
p_1p_2^2q_1-p_2^3 q_2+p_2p_{2,xx}-p_{2,x}^2),
\vspace{2mm}\\
\chi_{2,1}=
-\frac 1{3p_2^3}
(p_1p_2 p_{2,x}\kappa_{2,0}-p_{1,x}p_2^2 \kappa_{2,0} 
+3 p_1 p_2^2 \kappa_{2,1}- p_1p_2^2 q_1 -p_2^3  q_2 +p_2 p_{2,xx}-p_{2,x}^2
)
.
\ea \right.
\]
\end{lem}

\noindent {\it Proof}: 
We begin with 
the following three ansatzes:
\[
       \begin{array}{l}
\phi_{21}
\mathop{=}\limits_{\zeta\to  0}               
\kappa_{1,-1}\zeta^{-1} +\kappa_{1,0}+\kappa _{1,2}\zeta ^2
+O(\zeta^{3}),\ 
\phi_{31}
\mathop{=}\limits_{\zeta\to  0}               
\chi_{1,0} +\chi_{1,1}\zeta
+O(\zeta^{2}),\ 
\textrm{as}\ P\to  P_{\infty_1};
\vspace{2mm}
\\
           \phi_{21}
\mathop{=}\limits_{\zeta\to  0}       
\kappa_{2,0}+ \kappa _{2,1}\zeta  
+O(\zeta^{2}),
\ 
 \phi_{31}
\mathop{=}\limits_{\zeta\to  0}       
\chi_{2,-1}\zeta ^{-1}+ \chi _{2,0} +  \chi_{2,1}\zeta  
+O(\zeta^{2}),
\  \textrm{as}\ P\to  P_{\infty_2};
\vspace{2mm}\\
 \phi_{21}
\mathop{=}\limits_{\zeta\to  0}       
\kappa_{3,1}\zeta +\kappa_{3,2}\zeta^2 + \kappa _{3,3}\zeta^3 
+O(\zeta^{4}),
\ 
 \phi_{31}
\mathop{=}\limits_{\zeta\to  0}       
\chi_{3,1}\zeta + \chi _{3,2}\zeta^2 +  \chi_{3,3}\zeta ^3 
+O(\zeta^{4}),
\  \textrm{as}\ P\to  P_{\infty_3};
                                                     \end{array}
\]
where the coefficients, $\kappa_{ij}$ and $\chi_{ij}$, are functions to be determined.
Substituting those expansions into the Riccati type equations
\eqref{eq:phi_{ij}_x:pma-quasiperiodicsols} with $i=2,3$ and $j=1$, i.e., 
\be 
\phi_{21,x}=q_1+3\lambda \phi_{21}-p_1\phi_{21}^2-p_2\phi_{21}\phi_{31},
\ 
\phi_{31,x}=q_2+3\lambda \phi_{31}-p_1\phi_{21}\phi_{31}-p_2\phi_{31}^2,
\ee 
and comparing the three lowest powers $\zeta^i$ in each resulting equation, where $i$ goes either from $-2$ to $0$, or from $-1$ to $1$, or from $0$ to $2$, we obtain
a set of relations on the coefficient functions
$\kappa_{i,j}$ and $\chi_{i,j}$,
which yields the asymptotic properties in \eqref{eq:asymptoticbehavioursofphi_{21}:pma-quasiperiodicsols} and 
\eqref{eq:asymptoticbehavioursofphi_{31}:pma-quasiperiodicsols}.
 The proof is finished.
\hfill $\Box$

To determine asymptotic properties of the Baker-Akhiezer function $\psi_1$ at the points at infinity,
we now analyze  
\be 
J_r^{(1)}=U_{11} +U_{12}\phi_{21} +U_{13}\phi_{31}=
-2\lambda + p_1\phi_{21}+p_2 \phi_{31},
\label{eq:defofJ_{r}^{(1)}:pma-quasiperiodicsols}
\ee
and 
\be
I_r^{(1)}=V_{11}^{[r]}+V_{12}^{[r]}\phi_{21}+V_{13}^{[r]}\phi _{31}.
\label{eq:defofI_{r}^{(1)}:pma-quasiperiodicsols}
 \ee

\begin{lem}
\label{lem:asymptoticbehavioursofI_{r}^{(1)}anfJ_{r}^{(1)}:pma-quasiperiodicsols}
Let $u=(p_1,p_2,q_1,q_2)^T$ satisfy 
the $r$-th 
four-component 
AKNS equations \eqref{eq:4cAKNSsh:pma-quasiperiodicsols} and $\zeta=\lambda^{-1}$.
Suppose that $P\in {\cal K}_g\backslash \{P_{\infty_1},P_{\infty_2},P_{\infty_3}\}$ and $(x,t_r)\in \mathbb {C}^2$. Then 
\be 
J_{r}^{(1)}(P,x,t_r)
\mathop{=}\limits_{\zeta\to  0}\left\{
                                                     \begin{array}{ll}
 \zeta^{-1} +\frac {p_{1,x}}{p_1}
+\textrm{O}(\zeta),& \textrm{as}\ P\to  P_{\infty_1},
\vspace{2mm}
\\
  \zeta^{-1}+\frac {p_{2,x}}{p_2}
+\textrm{O}(\zeta),& \textrm{as}\ P\to  P_{\infty_2},
\vspace{2mm}\\
-2 \zeta ^{-1}
+\textrm{O}(\zeta),& \textrm{as}\ P\to  P_{\infty_3},
                                                     \end{array}
                                                 \right.
\label{eq:asymptoticbehavioursofJ_{r}^{[1]}:pma-quasiperiodicsols}
\ee
and 
\be 
I_r^{(1)}(P,x,t_r)
\mathop{=}\limits_{\zeta\to  0}\left\{
                                                     \begin{array}{ll}
\zeta ^{-r}+\frac {p_{1,t_r}}{p_1}
+\textrm{O}(\zeta),& \textrm{as}\ P\to  P_{\infty_1},
\vspace{2mm}
\\
\zeta^{-r}+\frac {p_{2,t_r}}{p_2}
+\textrm{O}(\zeta),& \textrm{as}\ P\to  P_{\infty_2},
\vspace{2mm}\\
-2\zeta^{-r}
+\textrm{O}(\zeta),& \textrm{as}\ P\to  P_{\infty_3}.
                                                     \end{array}
                                                 \right.
\label{eq:asymptoticbehavioursofI_{r}^{[1]}:pma-quasiperiodicsols}
\ee
\end{lem}

\noindent {\it Proof}:
First, 
based on \eqref{eq:defofJ_{r}^{(1)}:pma-quasiperiodicsols},
we obatin 
\eqref{eq:asymptoticbehavioursofJ_{r}^{[1]}:pma-quasiperiodicsols}
directly from Lemma \ref{lem:asymptoticbehavioursofphi_{21}andphi_{31}:pma-quasiperiodicsols}.

Second, 
 note that 
the first compatibility condition in   \eqref{eq:compatibilityconditions:pma-quasiperiodicsols}
reads
\be I_{r,x}^{(1)}=\bigl(\frac {\psi_{1,t_r}}{\psi_1}\Bigr)_x = 
\bigl(\frac {\psi_{1,x}}{\psi_1}\Bigr)_{t_r}=
J_{r,t_r}^{(1)},
\label{eq:1stcompatabilitycondition:pma-quasiperiodicsols}
\ee
and that from \eqref{eq:recursionrelationofV_{r}:pma-quasiperiodicsols},
we obtain
\[
V_{11}^{[r+1]}=\lambda V_{11}^{[r]}+a^{[r+1]},\ 
V_{12}^{[r+1]}=\lambda V_{12}^{[r]}+b_{1}^{[r+1]},\  
V_{13}^{[r+1]}=\lambda V_{13}^{[r]}+b_{2}^{[r+1]},
\]
and thus, we have 
\be I_{r+1}^{(1)}=\lambda I_r^{(1)}+a^{[r+1]}+b_1^{[r+1]}\phi_{21}+b_{2}^{[r+1]}\phi_{31}.
 \label{eq:recursionrelationof1stcolumnofV_{r}:pma-quasiperiodicsols}\ee
Now,
based on \eqref{eq:1stcompatabilitycondition:pma-quasiperiodicsols} and 
\eqref{eq:recursionrelationof1stcolumnofV_{r}:pma-quasiperiodicsols},
we can verify 
\eqref{eq:asymptoticbehavioursofI_{r}^{[1]}:pma-quasiperiodicsols}
from 
\eqref{eq:asymptoticbehavioursofJ_{r}^{[1]}:pma-quasiperiodicsols} 
by the mathematical induction.
 The proof is finished.
\hfill $\Box$

\begin{thm}
Let $u=(p_1,p_2,q_1,q_2)^T$ satisfy 
the $r$-th four-component AKNS equations \eqref{eq:4cAKNSsh:pma-quasiperiodicsols} and $\zeta=\lambda^{-1}$.
Suppose that $P\in {\cal K}_g\backslash \{P_{\infty_1},P_{\infty_2},P_{\infty_3}\}$ and $(x,t_r)\in \mathbb {C}^2$. Then we have
\bea &&
\psi_1(P,x,x_0,t_r,t_{0,r})
\nonumber 
\\
&&
\mathop{=}\limits_{\zeta\to  0}\left\{
                                                     \begin{array}{ll}
\frac {p_{1}(x,t_r)}{p_1(x_0,t_{0,r})}
{\rm{exp}}\bigl(\zeta^{-1}(x-x_0)+  \zeta^{-r} (t_r-t_{0,r}) 
+\textrm{O}(\zeta)\bigr),& \textrm{as}\ P\to P_{\infty_1},
\vspace{2mm}
\\
\frac {p_{2}(x,t_r)}{p_2(x_0,t_{0,r})}
{\rm{exp}}\bigl(
\zeta^{-1}(x-x_0)+  \zeta^{-r} (t_r-t_{0,r}) 
+ \textrm{O}(\zeta)\bigr),& \textrm{as}\ P\to P_{\infty_2},
\vspace{2mm}\\
{\rm{exp}}\bigl(-2\zeta^{-1}(x-x_0)-2 \zeta^{-r} (t_r-t_{0,r}) 
+\textrm{O}(\zeta)\bigr),& \textrm{as}\ P\to P_{\infty_3}.
                                                     \end{array}
                                                 \right.
\qquad \qquad 
\label{eq:asymptoticbehavioursofpsi_1:pma-quasiperiodicsols}
\eea
\end{thm}

\noindent {\it Proof}:
The first formula in
\eqref{eq:expressionsforpsi_i:pma-quasiperiodicsols} on the Baker-Akhiezer function
 $\psi_1$ gives
\[
\psi_1(P,x,x_0,t_r,t_{0,r})=\, {\rm{exp}}\Bigl(
\int_{x_0}^xJ_r^{(1)}(P,x',t_r)\, dx'
+\int_{t_{0,r}}^{t_r}I_r^{(1)}(P,x_0,t')\, dt'
\Bigr),
\]
where $J_r^{(1)}$ and $I_r^{(1)}$ are defined by 
\eqref{eq:defofJ_{r}^{(1)}:pma-quasiperiodicsols}
and \eqref{eq:defofI_{r}^{(1)}:pma-quasiperiodicsols}.
Based on 
Lemma 
\ref{lem:asymptoticbehavioursofI_{r}^{(1)}anfJ_{r}^{(1)}:pma-quasiperiodicsols},
this expression generates the asymptotic properties of 
$\psi_1$ in \eqref{eq:asymptoticbehavioursofpsi_1:pma-quasiperiodicsols}.
The proof is finished.
\hfill $\Box$

\subsection{Asymptotics of the second Baker-Akhiezer function}

We now start with determining asymptotic properties of the meromorphic functions 
$\phi_{12}$ and $\phi_{32}$ at the points at infinity.

\begin{lem}
\label{lem:asymptoticbehavioursofphi_{12}andphi_{32}:pma-quasiperiodicsols}
Let $u=(p_1,p_2,q_1,q_2)^T$ satisfy 
the $r$-th 
four-component AKNS equations \eqref{eq:4cAKNSsh:pma-quasiperiodicsols} and $\zeta=\lambda^{-1}$.
Suppose that $P\in {\cal K}_g\backslash \{P_{\infty_1},P_{\infty_2},P_{\infty_3}\}$ and $(x,t_r)\in \mathbb {C}^2$. Then 
\be 
\phi_{12}(P,x,t_r)
\mathop{=}\limits_{\zeta\to  0}\left\{
                                                     \begin{array}{ll}
\frac {p_1}3 \zeta +(\frac {p_2}3\chi_{1,1}-\frac 19 p_{1,x})\zeta^2+\kappa _{1,3}\zeta ^3 
+\textrm{O}(\zeta^{4}),& \textrm{as}\ P\to  P_{\infty_1},
\vspace{2mm}
\\
                                                            \frac{1}{3}
p_2\chi_{2,-1}+ \kappa _{2,1}\zeta +  \kappa_{2,2}\zeta ^2 
+\textrm{O}(\zeta^{3}),& \textrm{as}\ P\to  P_{\infty_2},
\vspace{2mm}\\
-\frac 3 {q_1} \zeta ^{-1}+
\frac {q_{1,x}}{q_1^2}+ \frac {q_1q_{1,xx}-q_{1,x}^2 -p_1q_1^3-p_2q_1^2q_2}{3q_1^3}\zeta 
+\textrm{O}(\zeta^{2}),& \textrm{as}\ P\to  P_{\infty_3},
                                                     \end{array}
                                                 \right.
\label{eq:asymptoticbehavioursofphi_{12}:pma-quasiperiodicsols}
\ee
and 
\be 
\phi_{32}(P,x,t_r)
\mathop{=}\limits_{\zeta\to  0}\left\{
                                                     \begin{array}{ll}
\chi_{1,1}\zeta +\chi_{1,2}\zeta^2 
+\textrm{O}(\zeta^{3}),& \textrm{as}\ P\to  P_{\infty_1},
\vspace{2mm}
\\
 \chi_{2,-1}\zeta^{-1}+ \chi_{2,0}+\chi_{2,1}\zeta   
+\textrm{O}(\zeta^{2}),& \textrm{as}\ P\to  P_{\infty_2},
\vspace{2mm}\\
\frac {q_2} {q_1} +
\frac 13 \bigl(\frac {q_{2}}{q_1}\bigr)_x \zeta 
+ 
\frac 19 \bigl[  
\bigl(\frac {q_2}{q_1}\bigr)_{xx}
+\frac {q_{1,x}}{q_1} \bigl(\frac {q_2}{q_1} \bigr)_x
\bigr]
\zeta ^2
+\textrm{O}(\zeta^{3}),& \textrm{as}\ P\to  P_{\infty_3},
                                                     \end{array}
                                                 \right.
\label{eq:asymptoticbehavioursofphi_{32}:pma-quasiperiodicsols}
\ee
where 
\[ 
\left\{
\ba {l}
\chi_{1,1,x}=\frac 13 p_1q_2,\ \chi_{1,2,x}=\frac 13 (p_2q_2-p_1q_1)\chi_{1,1}-\frac 19 p_{1,x}q_2,
\vspace{2mm}\\
\kappa_{1,3}=-\frac 19 p_{2,x}\chi_{1,1}+\frac 13 p_2 \chi_{1,2}
-\frac 1 {27}p_1(p_1q_1+p_2q_2)+\frac 1 {27}p_{1,xx},
\ea \right.
\]
and 
\[ 
\left\{\ba {l}
\chi_{2,-1,x}=-\frac 13 p_2q_1 \chi_{2,-1}^2,
\ 
\kappa_{2,1}
=-\frac 19 p_{2,x}\chi_{2,-1}+\frac 13 p_2 \chi_{2,0}+\frac 13 p_1,
\vspace{2mm}\\
\kappa_{2,2}=-\frac 1 {27} p_1p_2q_1 \chi_{2,-1} -\frac 1 {27} p_2^2q_2 \chi_{2,-1}+\frac 1 {27}
p_{2,xx}\chi_{2,-1} -\frac 19 p_{2,x}\chi_{2,0}+\frac 13 p_2\chi_{2,1}-\frac 19 p_{1,x},
\vspace{2mm}
\\
\chi_{2,0,x}+\frac 23 p_2q_1 \chi_{2,-1}\chi_{2,0}
-\frac 19 p_{2,x}q_1\chi_{2,-1}^2+\frac 13 (p_1q_1-p_2q_2)\chi_{2,-1}=0,
\vspace{2mm}\\
\chi_{2,1,x}+\frac 23 p_2q_1\chi_{2,-1}\chi_{2,1}
-\frac 1{27}p_1p_2  q_{1} ^{2}   
\chi_{2,-1}  ^{2}
-\frac1{27}
p_2^2 q_1 q_2  
  \chi_{2,-1} ^{2}
 +\frac 1{27}p_{2,xx}
q_{1}   \chi_{2,-1}  ^{2}
+\frac 13
p_2q_1\chi_{2,-1}  ^{2} 
\vspace{2mm}\\
\quad 
-\frac 29
p_{2,x}q_{1}
\chi_{2,-1}  \chi_{2,0}  
+\frac 13
p_1q_{1}  \chi_{2,0}  
-\frac 13
p_2q_{2}  \chi_{2,0} 
+\frac 19
p_{2,x}q_{2} 
  \chi_{2,-1}
-\frac 19
p_{1,x}q_{1} \chi_{2,-1}  
-\frac 13
p_1q_{2}=0
.
\ea \right.
\]
\end{lem}

\noindent {\it Proof}: 
We begin with 
the following three ansatzes:
\[
       \begin{array}{l}
\phi_{12}
\mathop{=}\limits_{\zeta\to  0}               
\kappa_{1,1}\zeta +\kappa_{1,2}\zeta^2+\kappa _{1,3}\zeta ^3 
+O(\zeta^{4}),\ 
\phi_{32}
\mathop{=}\limits_{\zeta\to  0}               
\chi_{1,1}\zeta +\chi_{1,2}\zeta^2
+O(\zeta^{3}),\ 
\textrm{as}\ P\to  P_{\infty_1};
\vspace{2mm}
\\
           \phi_{12}
\mathop{=}\limits_{\zeta\to  0}       
\kappa_{2,0}+ \kappa _{2,1}\zeta +  \kappa_{2,2}\zeta ^2 
+O(\zeta^{3}),
\ 
 \phi_{32}
\mathop{=}\limits_{\zeta\to  0}       
\chi_{2,-1}\zeta ^{-1}+ \chi _{2,0} +  \chi_{2,1}\zeta  
+O(\zeta^{2}),
\  \textrm{as}\ P\to  P_{\infty_2};
\vspace{2mm}\\
 \phi_{12}
\mathop{=}\limits_{\zeta\to  0}       
\kappa_{3,-1}\zeta ^{-1}+\kappa_{3,0}+ \kappa _{3,1}\zeta 
+O(\zeta^{2}),
\ 
 \phi_{32}
\mathop{=}\limits_{\zeta\to  0}       
\chi_{3,0}+ \chi _{3,1}\zeta +  \chi_{3,2}\zeta ^2 
+O(\zeta^{3}),
\  \textrm{as}\ P\to  P_{\infty_3};
                                                     \end{array}
\]
where the coefficients, $\kappa_{ij}$ and $\chi_{ij}$, are functions to be determined.
Substituting those expansions into the Riccati type equations
\eqref{eq:phi_{ij}_x:pma-quasiperiodicsols} with $i=1,3$ and $j=2$, i.e., 
\be
\phi_{12,x}=-3\lambda \phi_{12}+p_1+p_2\phi_{32}-q_1\phi_{12}^2,
\ 
\phi_{32,x}=q_2\phi_{12}-q_1\phi_{12}\phi_{32},
\ee
and comparing the three lowest powers $\zeta^i$ in each resulting equation, where $i$ goes either from $-2$ to $0$, or from $-1$ to $1$, or from $0$ to $2$, we obtain
a set of relations on the coefficient functions
$\kappa_{i,j}$ and $\chi_{i,j}$,
which leads to the asymptotic properties in \eqref{eq:asymptoticbehavioursofphi_{12}:pma-quasiperiodicsols} and 
\eqref{eq:asymptoticbehavioursofphi_{32}:pma-quasiperiodicsols}.
 This proves the lemma.
\hfill $\Box$

To determine asymptotic properties of the Baker-Akhiezer function $\psi_2$ at the points at infinity,
we now analyze 
\be 
J_r^{(2)}=U_{21}\phi_{12} +U_{22} +U_{23}\phi_{32}=
q_1\phi_{12}+\lambda ,
\label{eq:defofJ_{r}^{(2)}:pma-quasiperiodicsols}
\ee
and 
\be
I_r^{(2)}=V_{21}^{[r]}\phi_{12}+V_{22}^{[r]}+V_{23}^{[r]}\phi _{32}.
\label{eq:defofI_{r}^{(2)}:pma-quasiperiodicsols}
 \ee

\begin{lem}
\label{lem:asymptoticbehavioursofI_{r}^{(2)}anfJ_{r}^{(2)}:pma-quasiperiodicsols}
Let $u=(p_1,p_2,q_1,q_2)^T$ satisfy 
the $r$-th 
four-component
AKNS equations \eqref{eq:4cAKNSsh:pma-quasiperiodicsols} and $\zeta=\lambda^{-1}$.
Suppose that $P\in {\cal K}_g\backslash \{P_{\infty_1},P_{\infty_2},P_{\infty_3}\}$ and $(x,t_r)\in \mathbb {C}^2$. Then 
\be 
J_{r}^{(2)}(P,x,t_r)
\mathop{=}\limits_{\zeta\to  0}\left\{
                                                     \begin{array}{ll}
 \zeta^{-1} 
+\textrm{O}(\zeta),& \textrm{as}\ P\to  P_{\infty_1},
\vspace{2mm}
\\
  \zeta^{-1}+\rho_r ^{(2)}
+\textrm{O}(\zeta),& \textrm{as}\ P\to  P_{\infty_2},
\vspace{2mm}\\
-2 \zeta ^{-1}+
\frac {q_{1,x}}{q_1}
+\textrm{O}(\zeta),& \textrm{as}\ P\to  P_{\infty_3},
                                                     \end{array}
                                                 \right.
\label{eq:asymptoticbehavioursofJ_{r}^{(2)}:pma-quasiperiodicsols}
\ee
and 
\be 
I_r^{(2)}(P,x,t_r)
\mathop{=}\limits_{\zeta\to  0}\left\{
                                                     \begin{array}{ll}
\zeta ^{-r}
+\textrm{O}(\zeta),& \textrm{as}\ P\to  P_{\infty_1},
\vspace{2mm}
\\
\zeta^{-r}+ \sigma_{r}^{(2)}
+\textrm{O}(\zeta),& \textrm{as}\ P\to  P_{\infty_2},
\vspace{2mm}\\
-2\zeta^{-r}+
\frac {q_{1,t_r}}{q_1}  
+\textrm{O}(\zeta),& \textrm{as}\ P\to  P_{\infty_3},
                                                     \end{array}
                                                 \right.
\label{eq:asymptoticbehavioursofI_{r}^{(2)}:pma-quasiperiodicsols}
\ee
where 
$\rho_r^{(2)}=\frac 13 p_2q_1\chi_{2,-1}$ and $\sigma_{r,x}^{(2)}=\rho_{r,t_r}^{(2)}$, with $\chi_{2,-1}$ being defined in Lemma \ref{lem:asymptoticbehavioursofphi_{12}andphi_{32}:pma-quasiperiodicsols}.
\end{lem}

\noindent {\it Proof}:
First, 
based on \eqref{eq:defofJ_{r}^{(2)}:pma-quasiperiodicsols},
we obatin 
\eqref{eq:asymptoticbehavioursofJ_{r}^{(2)}:pma-quasiperiodicsols}
directly from Lemma \ref{lem:asymptoticbehavioursofphi_{12}andphi_{32}:pma-quasiperiodicsols}.

Second, 
 note that the second compatibility condition in 
\eqref{eq:compatibilityconditions:pma-quasiperiodicsols}
reads
\be I_{r,x}^{(2)}=\bigl(\frac {\psi_{2,t_r}}{\psi_2}\Bigr)_x = 
\bigl(\frac {\psi_{2,x}}{\psi_2}\Bigr)_{t_r}=
J_{r,t_r}^{(2)},
\label{eq:2ndcompatabilitycondition:pma-quasiperiodicsols}
\ee
and that from \eqref{eq:recursionrelationofV_{r}:pma-quasiperiodicsols},
we get
\[
V_{21}^{[r+1]}=\lambda V_{21}^{[r]}+c_1^{[r+1]},\ 
V_{22}^{[r+1]}=\lambda V_{22}^{[r]}+d_{11}^{[r+1]},\  
V_{23}^{[r+1]}=\lambda V_{23}^{[r]}+d_{12}^{[r+1]},
\]
and this leads to
\be I_{r+1}^{(2)}=\lambda I_r^{(2)}+c_1^{[r+1]}\phi_{12}+d_{11}^{[r+1]}+d_{12}^{[r+1]}\phi_{32}.
 \label{eq:recursionrelationof2ndcolumnofV_{r}:pma-quasiperiodicsols}\ee
Now,
based on \eqref{eq:2ndcompatabilitycondition:pma-quasiperiodicsols} and 
\eqref{eq:recursionrelationof2ndcolumnofV_{r}:pma-quasiperiodicsols},
we can prove
\eqref{eq:asymptoticbehavioursofI_{r}^{(2)}:pma-quasiperiodicsols}
from 
\eqref{eq:asymptoticbehavioursofJ_{r}^{(2)}:pma-quasiperiodicsols} 
by the mathematical induction.
 This competes the proof.
\hfill $\Box$

\begin{thm}
Let $u=(p_1,p_2,q_1,q_2)^T$ satisfy 
the $r$-th four-component AKNS equations \eqref{eq:4cAKNSsh:pma-quasiperiodicsols} and $\zeta=\lambda^{-1}$.
Suppose that $P\in {\cal K}_g\backslash \{P_{\infty_1},P_{\infty_2},P_{\infty_3}\}$ and $(x,t_r)\in \mathbb {C}^2$. Then we have
\bea &&
\psi_2(P,x,x_0,t_r,t_{0,r})
\nonumber 
\\
&&
\mathop{=}\limits_{\zeta\to  0}\left\{
                                                     \begin{array}{ll}
{\rm{exp}}\bigl(\zeta^{-1}(x-x_0)+  \zeta^{-r} (t_r-t_{0,r}) 
+\textrm{O}(\zeta)\bigr),& \textrm{as}\ P\to P_{\infty_1},
\vspace{2mm}
\\
{\rm{exp}}\bigl(
  \int_{x_0}^x \rho_r^{(2)} (P,x',t_r)\, dx'+\int_{t_{0,r}}^{t_r}
 \sigma_r^{(2)}(P,x_0,t')\, dt' & 
\vspace{2mm}\\
\quad\  +\zeta^{-1}(x-x_0)+  \zeta^{-r} (t_r-t_{0,r}) 
+ \textrm{O}(\zeta)\bigr),& \textrm{as}\ P\to P_{\infty_2},
\vspace{2mm}\\
\frac {q_1(x,t_r)}{q_1(x_0,t_{0,r})}
{\rm{exp}}\bigl(-2\zeta^{-1}(x-x_0)-2 \zeta^{-r} (t_r-t_{0,r}) 
+\textrm{O}(\zeta)\bigr),& \textrm{as}\ P\to P_{\infty_3},
                                                     \end{array}
                                                 \right.
\qquad \qquad 
\label{eq:asymptoticbehavioursofpsi_2:pma-quasiperiodicsols}
\eea
where $\rho_r^{(2)}$ and $\sigma_{r}^{(2)}$ are defined in
Lemma \ref{lem:asymptoticbehavioursofI_{r}^{(2)}anfJ_{r}^{(2)}:pma-quasiperiodicsols}.
\end{thm}

\noindent {\it Proof}:
The second formula in
\eqref{eq:expressionsforpsi_i:pma-quasiperiodicsols} 
presents
\[
\psi_2(P,x,x_0,t_r,t_{0,r})=\, {\rm{exp}}\Bigl(
\int_{x_0}^xJ_r^{(2)}(P,x',t_r)\, dx'
+\int_{t_{0,r}}^{t_r}I_r^{(2)}(P,x_0,t')\, dt'
\Bigr),
\]
where $J_r^{(2)}$ and $I_r^{(2)}$ are given by 
\eqref{eq:defofJ_{r}^{(2)}:pma-quasiperiodicsols}
and \eqref{eq:defofI_{r}^{(2)}:pma-quasiperiodicsols}.
This expression generates the asymptotic properties of 
the Baker-Akhiezer function 
$\psi_2$ in \eqref{eq:asymptoticbehavioursofpsi_2:pma-quasiperiodicsols}, based on 
Lemma 
\ref{lem:asymptoticbehavioursofI_{r}^{(2)}anfJ_{r}^{(2)}:pma-quasiperiodicsols}.
The proof is finished.
\hfill $\Box$

\subsection{Asymptotics of the third Baker-Akhiezer function}

We thirdly start with determining asymptotic properties of the meromorphic functions 
$\phi_{13}$ and $\phi_{23}$ at the points at infinity.

\begin{lem}
\label{lem:asymptoticbehavioursofphi_{13}andphi_{23}:pma-quasiperiodicsols}
Let $u=(p_1,p_2,q_1,q_2)^T$ satisfy 
the $r$-th 
four-component
AKNS equations \eqref{eq:4cAKNSsh:pma-quasiperiodicsols} and $\zeta=\lambda^{-1}$.
Suppose that $P\in {\cal K}_g\backslash \{P_{\infty_1},P_{\infty_2},P_{\infty_3}\}$ and $(x,t_r)\in \mathbb {C}^2$. Then 
\be 
\phi_{13}(P,x,t_r)
\mathop{=}\limits_{\zeta\to  0}\left\{
                                                     \begin{array}{ll}
                                                   \frac{1}{3}
p_1\chi_{1,-1}+ \kappa _{1,1}\zeta +  \kappa_{1,2}\zeta ^2
+\textrm{O}(\zeta^{3}),
& \textrm{as}\ P\to  P_{\infty_1},
\vspace{2mm}
\\
\frac {p_2}3 \zeta +(\frac {p_1}3\chi_{2,1}-\frac 19 p_{2,x})\zeta^2+\kappa _{2,3}\zeta ^3 
+\textrm{O}(\zeta^{4}),
& \textrm{as}\ P\to  P_{\infty_2},
\vspace{2mm}\\
-\frac 3 {q_2} \zeta ^{-1}+
\frac {q_{2,x}}{q_2^2}+ \frac {q_2q_{2,xx}-q_{2,x}^2 -p_2q_2^3-p_1q_1q_2^2}{3q_2^3}\zeta 
+\textrm{O}(\zeta^{2}),& \textrm{as}\ P\to  P_{\infty_3},
                                                     \end{array}
                                                 \right.
\label{eq:asymptoticbehavioursofphi_{13}:pma-quasiperiodicsols}
\ee
and 
\be 
\phi_{23}(P,x,t_r)
\mathop{=}\limits_{\zeta\to  0}\left\{
                                                     \begin{array}{ll}
 \chi_{1,-1}\zeta^{-1}+ \chi_{1,0}+\chi_{1,1}\zeta   
+\textrm{O}(\zeta^{2}),
& \textrm{as}\ P\to  P_{\infty_1},
\vspace{2mm}
\\
\chi_{2,1}\zeta +\chi_{2,2}\zeta^2 
+\textrm{O}(\zeta^{3}),
& \textrm{as}\ P\to  P_{\infty_2},
\vspace{2mm}\\
\frac {q_1} {q_2} +
\frac 13 \bigl(\frac {q_{1}}{q_2}\bigr)_x \zeta 
+ 
\frac 19 \bigl[  
\bigl(\frac {q_1}{q_2}\bigr)_{xx}
+\frac {q_{2,x}}{q_2} \bigl(\frac {q_1}{q_2} \bigr)_x
\bigr]
\zeta ^2
+\textrm{O}(\zeta^{3}),& \textrm{as}\ P\to  P_{\infty_3},
                                                     \end{array}
                                                 \right.
\label{eq:asymptoticbehavioursofphi_{23}:pma-quasiperiodicsols}
\ee
where 
\[ 
\left\{
\ba {l}
\chi_{2,1,x}=\frac 13 p_2q_1,\ \chi_{2,2,x}=\frac 13 (p_1q_1-p_2q_2)\chi_{2,1}-\frac 19 p_{2,x}q_1,
\vspace{2mm}\\
\kappa_{2,3}=-\frac 19 p_{1,x}\chi_{2,1}+\frac 13 p_1 \chi_{2,2}
-\frac 1 {27}p_2(p_1q_1+p_2q_2)+\frac 1 {27}p_{2,xx},
\ea \right.
\]
and 
\[
\left\{ \ba {l}
\chi_{1,-1,x}=-\frac 13 p_1q_2 \chi_{1,-1}^2,
\ 
\kappa_{1,1}
=-\frac 19 p_{1,x}\chi_{1,-1}+\frac 13 p_1 \chi_{1,0}+\frac 13 p_2,
\vspace{2mm}\\
\kappa_{1,2}=-\frac 1 {27} p_1p_2q_2 \chi_{1,-1} -\frac 1 {27} p_1^2q_1 \chi_{1,-1}+\frac 1 {27}
p_{1,xx}\chi_{1,-1} -\frac 19 p_{1,x}\chi_{1,0}+\frac 13 p_1\chi_{1,1}-\frac 19 p_{2,x},
\vspace{2mm}
\\
\chi_{1,0,x}+\frac 23 p_1q_2 \chi_{1,-1}\chi_{1,0}
-\frac 19 p_{1,x}q_2\chi_{1,-1}^2+\frac 13 (p_2q_2-p_1q_1)\chi_{1,-1}=0,
\vspace{2mm}\\
\chi_{1,1,x}+\frac 23 p_1q_2\chi_{1,-1}\chi_{1,1}
-\frac 1{27}p_1p_2  q_{2} ^{2}   
\chi_{1,-1}  ^{2}
-\frac1{27}
p_1^2 q_1 q_2  
  \chi_{1,-1} ^{2}
 +\frac 1{27}p_{1,xx}
q_{2}   \chi_{1,-1}  ^{2}
+\frac 13
p_1q_2\chi_{1,-1}  ^{2} 
\vspace{2mm}\\
\quad 
-\frac 29
p_{1,x}q_{2}
\chi_{1,-1}  \chi_{1,0}  
+\frac 13
p_2q_{2}  \chi_{1,0}  
-\frac 13
p_1q_{1}  \chi_{1,0} 
+\frac 19
p_{1,x}q_{1} 
  \chi_{1,-1}
-\frac 19
p_{2,x}q_{2} \chi_{1,-1}  
-\frac 13
p_2q_{1}=0
.
\ea \right.
\]
\end{lem}

\noindent {\it Proof}: 
Similarly, we begin with 
the following three ansatzes:
\[
       \begin{array}{l}
\phi_{13}
\mathop{=}\limits_{\zeta\to  0}               
\kappa_{1,0}+ \kappa _{1,1}\zeta +  \kappa_{1,2}\zeta ^2 
+O(\zeta^{3}),\ 
\phi_{23}
\mathop{=}\limits_{\zeta\to  0}               
\chi_{1,-1}\zeta ^{-1}+ \chi _{1,0} +  \chi_{1,1}\zeta  
+O(\zeta^{2}),\ 
\textrm{as}\ P\to  P_{\infty_1};
\vspace{2mm}
\\
           \phi_{13}
\mathop{=}\limits_{\zeta\to  0}       
\kappa_{2,1}\zeta +\kappa_{2,2}\zeta^2+\kappa _{2,3}\zeta ^3 
+O(\zeta^{4}),\ 
 \phi_{23}
\mathop{=}\limits_{\zeta\to  0}       
\chi_{2,1}\zeta +\chi_{2,2}\zeta^2
+O(\zeta^{3}),
\  \textrm{as}\ P\to  P_{\infty_2};
\vspace{2mm}\\
 \phi_{13}
\mathop{=}\limits_{\zeta\to  0}       
\kappa_{3,-1}\zeta ^{-1}+\kappa_{3,0}+ \kappa _{3,1}\zeta 
+O(\zeta^{2}),
\ 
 \phi_{23}
\mathop{=}\limits_{\zeta\to  0}       
\chi_{3,0}+ \chi _{3,1}\zeta +  \chi_{3,2}\zeta ^2 
+O(\zeta^{3}),
\  \textrm{as}\ P\to  P_{\infty_3};
                                                     \end{array}
\]
where the coefficients, $\kappa_{ij}$ and $\chi_{ij}$, are functions to be determined.
Substituting those expansions into the Riccati type equations
\eqref{eq:phi_{ij}_x:pma-quasiperiodicsols} with $i=1,2$ and $j=3$, i.e., 
\be
\phi_{13,x}=-3\lambda \phi_{13}+p_1\phi_{23}+p_2-q_2\phi_{13}^2,
\ 
\phi_{23,x}=q_1\phi_{13}-q_2\phi_{13}\phi_{23},
\ee
and comparing the three lowest powers $\zeta^i$ in each resulting equation, where $i$ goes either from $-2$ to $0$, or from $-1$ to $1$, or from $0$ to $2$, we get
a set of relations on the coefficient functions
$\kappa_{i,j}$ and $\chi_{i,j}$,
which engenders the asymptotic properties in \eqref{eq:asymptoticbehavioursofphi_{13}:pma-quasiperiodicsols} and 
\eqref{eq:asymptoticbehavioursofphi_{23}:pma-quasiperiodicsols}.
 The proof is finished.
\hfill $\Box$

Now,
in order to determine asymptotic properties of
the Baker-Akhiezer function $\psi_3$
 at the points at infinity,
we similarly analyze 
\be 
J_r^{(3)}=U_{31}\phi_{13} +U_{32}\phi_{23} +U_{33}=
q_2\phi_{13}+\lambda ,
\label{eq:defofJ_{r}^{(3)}:pma-quasiperiodicsols}
\ee
and 
\be
I_r^{(3)}=V_{31}^{[r]}\phi_{13}+V_{32}^{[r]}\phi_{23} +V_{33}^{[r]}.
\label{eq:defofI_{r}^{(3)}:pma-quasiperiodicsols}
 \ee

\begin{lem}
\label{lem:asymptoticbehavioursofI_{r}^{(3)}anfJ_{r}^{(3)}:pma-quasiperiodicsols}
Let $u=(p_1,p_2,q_1,q_2)^T$ satisfy 
the $r$-th 
four-component
AKNS equations \eqref{eq:4cAKNSsh:pma-quasiperiodicsols} and $\zeta=\lambda^{-1}$.
Suppose that $P\in {\cal K}_g\backslash \{P_{\infty_1},P_{\infty_2},P_{\infty_3}\}$ and $(x,t_r)\in \mathbb {C}^2$. Then 
\be 
J_{r}^{(3)}(P,x,t_r)
\mathop{=}\limits_{\zeta\to  0}\left\{
                                                     \begin{array}{ll}
 \zeta^{-1}+\rho_r ^{(3)}
+\textrm{O}(\zeta),
& \textrm{as}\ P\to  P_{\infty_1},
\vspace{2mm}
\\
\zeta^{-1} 
+\textrm{O}(\zeta),
  & \textrm{as}\ P\to  P_{\infty_2},
\vspace{2mm}\\
-2 \zeta ^{-1}+
\frac {q_{2,x}}{q_2}
+\textrm{O}(\zeta),& \textrm{as}\ P\to  P_{\infty_3},
                                                     \end{array}
                                                 \right.
\label{eq:asymptoticbehavioursofJ_{r}^{(3)}:pma-quasiperiodicsols}
\ee
and 
\be 
I_r^{(3)}(P,x,t_r)
\mathop{=}\limits_{\zeta\to  0}\left\{
                                                     \begin{array}{ll}
\zeta^{-r}+ \sigma_{r}^{(3)}
+\textrm{O}(\zeta),
& \textrm{as}\ P\to  P_{\infty_1},
\vspace{2mm}
\\
\zeta ^{-r}
+\textrm{O}(\zeta),
& \textrm{as}\ P\to  P_{\infty_2},
\vspace{2mm}\\
-2\zeta^{-r}+
\frac {q_{2,t_r}}{q_2}  
+\textrm{O}(\zeta),& \textrm{as}\ P\to  P_{\infty_3},
                                                     \end{array}
                                                 \right.
\label{eq:asymptoticbehavioursofI_{r}^{(3)}:pma-quasiperiodicsols}
\ee
where 
$\rho_r^{(3)}=\frac 13 p_1q_2\chi_{1,-1}$ and $\sigma_{r,x}^{(3)}=\rho_{r,t_r}^{(3)}$, with $\chi_{1,-1}$ being defined in Lemma \ref{lem:asymptoticbehavioursofphi_{13}andphi_{23}:pma-quasiperiodicsols}.
\end{lem}

\noindent {\it Proof}:
Similarly, first 
based on \eqref{eq:defofJ_{r}^{(3)}:pma-quasiperiodicsols},
we obatin 
\eqref{eq:asymptoticbehavioursofJ_{r}^{(3)}:pma-quasiperiodicsols}
directly from Lemma \ref{lem:asymptoticbehavioursofphi_{13}andphi_{23}:pma-quasiperiodicsols}.

Second, 
 note that 
the third compatibility condition 
reads
\be I_{r,x}^{(3)}=\bigl(\frac {\psi_{3,t_r}}{\psi_3}\Bigr)_x = 
\bigl(\frac {\psi_{3,x}}{\psi_3}\Bigr)_{t_r}=
J_{r,t_r}^{(3)},
\label{eq:3rdcompatabilitycondition:pma-quasiperiodicsols}
\ee
and that from \eqref{eq:recursionrelationofV_{r}:pma-quasiperiodicsols},
we obtain
\[
V_{31}^{[r+1]}=\lambda V_{31}^{[r]}+c_2^{[r+1]},\ 
V_{32}^{[r+1]}=\lambda V_{32}^{[r]}+d_{21}^{[r+1]},\  
V_{33}^{[r+1]}=\lambda V_{33}^{[r]}+d_{22}^{[r+1]},
\]
and this tells
\be I_{r+1}^{(3)}=\lambda I_r^{(3)}+c_2^{[r+1]}\phi_{13}+d_{21}^{[r+1]}\phi_{23}+d_{22}^{[r+1]}.
 \label{eq:recursionrelationof3rdcolumnofV_{r}:pma-quasiperiodicsols}\ee
Finally,
based on \eqref{eq:3rdcompatabilitycondition:pma-quasiperiodicsols} and 
\eqref{eq:recursionrelationof3rdcolumnofV_{r}:pma-quasiperiodicsols},
we can verify 
\eqref{eq:asymptoticbehavioursofI_{r}^{(3)}:pma-quasiperiodicsols}
from 
\eqref{eq:asymptoticbehavioursofJ_{r}^{(3)}:pma-quasiperiodicsols} 
by the mathematical induction.
 This completes the proof.
\hfill $\Box$

\begin{thm}
Let $u=(p_1,p_2,q_1,q_2)^T$ satisfy 
the $r$-th four-component AKNS equations \eqref{eq:4cAKNSsh:pma-quasiperiodicsols} and $\zeta=\lambda^{-1}$.
Suppose that $P\in {\cal K}_g\backslash \{P_{\infty_1},P_{\infty_2},P_{\infty_3}\}$ and $(x,t_r)\in \mathbb {C}^2$. Then we have
\bea &&
\psi_3(P,x,x_0,t_r,t_{0,r})
\nonumber 
\\
&&
\mathop{=}\limits_{\zeta\to  0}\left\{
                                                     \begin{array}{ll}
{\rm{exp}}\bigl(
  \int_{x_0}^x \rho_r^{(3)} (P,x',t_r)\, dx'+\int_{t_{0,r}}^{t_r}
 \sigma_r^{(3)}(P,x_0,t')\, dt' & 
\vspace{2mm}\\
\quad\  +\zeta^{-1}(x-x_0)+  \zeta^{-r} (t_r-t_{0,r}) 
+ \textrm{O}(\zeta)\bigr),
& \textrm{as}\ P\to P_{\infty_1},
\vspace{2mm}
\\
{\rm{exp}}\bigl(\zeta^{-1}(x-x_0)+  \zeta^{-r} (t_r-t_{0,r}) 
+\textrm{O}(\zeta)\bigr),
& \textrm{as}\ P\to P_{\infty_2},
\vspace{2mm}\\
\frac {q_2(x,t_r)}{q_2(x_0,t_{0,r})}
{\rm{exp}}\bigl(-2\zeta^{-1}(x-x_0)-2 \zeta^{-r} (t_r-t_{0,r}) 
+\textrm{O}(\zeta)\bigr),& \textrm{as}\ P\to P_{\infty_3},
                                                     \end{array}
                                                 \right.
\qquad \qquad 
\label{eq:asymptoticbehavioursofpsi_3:pma-quasiperiodicsols}
\eea
where $\rho_r^{(3)}$ and $\sigma_{r}^{(3)}$ are defined in
Lemma \ref{lem:asymptoticbehavioursofI_{r}^{(3)}anfJ_{r}^{(3)}:pma-quasiperiodicsols}.
\end{thm}

\noindent {\it Proof}:
The third formula in
\eqref{eq:expressionsforpsi_i:pma-quasiperiodicsols} reads
\[
\psi_3(P,x,x_0,t_r,t_{0,r})=\, {\rm{exp}}\Bigl(
\int_{x_0}^xJ_r^{(3)}(P,x',t_r)\, dx'
+\int_{t_{0,r}}^{t_r}I_r^{(3)}(P,x_0,t')\, dt'
\Bigr),
\]
where $J_r^{(3)}$ and $I_r^{(3)}$ are determined by 
\eqref{eq:defofJ_{r}^{(3)}:pma-quasiperiodicsols}
and \eqref{eq:defofI_{r}^{(3)}:pma-quasiperiodicsols}. Based on 
Lemma 
\ref{lem:asymptoticbehavioursofI_{r}^{(3)}anfJ_{r}^{(3)}:pma-quasiperiodicsols},
this expression generates the asymptotic properties of the Baker-Akhiezer function  
$\psi_3$ in \eqref{eq:asymptoticbehavioursofpsi_3:pma-quasiperiodicsols}.
The proof is finished.
\hfill $\Box$

Now, note that a meromorphic function on a compact Riemann surface has the same number of zeros and poles. 
Therefore, 
in view of Lemma 
\ref{lem:asymptoticbehavioursofphi_{21}andphi_{31}:pma-quasiperiodicsols},
Lemma \ref{lem:asymptoticbehavioursofphi_{12}andphi_{32}:pma-quasiperiodicsols} and Lemma \ref{lem:asymptoticbehavioursofphi_{13}andphi_{23}:pma-quasiperiodicsols},
and from the expressions in 
\eqref{eq:PropertyofMeromorphicFunctions:pma-quasiperiodicsols}
for the meromorphic functions $\phi_{ij},\ 1\le i,j\le 3$, we can assume that 
their 
divisors are given by
\bea 
&&
(\phi_{21}(P,x,t_r))={\cal D} _{P_{\infty_3},\hat {\nu}_{h_1}(x,t_r),\cdots,
\hat {\nu}_g(x,t_r)}-
{\cal D} _{P_{\infty_1},\hat {\mu}_{h_1}(x,t_r),\cdots,
\hat {\mu}_g(x,t_r)},
\label{eq:devisorforphi_{21}:pma-quasiperiodicsols}
\\
&& 
(\phi_{31}(P,x,t_r))={\cal D} _{P_{\infty_3},\hat {\xi}_{h_2}(x,t_r),\cdots,
\hat {\xi}_g(x,t_r)}-
{\cal D} _{P_{\infty_2},\hat {\mu}_{h_2}(x,t_r),\cdots,
\hat {\mu}_g(x,t_r)},
\label{eq:devisorforphi_{31}:pma-quasiperiodicsols}
\\
&&
(\phi_{12}(P,x,t_r))={\cal D} _{P_{\infty_1},\hat {\mu}_{h_1}(x,t_r),\cdots,
\hat {\mu}_g(x,t_r)}-
{\cal D} _{P_{\infty_3},\hat {\nu}_{h_1}(x,t_r),\cdots,
\hat {\nu}_g(x,t_r)},
\\
&&
(\phi_{32}(P,x,t_r))={\cal D} _{P_{\infty_1},\hat {\xi}_{h_3}(x,t_r),\cdots,
\hat {\xi}_g(x,t_r)}-
{\cal D} _{P_{\infty_2},\hat {\nu}_{h_3}(x,t_r),\cdots,
\hat {\nu}_g(x,t_r)},\label{eq:devisorforphi_{32}:pma-quasiperiodicsols}
\\
&&
(\phi_{13}(P,x,t_r))={\cal D} _{P_{\infty_2},\hat {\mu}_{h_2}(x,t_r),\cdots,
\hat {\mu}_g(x,t_r)}-
{\cal D} _{P_{\infty_3},\hat {\xi}_{h_2}(x,t_r),\cdots,
\hat {\xi}_g(x,t_r)},
\\
&&
(\phi_{23}(P,x,t_r))={\cal D} _{P_{\infty_2},\hat {\nu}_{h_3}(x,t_r),\cdots,
\hat {\nu}_g(x,t_r)}-
{\cal D} _{P_{\infty_1},\hat {\xi}_{h_3}(x,t_r),\cdots,
\hat {\xi}_g(x,t_r)},
\eea
for some natural numbers $h_i,\ 1\le i\le 3$.
The case of $h_i>1$ for some $  1\le i\le 3$ could happen, particularly when
$y=-\frac {A_{ij}^{[m]}}{W^{[n]}_{jk}}$, and $E_{ij}^{[m]}$ and $2(A_{ij}^{[m]})^2+W_{jk}^{[n]}B_{ij}^{[m]}$ have common zeros, or when
$y=-\frac {C_{ij}^{[m]}}{W^{[n]}_{ik}}$, and $F_{ij}^{[m]}$ and $2(C_{ij}^{[m]})^2+W_{ik}^{[n]}D_{ij}^{[m]}$ have common zeros,
where $1\le i,j,k\le 3$ and $i\ne j\ne k$.

\section{Algebro-geometric solutions}

\label{sec:Algebro-geometricSolutions:pma-quasiperiodicsols}

In order to straighten out the corresponding flows in the soliton hierarchy \eqref{eq:4cAKNSsh:pma-quasiperiodicsols}, we equip
$\mathcal{K}_{g}$ with a homology basis of
$\mathbbm{a}$-cycles:
$\mathbbm{a}_{1},\ldots,\mathbbm{a}_{g}$, and $\mathbbm{b}$-cycles:
$\mathbbm{b}_{1},\ldots,\mathbbm{b}_{g}$, which are independent and have
intersection numbers as follows:
\[
        \mathbbm{a}_{j}\circ \mathbbm{a}_{k}=0,\ 
        \mathbbm{b}_{j}\circ\mathbbm{b}_{k}=0,\ 
        \mathbbm{a}_{j}\circ \mathbbm{b}_{k}=\delta_{jk},\ 
1\le j,k\le g.
\]

In what follows, we will choose the following
set as our basis for 
the space of holomorphic differentials on ${\cal K}_g$
\cite{MillerEKL-CPAM1995,KatoH-JPAA1988}:
\begin{equation}
        \tilde{\omega}_{l}=\frac{1}{3y^2(P)+S_m}
\left\{\ba {ll}
\lambda^{l-1}d\lambda, & 1\le l\le \textrm{deg}(S_m)-1,
\vspace{2mm}\\
y(P) \lambda ^{l-\textrm{deg}(S_m)}d\lambda,& \textrm{deg}(S_m)\le l\le g,
\ea
\right.
\end{equation}
which are $g$ linearly independent holomorphic differentials on
$\mathcal{K}_{g}$. 
By using the above homology basis, 
the period matrices $A=(A_{jk})$ and $B=(B_{jk})$ can be
constructed as
\begin{equation}
        A_{kj}=\int_{\mathbbm{a}_{j}}\tilde{\omega}_{k},\ 
        B_{kj}=\int_{\mathbbm{b}_{j}}\tilde{\omega}_{k},\ 1\le j,k\le g.
\end{equation}
It is possible to show that matrices $A$ and $B$ are
invertible \cite{Dubrovin-RMS1981}. So, we can define the matrices $C$ and $\tau$ by
$C=A^{-1}$ and $\tau=A^{-1}B$. 
The matrix $\tau$ can be shown to be
symmetric ($\tau_{kj}=\tau_{jk}$), and it has a positive-definite
imaginary part (Im$\,\tau>0$) \cite{Mumford-book19834,FarkasK-book1992,GriffithsH-book1994}. If we normalize $\tilde{\omega}_{l},
1\le l\le g,$
into a new basis $\underline{\omega}=(\omega_{1},\cdots,\omega_g)$:
\begin{equation}
        \omega_{j}=\sum\limits_{l=1}^{g}C_{jl}\tilde{\omega}_{l},\ 1\leq j\leq g,
\end{equation}
where $C=(C_{ij})_{g\times g}$, 
then we obtain
\begin{equation}
        \int_{\mathbbm{a}_{k}}\omega_{j}=\sum\limits_{l=1}^{g}
C_{jl}\int_{\mathbbm{a}_{k}}\tilde{\omega}_{l}=\delta_{jk},\ \int_{\mathbbm{b}_{k}}\omega_{j}=\tau_{jk},\ 1\le j,k\le g.
\end{equation}
To compute the $\mathbbm{b}$-periods of Abelian differentials of the second kind, we assume that 
\be \omega_k\mathop{=}\limits_{\zeta \to 0}\sum_{l=0}^\infty \varrho_{k,l}(P_{\infty_j})\zeta^l\, d\zeta,\ \textrm{as}\ P\to P_{\infty_j},
\ 
1\le k\le g, \ 1\le j\le 3,
\ee
where $\varrho_{k,l}(P_{\infty_j})$, $ l\ge 0$, are constants. 

Now, let $\mathcal{T}_{g}$ be the period lattice
$\mathcal{T}_{g}=\{\underline{z}\in\mathbb{C}^{g}\,|\, \underline{z}=\underline{N}+
\underline{L}\tau ,\ \underline{N},\underline{L}\in\mathbb{Z}^{g}\}$. The
complex torus $\mathscr{T}_g=\mathbb{C}^{g}/\mathcal{T}_{g}$
  is called the Jacobian variety of
$\mathcal{K}_{g}$. 
The Abel map
$\underline{\mathcal{A}}:\mathcal{K}_{g}\to 
\mathscr{T}_g$ is defined as follows:
\begin{equation}
       \underline{\mathcal{A}}(P)=\Bigl(\int_{Q_{0}}^{P}{\omega_1},
\cdots, \int_{Q_0}^P\omega_{g}\Bigr)\ 
(\mathrm{mod}\ \mathcal{T}_{g}),
\label{eq:defofAbelmap:pma-quasiperiodicsols}
\end{equation}
where $Q_0\in {\cal K}_g$ is a fixed base point. We take
the natural linear extension of the Abel map to the space of divisors
$\mathrm{Div}(\mathcal {K}_g)$:
\be 
        \underline{\mathcal{A}}\Bigl(\sum n_{k}P_{k}\Bigr)=\sum n_{k}\underline{\mathcal{A}}(P_{k}),
\ee
where $P,P_{k}\in\mathcal{K}_{g}$.

Let $\omega^{(2)}_{\infty_j,l}(P)$, $1\le j\le 3$ and $l\ge 2$, denote
the normalized Abelian differential of the second kind, being holomorphic on ${\cal K}_g\backslash 
\{P_{\infty_j}\}$ and possessing the asymptotic property: 
\be 
\omega^{(2)}_{\infty_j,l}(P)\mathop{=}\limits_{\zeta\to 0} \bigl(\zeta ^{-l}+\textrm{O}(1) \bigr) d\zeta, \ \textrm{as}\ P\to P_{\infty_j},
\ 1\le j\le 3,\ l\ge 2.\label{eq:conditionsof2rdAbeldifferentials:pma-quasiperiodicsols}
\ee
The adopted normalization condition is 
\be 
\int_{\mathbbm{a}_k}\omega^{(2)}_{\infty_j,l}=0, \ 1\le k\le g,
\ 1\le j\le 3,\ l\ge 2,
\ee  
and \eqref{eq:conditionsof2rdAbeldifferentials:pma-quasiperiodicsols} implies that the residues of $\omega^{(2)}_{\infty_j,l}$ at $P_{\infty_j}$ are all zero.
Based on
the asymptotic properties of the Baker-Akhiezer functions $\psi_j, \ 1\le j\le 3$, we introduce the following 
Abelian differentials of the second kind:
\bea &&
\Omega_{2}^{(2)}= \omega^{(2)}_{P_{\infty_1},2}+
  \omega^{(2)}_{P_{\infty_2},2}
- 2\omega^{(2)}_{P_{\infty_3},2},\\
&& 
{\tilde \Omega}_{r}^{(2)}=
r \omega^{(2)}_{P_{\infty_1},r+1}+
  r\omega^{(2)}_{P_{\infty_2},r+1}- 2r \omega^{(2)}_{P_{\infty_3},r+1}.
\eea
Then for $\Omega_{2}^{(2)}$, we have the asymptotic expansions:
\be 
\int_{Q_0}^P 
\Omega_{2}^{(2)}\mathop{=}\limits_{\zeta\to 0}
\left\{\ba {ll}
-\zeta^{-1} +e_{2,1}^{(2)}(Q_0)+\textrm{O}(\zeta), & \textrm{as} \ P\to P_{\infty _1},
\vspace{2mm}\\
-\zeta^{-1} +e_{2,2}^{(2)}(Q_0)+\textrm{O}(\zeta), & \textrm{as} \ P\to P_{\infty _2},
\vspace{2mm}\\
2\zeta^{-1} +e_{2,3}^{(2)}(Q_0)+\textrm{O}(\zeta), & \textrm{as} \ P\to P_{\infty _3},
\ea \right. 
\label{eq:asymptoticpropertyforOmega_2^{(2)}:pma-quasiperiodicsols}
\ee
and for ${\tilde \Omega}_{r}^{(2)}$, we have the asymptotic expansions:
\be 
\int_{Q_0}^P 
{\tilde \Omega}_{r}^{(2)}\mathop{=}\limits_{\zeta\to 0}
\left\{\ba {ll}
-\zeta^{-r} +{\tilde e}_{r,1}^{(2)}(Q_0)+\textrm{O}(\zeta), & \textrm{as} \ P\to P_{\infty _1},
\vspace{2mm}\\
-\zeta^{-r} +{\tilde e}_{r,2}^{(2)}(Q_0)+\textrm{O}(\zeta), & \textrm{as} \ P\to P_{\infty _2},
\vspace{2mm}\\
2\zeta^{-r} +{\tilde e}_{r,3}^{(2)}(Q_0)+\textrm{O}(\zeta), & \textrm{as} \ P\to P_{\infty _3},
\ea \right. 
\label{eq:asymptoticpropertyforOmega_r^{(2)}:pma-quasiperiodicsols}
\ee
where the paths of integration 
are chosen to be the same as the one in 
the Abel map \eqref{eq:defofAbelmap:pma-quasiperiodicsols}.
Denote the $\mathbbm{b}$-periods of the differentials $\Omega_2^{(2)}$ and $\tilde {\Omega}_r^{(2)}$ by 
\be 
\underline{U}_2^{(2)}=(U_{2,1}^{(2)},\cdots,U_{2,g}^{(2)}),\ U_{2,k}^{(2)}=
\frac 1 {2\pi i}\int_{\mathbbm{b}_k}\Omega_2^{(2)},\ 1\le k\le g,
\ee
and 
\be 
\underline{\tilde {U}}_r^{(2)}=(\tilde {U}_{r,1}^{(2)},\cdots,\tilde {U}_{r,g}^{(2)}),\ \tilde{U}_{r,k}^{(2)}=
\frac 1 {2\pi i}\int_{\mathbbm{b}_k}\tilde{\Omega}_r^{(2)},\ 1\le k\le g.
\ee
Through the relationship between the normalized meromorphic differential of the second kind and the normalized holomorphic differentials $\omega_k, \ 1\le k\le g$, we can derive that 
\be 
U_{2,k}^{(2)}=\varrho_{k,0}(P_{\infty_1}) +
\varrho_{k,0}(P_{\infty_2})-2 \varrho_{k,0}(P_{\infty_3}),\ 1\le k\le g,
\ee
and 
\be 
\tilde{U}_{r,k}^{(2)}=\varrho_{k,r}(P_{\infty_1}) +
\varrho_{k,r}(P_{\infty_2})-2 \varrho_{k,r}(P_{\infty_3}),\ 1\le k\le g.
\ee

Let $\omega^{(3)}_{Q_1,Q_2}$ stand for the normalized Abelian differential of the third kind, holomorphic on ${\cal K}_g\backslash \{Q_1,Q_2\}$ and with simple poles at $Q_l$ with residues $(-1)^{l+1},\ l=1,2$.
The adopted normalization condition reads 
\be 
\int_{\mathbbm{a}_k}\omega^{(3)}_{Q_1,Q_2}=0,\ 1\le k\le g,
\ee
and thus,
\be 
\int_{\mathbbm{b}_k}\omega^{(3)}_{Q_1,Q_2}=2\pi i 
\int_{Q_2}^{Q_1}\omega_k,\ 1\le k\le g,
\ee 
where the path of integration from $Q_2$ to $Q_1$ does not intersect the cycles $\mathbbm{a}_1,\cdots,\mathbbm{a}_g,\mathbbm{b}_1,\cdots,\mathbbm{b}_g$.
We then set 
\bea
&& 
{e} _{2,j}^{(3)}(Q_0)=
 {e} _{2,j}^{(3)}(Q_0,x,x_0,t_{r},t_{0,r})
=\int_{Q_0}^{P_{\infty_j}} \omega^{(3)}_{\hat {\nu }_0(x_0,t_{0,r}),\hat {\nu }_0(x,t_{r})},\ 1\le j\le 3,
\\
&&
{e} _{3,j}^{(3)}(Q_0)= {e} _{3,j}^{(3)}(Q_0,x,x_0,t_{r},t_{0,r})
=\int_{Q_0}^{P_{\infty_j}} \omega^{(3)}_{\hat {\xi }_0(x_0,t_{0,r}),\hat {\xi }_0(x,t_{r})},\ 1\le j\le 3,
\eea
where the paths of integration 
are chosen to be the same as the one in the Abel map \eqref{eq:defofAbelmap:pma-quasiperiodicsols}.

Denote by $\theta(\underline{z})$ the Riemann theta function associated with 
${\cal K}_g$ equipped with the above homology basis \cite{FarkasK-book1992}:
\be 
\theta(\underline{z})=
\sum_{\underline {N}\in \mathbb{Z}^g}\textrm{exp}\bigl(\pi i \langle \underline{N}\tau , \underline {N}\rangle+2\pi i 
\langle \underline{N}, \underline {z}\rangle\bigr),
\ee
where $\underline{z}=(z_1,\cdots,z_g)\in \mathbb{C}^g$ is a complex vector, and $\langle\cdot,\cdot\rangle$ stands for the Hermitian inner product on $\mathbb{C}^g$: 
\be \langle \underline{z},\underline{w}\rangle=\sum_{j=1}^g z_i\bar {w}_j,\ \underline{z}=(z_1,\cdots,z_g)
\in \mathbb{C}^g,\ \underline{w}=(w_1,\cdots,w_g)\in \mathbb{C}^g
.\ee The Riemann theta function is even and quasi-periodic. More precisely, it satisfies 
\be 
\theta(z_1,\cdots,z_{j-1},-z_j,z_{j+1},\cdots,z_g)=\theta(\underline{z}),
\ 1\le j\le g,\ee
and 
\be  
\theta(\underline{z}+\underline{N}+\underline{L}\tau )=\textrm{exp}\bigl(-\pi i 
\langle \underline{L}\tau , \underline {L}\rangle
-2\pi i \langle \underline{L}, \underline {z}\rangle
\bigr)
\theta (\underline{z}),
\ee
where
$\underline{z}=(z_1,\cdots,z_g)\in \mathbb{C}^g$, 
$\underline{N}=(N_1,\cdots ,N_g)\in \mathbb{Z}^g$ and $\underline{L}=(L_1,\cdots ,L_g)\in \mathbb{Z}^g$.
For brevity, define the function 
$ \underline{z}:{\cal K}_g \times \sigma^g {\cal K}_g\to \mathbb{C}^g$ by
\be
 \underline{z}(P, \underline{Q}) =\underline{M}-
\underline{\mathcal {A}}(P) 
+\sum_{j=1}^g {\cal D}_{Q_1,\cdots,Q_g}(Q_j)\underline{\mathcal {A}}(Q_j) ,
\ee 
where $P\in {\cal K}_g$, $\underline{Q} =(Q_1,\cdots,Q_g)\in 
\sigma^g {\cal K}_g
$,
$\sigma^g {\cal K}_g$ denotes the $g$-th symmetric power of ${\cal K}_g$ \cite{GriffithsH-book1994}, and 
 $\underline{M}=(M_1,\cdots ,M_g)$ is a vector 
of Riemann constants \cite{FarkasK-book1992,DicksonGU-RMP1999}:  
\be 
M_j=\frac 1 2(1+\tau_{jj})-\sum_{l=1,\,l\ne j}^g \int_{\mathbbm{a}_l}\omega_l(P) \int_{Q_0}^P \omega_j,\ 1\le j\le g.
\ee 
By Riemann's vanishing theorem \cite{BolokolosBEIM-book1994,DicksonGU-RMP1999}, 
the function $\theta (\underline{z}(P, \underline{Q}) )$ has exactly $g$ zeros $
Q_1,\cdots, Q_g$, if the divisor 
${\cal D}=Q_1+\cdots Q_g$ is nonspecial. 

Introduce three particular points in the $g$-th symmetric power $\sigma^g{\cal K}_g$: 
\bea 
&& 
\underline{\hat{\mu}}(x,t_r)=(\hat{\mu}_1(x,t_r),\cdots,\hat{\mu}_g(x,t_r)),
\\
&& \underline{\hat{\nu}}(x,t_r)=(\hat{\nu}_1(x,t_r),\cdots,\hat{\nu}_g(x,t_r)),
\\
&& \underline{\hat{\xi}}(x,t_r)=(\hat{\xi}_1(x,t_r),\cdots,\hat{\xi}_g(x,t_r)),
\eea 
and denote the corresponding three
particular divisors in $\textrm{Div}({\cal K}_g)$ by 
\be
{\cal D}_{\underline{\hat {\mu}}(x,t_r)}
=\sum\limits_{j=1}^{g}\hat{\mu}_{j }(x,t_r),\  
{\cal D}_{\underline{\hat {\nu}}(x,t_r)}
=\sum\limits_{j=1}^{g}\hat{\nu}_{j }(x,t_r),\  
{\cal D}_{\underline{\hat {\xi}}(x,t_r)}
=\sum\limits_{j=1}^{g}\hat{\xi}_{j }(x,t_r). \ee

\begin{thm}
\label{thm:theta-functionRepresentationsoftheBAfunctions:pma-quasiperiodicsols}
(Theta function representations of the Baker-Akhiezer functions)
Let $\Omega_\mu \subset \mathbb{C}^2$ be an open and connected set, 
$(x_0,t_{0,r}),(x,t_r)\in \Omega_\mu$, and $P=(\lambda ,y)\in \mathcal {K}_g \backslash \{P_{\infty_1},P_{\infty_2},P_{\infty_3}\}$.
Suppose that
${\cal K}_g$ is nonsingular and 
 ${\cal D}_{\underline{\hat{\mu}}(x,t_r)}$
or ${\cal D}_{\underline{\hat{\nu}}(x,t_r)}$ or 
${\cal D}_{\underline{\hat{\xi}}(x,t_r)}$ is nonspecial for 
$ (x,t_r)\in \Omega_\mu $. 
Then 
the Baker-Akhiezer functions 
have the following theta function representations:
\bea
&& 
\psi_1(P,x,x_0,t_r,t_{0,r})
\nonumber 
\\&&
=\frac {\theta (\underline{z} (P, \underline{\hat{\mu}}(x,t_r) ) )
\theta (\underline{z} (P_{\infty_3}, \underline{\hat{\mu}}(x_0,t_{0,r}) ) )
}
{\theta (\underline{z} (P_{\infty_3}, \underline{\hat{\mu}}(x,t_r) ) )
\theta (\underline{z} (P, \underline{\hat{\mu}}(x_0,t_{0,r}) ) )
}
\mathrm{exp}\Bigl(
\bigl( e_{2,3}^{(2)}(Q_0)-\int_{Q_0}^P \Omega _{2}^{(2)} \bigr)(x-x_0)
\nonumber
\\
&&
\quad + \Bigl.
\bigl( \tilde {e}_{r,3}^{(2)}(Q_0)-\int_{Q_0}^P {\tilde {\Omega}} _{r}^{(2)} \bigr)(t_r-t_{0,r})
\Bigr),\label{eq:thetapresentationforpsi_1:pma-quasiperiodicsols}
\\
 && 
\psi_2(P,x,x_0,t_r,t_{0,r})
\nonumber 
\\
&& 
=\frac {\theta (\underline{z} (P, \underline{\hat{\nu}}(x,t_r) ) )
\theta (\underline{z}(P_{\infty_1}, \underline{\hat{\nu}}(x_0,t_{0,r}) ) )
}
{\theta (\underline{z} (P_{\infty_1}, \underline{\hat{\nu}}(x,t_r) ) )
\theta (\underline{z}(P, \underline{\hat{\nu}}(x_0,t_{0,r}) ) )
}
\mathrm{exp}\Bigl(
\bigl( e_{2,1}^{(2)}(Q_0)-\int_{Q_0}^P \Omega _{2}^{(2)} \bigr)(x-x_0)
\nonumber
\\
&&
\quad + \Bigl.
\bigl( \tilde {e}_{r,1}^{(2)}(Q_0)-\int_{Q_0}^P {\tilde {\Omega}} _{r}^{(2)} \bigr)(t_r-t_{0,r})
+\bigl( {e} _{2,1}^{(3)}(Q_0)
-\int_{Q_0}^P \omega^{(3)}_{\hat {\nu }_0(x_0,t_{0,r}),\hat {\nu }_0(x,t_{r})}
 \bigr)
\Bigr),\label{eq:thetapresentationforpsi_2:pma-quasiperiodicsols}
\\
 && 
\psi_3(P,x,x_0,t_r,t_{0,r})
\nonumber 
\\
&& 
=\frac {\theta (\underline{z} (P, \underline{\hat{\xi}}(x,t_r) ) )
\theta (\underline{z} (P_{\infty_2}, \underline{\hat{\xi}}(x_0,t_{0,r}) ) )
}
{\theta (\underline{z} (P_{\infty_2}, \underline{\hat{\xi}}(x,t_r) ) )
\theta (\underline{z} (P, \underline{\hat{\xi}}(x_0,t_{0,r}) ) )
}
\mathrm{exp}\Bigl(
\bigl( e_{2,2}^{(2)}(Q_0)-\int_{Q_0}^P \Omega _{2}^{(2)} \bigr)(x-x_0)
\nonumber
\\
&&
\quad + \Bigl.
\bigl( \tilde {e}_{r,2}^{(2)}(Q_0)-\int_{Q_0}^P {\tilde {\Omega}} _{r}^{(2)}\bigr)(t_r-t_{0,r})
+\bigl(  {e} _{3,2}^{(3)}(Q_0)
-\int_{Q_0}^P \omega^{(3)}_{\hat {\xi }_0(x_0,t_{0,r}),\hat {\xi }_0(x,t_{r})}
 \bigr)
\Bigr),\qquad\qquad
\label{eq:thetapresentationforpsi_3:pma-quasiperiodicsols}
\eea
 where 
the paths of integration are the same as the one in the Abel map
\eqref{eq:defofAbelmap:pma-quasiperiodicsols}.   
\end{thm}

\noindent {\it Proof}:
Let $\Psi_1,\Psi_2$ and $\Psi_3$ denote the right-hand sides of 
\eqref{eq:thetapresentationforpsi_1:pma-quasiperiodicsols},
\eqref{eq:thetapresentationforpsi_2:pma-quasiperiodicsols} and \eqref{eq:thetapresentationforpsi_3:pma-quasiperiodicsols}, respectively.
By Theorem \ref{thm:zerosandpolesofpsi_i:pma-quasiperiodicsols},
$\psi_1$ has 
the simple zeros $\hat{\mu}_1(x,t_r),\cdots,\hat{\mu}_g(x,t_r)$ and 
the simple poles 
$\hat{\mu}_1(x_0,t_{0,r}),\cdots,\hat{\mu}_g(x_0,t_{0,r})$,
$\psi_2$ has 
the simple zeros $\hat{\nu}_0(x,t_r),\hat{\nu}_1(x,t_r),\cdots,\hat{\nu}_g(x,t_r)$ and 
the simple poles 
$ \hat{\nu}_0(x_0,t_{0,r}),\hat{\nu}_1(x_0,t_{0,r}),\cdots,\hat{\nu}_g(x_0,t_{0,r})$,
and 
$\psi_3$ has 
the simple zeros $\hat{\xi}_0(x,t_{r}),\hat{\xi}_1(x,t_r),\cdots,\hat{\xi}_g(x,t_r)$ and 
the simple poles 
$\hat{\xi}_0(x_0,t_{0,r}),\hat{\xi}_1(x_0,t_{0,r}),\cdots,\hat{\xi}_g(x_0,t_{0,r})$.
They all have three essential singularities at $P_{\infty_1},P_{\infty_2},P_{\infty_3}$.
By Riemann's vanishing theorem \cite{DicksonGU-RMP1999}, we know that 
$\Psi_i$, $1\le i\le 3$, have the same properties as $\psi_i,\ 1\le i\le 3$, respectively. 
Therefore, the Riemann-Roch theorem tells that 
\[ \frac {\Psi_i}{\psi_i}=\gamma_i, \  1\le i\le 3,\]
where $\gamma_i$, $1\le i\le 3$, are constants depending on $P$. 
Using the asymptotic properties of $\psi_i$ and $\Psi_i$, $1\le i\le 3$, one has
\[ 
\ba {l}
\D \frac {\Psi_1}{\psi_1} 
\mathop{=}\limits_{\zeta\to  0}
\frac {\textrm{exp}\bigl( -2\zeta^{-1}(x-x_0)-2 \zeta^{-r}(t_r-t_{0,r})+\textrm{O}(\zeta)\bigr)
 \bigl(1+\textrm{O}(\zeta )\bigr)}
{ 
\textrm{exp}\bigl( -2\zeta^{-1}(x-x_0)-2  \zeta^{-r}(t_r-t_{0,r})+\textrm{O}(\zeta)\bigr)
}
\mathop{=}\limits_{\zeta\to  0}
1+\textrm{O}(\zeta) \ \textrm{as}\ P\to P_{\infty_3},
\vspace{2mm}\\
\D \frac {\Psi_2}{\psi_2} 
\mathop{=}\limits_{\zeta\to  0}
\frac {\textrm{exp}\bigl( \zeta^{-1}(x-x_0)+
 \zeta^{-r}(t_r-t_{0,r})+\textrm{O}(\zeta)\bigr)
\bigl (1+\textrm{O}(\zeta )\bigr)}
{ 
\textrm{exp}\bigl( \zeta^{-1}(x-x_0)+
 \zeta^{-r}(t_r-t_{0,r})+\textrm{O}(\zeta)\bigr)
}
\mathop{=}\limits_{\zeta\to  0}
1+\textrm{O}(\zeta) \ \textrm{as}\ P\to P_{\infty_1},
\vspace{2mm}\\
\D \frac {\Psi_3}{\psi_3} 
\mathop{=}\limits_{\zeta\to  0}
\frac {\textrm{exp}\bigl(\zeta^{-1}(x-x_0)+
\zeta^{-r}(t_r-t_{0,r})+\textrm{O}(\zeta)\bigr)
 \bigl(1+\textrm{O}(\zeta )\bigr)}
{ 
\textrm{exp}\bigl( \zeta^{-1}(x-x_0)+
\zeta^{-r}(t_r-t_{0,r})+\textrm{O}(\zeta)\bigr)
}
\mathop{=}\limits_{\zeta\to  0}
1+\textrm{O}(\zeta) \ \textrm{as}\ P\to P_{\infty_2}.
\ea \]
These concludes that $\gamma_i=1$, $1\le i\le 3$. Therefore, 
$\Psi_i=\psi_i$, $1\le i\le 3$. This completes the proof of the 
 theorem. 
\hfill $\Box$

Using 
the linear equivalences (see, e.g., \cite{BolokolosBEIM-book1994,Schlichenmaier-book1989}):
\[\ba {l}
{\cal D}_{P_{\infty_3},\hat{\nu}_{h_1}(x,t_r),\cdots,\hat{\nu}_g(x,t_r)}\sim
{\cal D}_{P_{\infty_1},\hat{\mu}_{h_1}(x,t_r),\cdots,\hat{\mu}_g(x,t_r)},
\vspace{2mm}
\\
{\cal D}_{P_{\infty_3},\hat{\xi}_{h_2}(x,t_r),\cdots,\hat{\xi}_g(x,t_r)}\sim
{\cal D}_{P_{\infty_2},\hat{\mu}_{h_2}(x,t_r),\cdots,\hat{\mu}_g(x,t_r)},
\vspace{2mm}
\\
{\cal D}_{P_{\infty_1},\hat{\xi}_{h_3}(x,t_r),\cdots,\hat{\xi}_g(x,t_r)}\sim
{\cal D}_{P_{\infty_2},\hat{\nu}_{h_3}(x,t_r),\cdots,\hat{\nu}_g(x,t_r)},
\ea \]
which are due to \eqref{eq:devisorforphi_{21}:pma-quasiperiodicsols}, \eqref{eq:devisorforphi_{31}:pma-quasiperiodicsols} and \eqref{eq:devisorforphi_{32}:pma-quasiperiodicsols},
we obtain
\[\ba {l}
\D \underline{{\cal A}}(P_{\infty_3})+ 
\sum_{j=h_1}^g
\underline{{\cal A}}(\hat{\nu}_j(x,t_r))
=
\underline{{\cal A}}(P_{\infty_1})+
\sum_{j=h_1}^g
\underline{{\cal A}}(\hat{\mu}_j(x,t_r))
,\vspace{2mm}\\
\D \underline{{\cal A}}(P_{\infty_3})+\sum_{j=h_2}^g \underline{{\cal A}}(\hat{\xi}_j(x,t_r))
=
\underline{{\cal A}}(P_{\infty_2})
+\sum_{j=h_2}^g \underline{{\cal A}}(\hat{\mu}_j(x,t_r))
,\vspace{2mm}\\
\D \underline{{\cal A}}(P_{\infty_1})+\sum_{j=h_3}^g \underline{{\cal A}}(\hat{\xi}_j(x,t_r))
=
\underline{{\cal A}}(P_{\infty_2})
+\sum_{j=h_3}^g \underline{{\cal A}}(\hat{\nu}_j(x,t_r))
,
\ea
\]
respectively. 
Define the Abel-Jacobi coordinates
\bea
&& 
\underline{\rho}^{(1)}(x,t_r)=
\underline{\mathcal {A}}({\cal D}_{\underline{\hat {\mu}}(x,t_r)})= 
\sum_{j=1}^g \int_{Q_0}^{\hat{\mu}_j(x,t_r)}\underline {\omega},
\\ 
&& 
\underline{\rho}^{(2)}(x,t_r)=
\underline{\mathcal {A}}({\cal D}_{\underline{\hat {\nu}}(x,t_r)})= 
\sum_{j=1}^g \int_{Q_0}^{\hat{\nu}_j(x,t_r)}\underline {\omega},
\\ 
&& 
\underline{\rho}^{(3)}(x,t_r)=
\underline{\mathcal {A}}({\cal D}_{\underline{\hat {\xi}}(x,t_r)})= 
\sum_{j=1}^g \int_{Q_0}^{\hat{\xi}_j(x,t_r)}\underline {\omega},
\eea
and then we have 
\[\ba {l}
\theta (\underline{z}(P,\underline{\hat{\mu}}(x,t_r)))
=\theta (\underline{M}-\underline{{\cal A}}(P)+
\underline{\rho}^{(1)}(x,t_r)), 
\vspace{2mm}\\
\theta (\underline{z}(P,\underline{\hat{\nu}}(x,t_r)))
=\theta (\underline{M}-\underline{{\cal A}}(P)+
\underline{\rho}^{(2)}(x,t_r)), 
\vspace{2mm}\\
\theta (\underline{z}(P,\underline{\hat{\xi}}(x,t_r)))
=\theta (\underline{M}-\underline{{\cal A}}(P)+
\underline{\rho}^{(3)}(x,t_r)).
\ea
\]
The Abel-Jacobi coordinates can be linearized on the Riemann surface ${\cal K}_g$ as follows.

\begin{thm} (Straightening out of the flows)
Let $(x,t_r),(x_0,t_{0,r})\in \mathbb{C}^2$, and $u=(p_1,p_2,q_1,q_2)^T$ solve the $r$-th four-component AKNS  equations \eqref{eq:4cAKNSsh:pma-quasiperiodicsols}. Suppose that 
${\cal K}_g$ is nonsingular and 
${\cal D}_{\underline{\hat {\mu}}(x,t_r)} $ or ${\cal D}_{\underline{\hat {\nu}}(x,t_r)} $ or ${\cal D}_{\underline{\hat {\xi}}(x,t_r)} $ is nonspecial. Then we have 
\bea
&& 
 \underline{\rho}^{(1)}(x,t_r)= \underline{\rho}^{(1)}(x_0,t_{0,r})
+\underline{U}_2^{(2)}(x-x_0)+\underline{\tilde{U}}_{2,r}^{(2)}(t-t_{0,r})\ (\textrm{mod}\ {\cal T}_g),
\label{eq:lineareqnforrho_1:pma-quasiperiodicsols}
\\
&&
\underline{{\cal A}}(\hat {\nu}_0(x,t_r))+\underline{\rho}^{(2)}(x,t_r)= 
\underline{{\cal A}}(\hat {\nu}_0(x_0,t_{0,r}))+
\underline{\rho}^{(2)}(x_0,t_{0,r})
\nonumber \\
&& \qquad \qquad \qquad \qquad \qquad \quad
+\underline{U}_2^{(2)}(x-x_0)+\underline{\tilde{U}}_{2,r}^{(2)}(t-t_{0,r})\ (\textrm{mod}\ {\cal T}_g),
\label{eq:lineareqnforrho_2:pma-quasiperiodicsols}
\\
&&
\underline{{\cal A}}(\hat {\xi}_0(x,t_r))+
\underline{\rho}^{(3)}(x,t_r)= 
\underline{{\cal A}}(\hat {\xi}_0(x_0,t_{0,r}))+
\underline{\rho}^{(3)}(x_0,t_{0,r})
\nonumber \\
&& \qquad \qquad \qquad \qquad \qquad \quad
+\underline{U}_2^{(2)}(x-x_0)+\underline{\tilde{U}}_{2,r}^{(2)}(t-t_{0,r})\ (\textrm{mod}\ {\cal T}_g).\label{eq:lineareqnforrho_3:pma-quasiperiodicsols}
\eea
\end{thm}

\noindent {\it Proof}:  In order to prove the theorem,
we introduce three meromorphic differentials
\be 
\Omega _j(x,x_0,t_r,t_{0,r})=\frac {\partial }{\partial \lambda }
\ln (\psi_j(P,x,x_0,t_r,t_{0,r}))\, d\lambda, \ 1\le j\le 3.
\ee

Let us first prove \eqref{eq:lineareqnforrho_1:pma-quasiperiodicsols}.
From the theta function representation \eqref{eq:thetapresentationforpsi_1:pma-quasiperiodicsols} for $\psi_1$, 
one infers 
\be 
\Omega_1 (x,x_0,t_r,t_{0,r})=-(x-x_0)\Omega_2^{(2)}-(t_r-t_{0,r})\tilde {\Omega}_r^{(2)}
+\sum_{j=1}^g \omega^{(3)}_{\hat{\mu}_j(x,t_r),\hat{\mu}_j(x_0,t_{0,r})}+\tilde \omega,
\label{eq:expforOmega_1:pma-quasiperiodicsols}
\ee
where $\tilde \omega$ is a holomorphic differential on ${\cal K}_g$, which can be expressed by 
\be  \tilde \omega= \sum_{j=1}^g h_j\omega_j,  \ee
$h_j\in \mathbb{C}$ being constants, $1\le j\le g$.

Since $\psi_1(P,x,x_0,t_r,t_{0,r})$ is single-valued on ${\cal K}_g$, all $\mathbbm{a}$- and $\mathbbm{b}$-periods of $\Omega_1$ are integer multiples  of $2\pi i$ and thus
\[
2\pi il_k=\int_{\mathbbm{a}_k}\Omega_1(x,x_0,t_r,t_{0,r})=\int_{\mathbbm{a}_k}
\tilde \omega=h_k, \ 1\le k\le g,
\]
for some $l_k\in \mathbb{Z}$. Similarly, for some $n_k\in \mathbb{Z}$, we have
 \[
\ba {l}
\D 2\pi i n_k=\int_{\mathbbm{b}_k}\Omega_1(x,x_0,t_r,t_{0,r})
\vspace{2mm}\\
\D = -(x-x_0)\int_{\mathbbm{b}_k}\Omega_2^{(2)}-(t_r-t_{0,r})
\int_{\mathbbm{b}_k}\tilde{\Omega}_r^{(2)}
+\sum_{j=1}^g \int_{\mathbbm{b}_k}\omega^{(3)}_{\hat{\mu}_j(x,t_r),\hat {\mu}_j(x_0,t_{0,r})}+
\int_{\mathbbm{b}_k}\tilde{\omega}
\vspace{2mm}
\\ 
\D =
-(x-x_0)\int_{\mathbbm{b}_k}\Omega_2^{(2)}-(t_r-t_{0,r})
\int_{\mathbbm{b}_k}\tilde{\Omega}_r^{(2)}
\vspace{2mm}\\
\D \quad + 2\pi i \sum_{j=1}^g \int_{\hat{\mu}_j(x_0,t_{0,r})}^{\hat{\mu}_j(x,t_r)}\omega_k
+2\pi i \sum_{j=1}^g l_j 
\int_{\mathbbm{b}_k}\omega_j
\vspace{2mm}
\\
\D =-2\pi i(x-x_0)U_{2,k}^{(2)}-2\pi i(t_r-t_{0,r})\tilde {U}_{r,k}^{(2)}
\vspace{2mm}\\
\D \quad +
2\pi i\Bigl(
\sum_{j=1}^g \int_{Q_0}^{\hat{\mu}_j(x,t_r)}\omega_k-
\sum_{j=1}^g \int_{Q_0}^{\hat{\mu}_j(x_0,t_{0,r})}\omega_k
\Bigr)+
2\pi i \sum_{j=1}^g l_j\tau_{jk},\ 1\le k\le g.
\ea
\]
Thus, we arrive at
\be 
\underline{N}=
-(x-x_0)\underline{U}_2^{(2)}-(t_r-t_{0,r})\underline{\tilde {U}}_{r}^{(2)}
+\sum_{j=1}^g \int_{Q_0}^{\hat{\mu}_j(x,t_r)}\underline{\omega}-
\sum_{j=1}^g \int_{Q_0}^{\hat{\mu}_j(x_0,t_{0,r})}\underline{\omega}
+\underline{L}\tau,
\label{eq:arelationforrho_1:pma-quasiperiodicsols}
\ee
where $\underline{N}=(n_1,\cdots,n_g)\in \mathbb{Z}^g$ and $\underline{L}=(l_1,\cdots,l_g)\in \mathbb{Z}^g$.
The equation \eqref{eq:arelationforrho_1:pma-quasiperiodicsols}
exactly tells the first equality in 
\eqref{eq:lineareqnforrho_1:pma-quasiperiodicsols}.

Similarly, we can prove 
\eqref{eq:lineareqnforrho_2:pma-quasiperiodicsols} and \eqref{eq:lineareqnforrho_3:pma-quasiperiodicsols}
by using 
the other two meromorphic differentials $\Omega_2$ and $\Omega_3$, respectively. 
Only a difference is to change $
\sum_{j=1}^g \omega^{(3)}_{\hat{\mu}_j(x,t_r),\hat{\mu}_j(x_0,t_{0,r})}$
into 
 $
\sum_{j=0}^g \omega^{(3)}_{\hat{\nu}_j(x,t_r),\hat{\nu}_j(x_0,t_{0,r})}$
or  $
\sum_{j=0}^g \omega^{(3)}_{\hat{\xi}_j(x,t_r),\hat{\xi}_j(x_0,t_{0,r})}$
on the right hand side of
\eqref{eq:expforOmega_1:pma-quasiperiodicsols}, which brings the terms
$\underline{{\cal A}}(\hat {\nu}_0(x,t_r))$ and $\underline{{\cal A}}(\hat {\nu}_0(x_0,t_{0,r}))$ 
 in 
\eqref{eq:lineareqnforrho_2:pma-quasiperiodicsols}, and
$\underline{{\cal A}}(\hat {\xi}_0(x,t_r))$ and $\underline{{\cal A}}(\hat {\xi}_0(x_0,t_{0,r}))$ in \eqref{eq:lineareqnforrho_3:pma-quasiperiodicsols}.
The proof is finished. \hfill $\Box$

Now, we are able to present theta function representations of solutions of
the $r$-th four-component AKNS equations \eqref{eq:4cAKNSsh:pma-quasiperiodicsols}.

\begin{thm}
\label{thm:theta-functionRepresentationsofthesolution:pma-quasiperiodicsols}
 (Theta function representations of solutions)
Let $\Omega_\mu \subset \mathbb{C}^2$ be an open and connected set, 
$(x_0,t_{0,r}),(x,t_r)\in \Omega_\mu$, and $P=(\lambda ,y)\in \mathcal {K}_g \backslash \{P_{\infty_1},P_{\infty_2},P_{\infty_3}\}$. 
Suppose that 
${\cal K}_g$ is nonsingular and 
${\cal D}_{\underline{\hat{\mu}}(x,t_r)}$
or ${\cal D}_{\underline{\hat{\nu}}(x,t_r)}$ or 
${\cal D}_{\underline{\hat{\xi}}(x,t_r)}$ is nonspecial for 
$ (x,t_r)\in \Omega_\mu $. 
Then 
the solution
$u=(p_1,p_2,q_1,q_2)^T$ of
the $r$-th four-component AKNS equations \eqref{eq:4cAKNSsh:pma-quasiperiodicsols}
has the following theta function representations:
\bea
&&
p_1(x,t_r)=
p_1(x_0,t_{0,r})
\frac {\theta (\underline{z} (P_{\infty_1}, \underline{\hat{\mu}}(x,t_r) ) )
\theta (\underline{z} (P_{\infty_3}, \underline{\hat{\mu}}(x_0,t_{0,r}) ) )
}
{\theta (\underline{z} (P_{\infty_3}, \underline{\hat{\mu}}(x,t_r) ) )
\theta (\underline{z} (P_{\infty_1}, \underline{\hat{\mu}}(x_0,t_{0,r}) ) )
}
\nonumber 
\\ 
&&
\qquad \times 
\mathrm{exp}\Bigl(
\bigl( e_{2,3}^{(2)}(Q_0)-e_{2,1}^{(2)}(Q_0)
\bigr)(x-x_0)
 + 
\bigl( \tilde {e}_{r,3}^{(2)}(Q_0)-
\tilde {e}_{r,1}^{(2)}(Q_0)
\bigr)(t_r-t_{0,r})
\Bigr),
\label{eq:thetapresentationforp_1:pma-quasiperiodicsols}
\\
&&
p_2(x,t_r)=
p_2(x_0,t_{0,r})
\frac {\theta (\underline{z} (P_{\infty_2}, \underline{\hat{\mu}}(x,t_r) ) )
\theta (\underline{z} (P_{\infty_3}, \underline{\hat{\mu}}(x_0,t_{0,r}) ) )
}
{\theta (\underline{z} (P_{\infty_3}, \underline{\hat{\mu}}(x,t_r) ) )
\theta (\underline{z} (P_{\infty_2}, \underline{\hat{\mu}}(x_0,t_{0,r}) ) )
}
\nonumber 
\\ 
&&\qquad  \times 
\mathrm{exp}\Bigl(
\bigl( e_{2,3}^{(2)}(Q_0)-e_{2,2}^{(2)}(Q_0)
\bigr)(x-x_0)
 + 
\bigl( \tilde {e}_{r,3}^{(2)}(Q_0)-
\tilde {e}_{r,2}^{(2)}(Q_0)
\bigr)(t_r-t_{0,r})
\Bigr),\qquad \quad
\label{eq:thetapresentationforp_2:pma-quasiperiodicsols}
\eea
and 
\bea 
&&
q_1(x,t_r)=
q_1(x_0,t_{0,r})
\frac {\theta (\underline{z} (P_{\infty_3}, \underline{\hat{\nu}}(x,t_r) ) )
\theta (\underline{z} (P_{\infty_1}, \underline{\hat{\nu}}(x_0,t_{0,r}) ) )
}
{\theta (\underline{z} (P_{\infty_1}, \underline{\hat{\nu}}(x,t_r) ) )
\theta (\underline{z} (P_{\infty_3}, \underline{\hat{\nu}}(x_0,t_{0,r}) ) )
}
\nonumber 
\\ &&
\qquad 
\times \mathrm{exp}\Bigl(
\bigl( e_{2,1}^{(2)}(Q_0)-e_{2,3}^{(2)}(Q_0)
\bigr)(x-x_0)
 + 
\bigl( \tilde {e}_{r,1}^{(2)}(Q_0)-
\tilde {e}_{r,3}^{(2)}(Q_0)
\bigr)(t_r-t_{0,r})
\Bigr.
\nonumber 
\\
&& \qquad 
\qquad \ \Bigl.
+ e^{(3)}_{2,1}(Q_0,x,x_0,t_{r},t_{0,r})-e^{(3)}_{2,3}(Q_0,x,x_0,t_{r},t_{0,r})
\Bigr)
,\label{eq:thetapresentationforq_1:pma-quasiperiodicsols}
\\
&&
q_2(x,t_r)=
q_2(x_0,t_{0,r})
\frac {\theta (\underline{z} (P_{\infty_3}, \underline{\hat{\xi}}(x,t_r) ) )
\theta (\underline{z} (P_{\infty_2}, \underline{\hat{\xi}}(x_0,t_{0,r}) ) )
}
{\theta (\underline{z} (P_{\infty_2}, \underline{\hat{\xi}}(x,t_r) ) )
\theta (\underline{z} (P_{\infty_3}, \underline{\hat{\xi}}(x_0,t_{0,r}) ) )
}
\nonumber 
\\ && 
\qquad \times \mathrm{exp}\Bigl(
\bigl( e_{2,2}^{(2)}(Q_0)-e_{2,3}^{(2)}(Q_0)
\bigr)(x-x_0)
 + 
\bigl( \tilde {e}_{r,2}^{(2)}(Q_0)-
\tilde {e}_{r,3}^{(2)}(Q_0)
\bigr)(t_r-t_{0,r})
\Bigr.
\nonumber 
\\
&& \qquad 
\qquad\  \Bigl.
+ e^{(3)}_{3,2}(Q_0,x,x_0,t_{r},t_{0,r})-e^{(3)}_{3,3}(Q_0,x,x_0,t_{r},t_{0,r})
\Bigr)
.\label{eq:thetapresentationforq_2:pma-quasiperiodicsols}
\eea
\end{thm}

\noindent {\it Proof}:
Based on the asymptotic properties of $\Omega^{(2)}_2$ and 
$\tilde {\Omega}_r^{(2)}$
in \eqref{eq:asymptoticpropertyforOmega_2^{(2)}:pma-quasiperiodicsols} and 
\eqref{eq:asymptoticpropertyforOmega_r^{(2)}:pma-quasiperiodicsols},
and following Theorem \ref{thm:theta-functionRepresentationsoftheBAfunctions:pma-quasiperiodicsols}, we can 
expand 
the Baker-Akhiezer functions near the indicated points at infinity as follows: 
\bea
&& 
\psi_1 \mathop{=}\limits_{\zeta\to 0}
\frac {\theta (\underline{z} (P_{\infty_1}, \underline{\hat{\mu}}(x,t_r) ) )
\theta (\underline{z} (P_{\infty_3}, \underline{\hat{\mu}}(x_0,t_{0,r}) ) )
}
{\theta (\underline{z} (P_{\infty_3}, \underline{\hat{\mu}}(x,t_r) ) )
\theta (\underline{z} (P_{\infty_1}, \underline{\hat{\mu}}(x_0,t_{0,r}) ) )
}\times
\nonumber
\\
&&
\qquad 
\mathrm{exp}\Bigl(
\bigl( e_{2,3}^{(2)}(Q_0)-
e_{2,1}^{(2)}(Q_0)
\bigr)(x-x_0)
+
\bigl( \tilde {e}_{r,3}^{(2)}(Q_0)-
\tilde {e}_{r,1}^{(2)}(Q_0)
\bigr)(t_r-t_{0,r})
\Bigr.
\nonumber
\\
&&
\qquad 
\Bigl.
+\zeta^{-1}(x-x_0)+\zeta^{-r}(t_r-t_{0,r})+\textrm{O}(\zeta)
\Bigr)\bigl(1+\textrm{O}(\zeta)\bigr),\ \textrm{as}\ P\to P_{\infty_1},
\nonumber 
\\
&& 
\psi_1 \mathop{=}\limits_{\zeta\to 0}
\frac {\theta (\underline{z} (P_{\infty_2}, \underline{\hat{\mu}}(x,t_r) ) )
\theta (\underline{z} (P_{\infty_3}, \underline{\hat{\mu}}(x_0,t_{0,r}) ) )
}
{\theta (\underline{z} (P_{\infty_3}, \underline{\hat{\mu}}(x,t_r) ) )
\theta (\underline{z} (P_{\infty_2}, \underline{\hat{\mu}}(x_0,t_{0,r}) ) )
}\times
\nonumber
\\
&&
\qquad 
\mathrm{exp}\Bigl(
\bigl( e_{2,3}^{(2)}(Q_0)-
e_{2,2}^{(2)}(Q_0)
\bigr)(x-x_0)
+
\bigl( \tilde {e}_{r,3}^{(2)}(Q_0)-
\tilde {e}_{r,2}^{(2)}(Q_0)
\bigr)(t_r-t_{0,r})
\Bigr.
\nonumber
\\
&&
\qquad 
\Bigl.
+\zeta^{-1}(x-x_0)+\zeta^{-r}(t_r-t_{0,r})+\textrm{O}(\zeta)
\Bigr)\bigl(1+\textrm{O}(\zeta)\bigr),\ \textrm{as}\ P\to P_{\infty_2},
\nonumber 
\eea
and 
\bea 
&& 
\psi_2 \mathop{=}\limits_{\zeta\to 0}
\frac {\theta (\underline{z} (P_{\infty_3}, \underline{\hat{\nu}}(x,t_r) ) )
\theta (\underline{z} (P_{\infty_1}, \underline{\hat{\nu}}(x_0,t_{0,r}) ) )
}
{\theta (\underline{z} (P_{\infty_1}, \underline{\hat{\nu}}(x,t_r) ) )
\theta (\underline{z} (P_{\infty_3}, \underline{\hat{\nu}}(x_0,t_{0,r}) ) )
}
\mathrm{exp}\Bigl(
\bigl( e_{2,1}^{(2)}(Q_0)-
e_{2,3}^{(2)}(Q_0)
\bigr)(x-x_0)\Bigr.\qquad \qquad 
\nonumber
\\
&&
\qquad 
+
\bigl( \tilde {e}_{r,1}^{(2)}(Q_0)-
\tilde {e}_{r,3}^{(2)}(Q_0)
\bigr)(t_r-t_{0,r})+\bigl({e}_{2,1}^{(3)}(Q_0) - {e}_{2,3}^{(3)}(Q_0) \bigr)
\nonumber
\\
&&
\qquad 
\Bigl.
-2\zeta^{-1}(x-x_0)-2 \zeta^{-r}(t_r-t_{0,r})+\textrm{O}(\zeta)
\Bigr)\bigl(1+\textrm{O}(\zeta)\bigr),\ \textrm{as}\ P\to P_{\infty_3},
\nonumber 
\\
&& 
\psi_3 \mathop{=}\limits_{\zeta\to 0}
\frac {\theta (\underline{z} (P_{\infty_3}, \underline{\hat{\xi}}(x,t_r) ) )
\theta (\underline{z} (P_{\infty_2}, \underline{\hat{\xi}}(x_0,t_{0,r}) ) )
}
{\theta (\underline{z} (P_{\infty_2}, \underline{\hat{\xi}}(x,t_r) ) )
\theta (\underline{z} (P_{\infty_3}, \underline{\hat{\xi}}(x_0,t_{0,r}) ) )
}
\mathrm{exp}\Bigl(
\bigl( e_{2,2}^{(2)}(Q_0)-
e_{2,3}^{(2)}(Q_0)
\bigr)(x-x_0) \Bigr.
\nonumber
\\
&&
\qquad 
+
\bigl( \tilde {e}_{r,2}^{(2)}(Q_0)-
\tilde {e}_{r,3}^{(2)}(Q_0)
\bigr)(t_r-t_{0,r})
+\bigl({e}_{3,2}^{(3)}(Q_0) - {e}_{3,3}^{(3)}(Q_0) \bigr)
\nonumber
\\
&&
\qquad 
\Bigl.
-2\zeta^{-1}(x-x_0)-2 \zeta^{-r}(t_r-t_{0,r})+\textrm{O}(\zeta)
\Bigr)\bigl(1+\textrm{O}(\zeta)\bigr),\ \textrm{as}\ P\to P_{\infty_3}.
\nonumber
\eea
Now, comparing with the asymptotic behaviors of $\psi_1$ and $\psi_2$ and $\psi_3$ established in \eqref{eq:asymptoticbehavioursofpsi_1:pma-quasiperiodicsols}, \eqref{eq:asymptoticbehavioursofpsi_2:pma-quasiperiodicsols} and \eqref{eq:asymptoticbehavioursofpsi_3:pma-quasiperiodicsols}, respectively, we obtain 
the Riemann theta function presentations of $p_1,p_2,q_1$ and $q_2$ 
in \eqref{eq:thetapresentationforp_1:pma-quasiperiodicsols}-\eqref{eq:thetapresentationforq_2:pma-quasiperiodicsols}
immediately.  
This completes the proof of the theorem.
\hfill $\Box$

\section{Concluding remarks}

The paper is dedicated to development of explicit Riemann theta function representations of algebro-geometric solutions to entire soliton hierarchies.
We introduced a class of trigonal curves
based on linear combinations of Lax matrices in the zero curvature formulation, 
 and analyzed general properties of their meromorphic functions, including derivative relations between derivatives 
of the characteristic variables
with respect to time and space. 
We straightened out all
soliton flows under the Abel-Jacobi coordinates through
determining zeros and poles of the Baker-Akhiezer functions, and constructed the Riemann theta function representations for algebro-geometric solutions to the four-component AKNS equations
 from checking 
 asymptotic behaviors of the Baker-Akhiezer functions at the points at infinity.

We point out that we can similarly construct algebro-geometric solutions to a linear combination of different AKNS equations in the four-component AKNS soliton hierarchy, which just increases asymptotic complexity
(see, e.g., \cite{GengZD-AM2014}).
Various choices of linear combinations of Lax matrices lead to different algebro-geometric solutions to soliton hierarchies. 
However, it needs further investigation how to apply higher-order algebraic curves in finding algebro-geometric solutions to soliton equations. Higher-order matrix spectral problems  
 lead to tremendous difficulty in computing algebro-geometric solutions. More components in the vector of eigenfunctions will cause complicated situations while deriving asymptotic expansions for the Baker-Akhiezer functions. 

Reducing algebro-geometric solutions tells various classes of exact solutions to soliton equations \cite{BolokolosBEIM-book1994}. Two such classes of 
analytical solutions on the real field are quasi-periodic wave solutions 
\cite{MaZG-MPLA2009}
and lump solutions \cite{SatsumaA-JMP1979}. The study of lump solutions 
by bilinear techniques brings us the following open questions.
First, how can one determine 
positive definiteness (or positive semidefiniteness) for hypermatrices of even orders? 
For example, if one has a hypermatrix of order 4: 
$A=(a_{ijkl})_{n\times n\times n\times n}$, when does it satisfy
\[ \sum_{i,j,k,l=1}^n a_{ijkl} x_{i}x_{j} x_{k}x_l>0\  \, (\textrm{or}\,\ge 0)\] 
for all non-zero vector $x=(x_1,\cdots,x_n)\in \mathbb{R}^n$?
There is the same question on Hermitian positive definiteness (or Hermitian positive semidefiniteness) for hypermatrices of even orders.
Note that a multivariate polynomial $P$ has infinitely many zeros, if $P$ changes sign, i.e., 
there exist two points $x,y\in \mathbb{R}^n$ such that
$P(x)>0$ and $P(y)<0$. All multivariate polynomilas of odd orders are such examples. That is why we restrict orders of hypermatrices to even orders. Now a more general question is: 
When is a 
multivariate polynomial of even order positive (or non-negative)?
A more specific question is:
When does a multivariate polynomial of even order have a unique zero \cite{MaZTT-AMC2012}? Equivalently,  
how can one judge if
\[ P(x)\ge 0, \ \forall x\in    \mathbb{R}^n,\]
but there is just one zero $x_0\in \mathbb{R}^n$:
\[ P(x_0)=0,\ x_0\in \mathbb{R}^n,\]
for a multivariate polynomial $P$ of even order in $x=(x_1,\cdots,x_n)$?

There is also a conjecture on integrability of commuting soliton equations in a soliton hierarchy \cite{Ma-book2013}. It is known that there are infinitely many functionally independent Lie symmetries inherited from a recursion operator of a soliton hierarchy.
Those Lie symmetries yield an infinite number of one-parameter Lie groups of solutions to each equation in the underlying soliton hierarchy \cite{Olver-book1993}.
 It is conjectured \cite{Ma-book2013} that
those infinitely many one-parameter Lie groups of solutions form a dense subset of solutions in the whole solution set to each equation in the soliton hierarchy.
More specifically, let us denote a soliton hierarchy by
$u_{t_n}=K_n(u),\ n\ge 0.$ For a given equation $u_{t_r}=K_r(u)$,
we assume that
\[
\textrm{symmetry} \ K_n \ \Rightarrow \ \textrm{Lie group of solutions}\ S_n(\varepsilon_n),\  \varepsilon_n\in I_n=(a_n,b_n)\subseteq \mathbb{R},\]
where the existence intervals $I_n$, $n\ge 0$, might be small.  
Denote by $T_r$ the set of solutions to the $r$-th equation $u_{t_r}=K_r$, and
make a metric space $(T_r({D}),d)$ with a bounded domain ${D}$:
\[T_r({D})=\{f|_ {D}\,| \, f \in T_r\},\ d(f,g)= \textrm{sup}_{(x,t_r)\in {D}}|f(x,t_t)-g(x,t_r)|.\]
Is the union $\cup_{n=0}^\infty S_n(\varepsilon_n)$ dense in the metric space $(T_r({D}),d)$ with any bounded domain ${D}$ for each equation $u_{t_r}=K_r$ in the underlying soliton hierarchy?
If the answer is yes, the solution to a Cauchy problem of a soliton equation in a hierarchy can be approximated by the solutions constructed from those Lie commuting symmetries. Thus, soliton hierarchies present good models of integrable nonlinear
 partial differential equations from a computational point of view, indeed.

\vspace{0.3cm}
\noindent{\bf Acknowledgments}

The work was supported in
part by
 NSFC under the grants 11371326 and 11271008,
  and the Distinguished Professorships by both Shanghai University of Electric Power and Shanghai Second Polytechnic University. 
The authors would also like to thank M. Adler, S. T. Chen, X. Gu, S. Manukure, M. Mcanally, E. Previato, Y. Sun, F. D. Wang, J. Yu and Y. Zhou for their valuable discussions.

\end{document}